\documentclass[authoryear,5p,twocolumn]{elsarticle}
\usepackage{epsfig}
\usepackage{color}

\usepackage{amssymb}
\usepackage{amsmath}
\usepackage{amsthm}
\usepackage{multirow}
\usepackage{twoopt}
\usepackage{graphicx}

\newcommand{\proton}{\ensuremath{^1}\textrm{H}}
\newcommand{\deut}{\ensuremath{^2}\textrm{H}}

\newcommand{\het}{\ensuremath{^3}\textrm{He}}
\newcommand{\hef}{\ensuremath{^4}\textrm{He}}

\usepackage[pdftex,             
            breaklinks=true,%
            colorlinks=true,%
            pdfauthor={Cheminet et al.},%
            pdftitle={Template for manuscripts in Advances in Space Research}%
           ]{hyperref}

\makeatletter
\newcommandtwoopt{\citeads}[3][][]{\href{http://adsabs.harvard.edu/abs/#3}{\def\hyper@linkstart##1##2{}\let\hyper@linkend\@empty\citealp[#1][#2]{#3}}}
\newcommandtwoopt{\citepads}[3][][]{\href{http://adsabs.harvard.edu/abs/#3}{\def\hyper@linkstart##1##2{}\let\hyper@linkend\@empty\citep[#1][#2]{#3}}}
\newcommandtwoopt{\citetads}[3][][]{\href{http://adsabs.harvard.edu/abs/#3}{\def\hyper@linkstart##1##2{}\let\hyper@linkend\@empty\citet[#1][#2]{#3}}}
\newcommandtwoopt{\citealpads}[3][][]{\href{http://adsabs.harvard.edu/abs/#3}{\def\hyper@linkstart##1##2{}\let\hyper@linkend\@empty\citealp[#1][#2]{#3}}}
\newcommandtwoopt{\citealtads}[3][][]{\href{http://adsabs.harvard.edu/abs/#3}{\def\hyper@linkstart##1##2{}\let\hyper@linkend\@empty\citealt[#1][#2]{#3}}}
\newcommandtwoopt{\citeyearads}[3][][]{\href{http://adsabs.harvard.edu/abs/#3}{\def\hyper@linkstart##1##2{}\let\hyper@linkend\@empty\citeyear[#1][#2]{#3}}}
\newcommandtwoopt{\citeadsstar}[3][][]{\href{http://adsabs.harvard.edu/abs/#3}{\def\hyper@linkstart##1##2{}\let\hyper@linkend\@empty\citealp*[#1][#2]{#3}}}
\newcommandtwoopt{\citepadsstar}[3][][]{\href{http://adsabs.harvard.edu/abs/#3}{\def\hyper@linkstart##1##2{}\let\hyper@linkend\@empty\citep*[#1][#2]{#3}}}
\newcommandtwoopt{\citetadsstar}[3][][]{\href{http://adsabs.harvard.edu/abs/#3}{\def\hyper@linkstart##1##2{}\let\hyper@linkend\@empty\citet*[#1][#2]{#3}}}
\newcommandtwoopt{\citeyearadsstar}[3][][]{\href{http://adsabs.harvard.edu/abs/#3}{\def\hyper@linkstart##1##2{}\let\hyper@linkend\@empty\citeyear*[#1][#2]{#3}}}
\newcommandtwoopt{\citeauthoradsstar}[3][][]{\href{http://adsabs.harvard.edu/abs/#3}{\def\hyper@linkstart##1##2{}\let\hyper@linkend\@empty\citeauthor*[#1][#2]{#3}}}
\newcommandtwoopt{\citepthesis}[3][][]{\href{http://tel.archives-ouvertes.fr/docs/#3}{\def\hyper@linkstart##1##2{}\let\hyper@linkend\@empty\citep[#1][#2]{#3}}}
\newcommandtwoopt{\citetthesis}[3][][]{\href{http://tel.archives-ouvertes.fr/docs/#3}{\def\hyper@linkstart##1##2{}\let\hyper@linkend\@empty\citet[#1][#2]{#3}}}
\makeatother

\journal{Advances in Space Research}

\begin{document}

\begin{frontmatter}



\title{Neutron monitors and muon detectors for solar modulation studies: Interstellar flux, yield function, and assessment of critical parameters in count rate calculations}

\author[lpsc]{D.Maurin\corref{cor}}
\cortext[cor]{Corresponding author; Tel.: +33476284082; fax: +33476284004}
\ead{dmaurin@lpsc.in2p3.fr}


\author[onera,irsn]{A. Cheminet}
\ead{Adrien.Cheminet@onera.fr}

\author[lpsc]{L. Derome}
\ead{derome@lpsc.in2p3.fr}

\author[lpsc]{A. Ghelfi}
\ead{ghelfi@lpsc.in2p3.fr}

\author[onera]{G. Hubert}
\ead{Guillaume.Hubert@onera.fr}

\address[lpsc]{LPSC, Universit\'e Grenoble-Alpes, CNRS/IN2P3,
      53 avenue des Martyrs, 38026 Grenoble, France}
\address[onera]{ONERA (French Aerospace Lab), 2 avenue Edouard Belin, 31055 Toulouse Cedex~4, France}
\address[irsn]{IRSN (Institute for Radiological Protection and Nuclear Safety),
CE Cadarache, B\^at. 159, 13115 Saint-Paul-Lez-Durance Cedex, France\vspace{-0.4cm}}

\begin{abstract}
Particles count rates at given Earth location and altitude result from the convolution of (i) the interstellar (IS) cosmic-ray fluxes outside the solar cavity, (ii) the time-dependent modulation of IS into Top-of-Atmosphere (TOA) fluxes, (iii) the rigidity cut-off (or geomagnetic transmission function) and grammage at the counter location, (iv) the atmosphere response to incoming TOA cosmic rays (shower development), and (v) the counter response to the various particles/energies in the shower. Count rates from neutron monitors or muon counters are therefore a proxy to solar activity. In this paper, we review all ingredients, discuss how their uncertainties impact count rate calculations, and how they translate into variation/uncertainties on the level of solar modulation $\phi$ (in the simple Force-Field approximation). The main uncertainty for neutron monitors is related to the yield function. However, many other effects have a significant impact, at the $5-10\%$ level on $\phi$ values. We find no clear ranking of the dominant effects, as some depend on the station position and/or the weather and/or the season. An abacus to translate any variation of count rates (for neutron and $\mu$ detectors) to a variation of the solar modulation $\phi$ is provided.
\end{abstract}


\end{frontmatter}

\parindent=0.5 cm

\section{Introduction}

After the discovery of cosmic rays (CR) by Hess in 1912, ground-based CR detectors located at various latitudes, longitudes and altitudes, played a major role to determine the CR composition and spectrum (see \citealtads{2009AdSpR..44.1081S} for a historical perspective). From the 50's, networks of neutron monitors \citepads{2000SSRv...93...11S} and muon telescopes \citepads{2000SSRv...93..207D} were developed. They provide today one of the most valuable data to inspect time variations of the integrated CR flux in the $10-100$ GeV/n range.

The formal link between these variations and the Sun activity was established in the mid-fifties, by means of a transport equation of CR fluxes in the solar cavity (\citeads{1965P&SS...13....9P}, \citealtads{1966ApJ...146..480J}). In the 80's, the effect of particle drift was shown to be responsible to a charge-sign dependent modulation \citepads{2013LRSP...10....3P}, following the Sun polarity cycle\footnote{A 11-yr average periodicity was established $\sim 250$ yrs ago from sunspot series \citep[see][for a review]{2007AdSpR..40..929V,2013LRSP...10....1U}. The now well observed 22-yr cycle (polarity reversal every 11 yrs) was first hinted at from magnetograph observations by \citetads{1961ApJ...133..572B}.}. However, the Force-Field approximation \citepads{1967ApJ...149L.115G,1968ApJ...154.1011G} has remained widely used thanks to its simplicity: this approximation, used in this work, has only one parameter $\phi(t)$.

Several strategies have been developed for time series reconstruction of the modulation level $\phi(t)$, and/or CR TOA fluxes at any time (of interest for many applications):
\begin{itemize}
  \item Using spacecraft measurements \citepads[e.g.,][]{2001JGR...10629979D,2011P&SS...59..355B,2013SoPh..284..599B}: it is the most direct approach, but the time coverage is limited to a few decades with a poor sampling;
  
  \item Comparison of calculated and measured count rates in ground-based detectors~\citepads{1999CzJPh..49.1743U,2002SoPh..207..389U,2005JGRA..11012108U,2011JGRA..116.2104U}: it covers a larger period (60~yrs), with a very good time resolution (a few minutes)\footnote{In the same spirit, the concentration of the cosmogenic radionuclide $^{10}$Be in ice cores \citepads{2003JGRA..108.1355W,2010JGRA..11500I20H} covers several thousands of years, but with a poor time resolution.};
  
  \item Extracting relationships between the modulation level and solar activity proxies, based on empirical (\citealtads{1994AdSpR..14..749B,1996AdSpR..17....7B}; \citealtads{2006AdSpR..37.1727O,2010ITNS...57.3148O}) or semi-empirical (\citealt{R.A-1992}; \citealtads{1994AdSpR..14..759N,1996AdSpR..17...19N}; \citealtads{1997ITNS...44.2150T}; \citealtads{2007AdSpR..40..313N}; \citealtads{2013AdSpR..52.2112A}) approaches.  
  \end{itemize}
All these strategies provide a satisfactory description of CR fluxes, though some fare better than others (for comparisons, see \citealtads{2010AdSpR..45.1026B,2012JGRA..117.8109M,2013JGRA..118.1837Z,2013AdSpR..51..329M}). Note also that empirical methods are expected to provide effective and less meaningful values for $\phi$  \citepads{2006AdSpR..37.1727O}.

In this paper, we focus on the second strategy, for a systematic study of the main uncertainties affecting the calculation of expected count rates in NM and muon detectors. This requires the description of the atmosphere and of the ground-based detector responses to incoming CRs \citep[e.g.,][]{2000SSRv...93..335C}. The various uncertainties, described in the \citetads{1974crvs.book.....D,2004ASSL..303.....D,2009crme.book.....D} textbooks, are generally discussed separately in research articles (uncertainty on the yield function, geomagnetic rigidity cutoff, seasonal effects\dots), and not propagated to the modulation parameter. For this reason, we believe it is useful to  recap and  gather them in a single study, re-assess which ones are the most important, and link these uncertainties to the expected level of variation/uncertainty they imply on the modulation level $\phi(t)$. The complementarity (different uncertainties and time coverage) of NM count rates and TOA CR flux measurements to obtain time-series of the solar modulation parameters is left to a second study\footnote{Recent CR instruments such as \href{http://pamela.roma2.infn.it}{PAMELA} and the \href{http://www.ams02.org}{AMS-02} on the International Space Station are or will be providing high-statistics fluxes on an unprecedented time frequency, which renders this comparison even more appealing.}.  

The paper is organised as follows: we start with a general presentation of the ingredients involved in the count rate calculations (Sect.~\ref{sec:count_rate}), and discuss a new fit for the IS fluxes (Sect.~\ref{sec:ref_is}). We then detail the calculation of the propagation in the atmosphere, providing a new yield function parametrisation (Sect.~\ref{sec:Yield}). Combining these inputs allows us to link the count rate variation with the solar modulation parameter, and to study the various sources of uncertainties (Sect.~\ref{sec:impact_factors}). The final ranking of the uncertainties in terms of both count rates and $\phi$ concludes this study (Sect.~\ref{sec:concl}).

\section{From IS fluxes to ground-based detector count rates}
\label{sec:count_rate}

A ground-based detector ${\cal D}$ at geographical coordinate $\vec{r}=(\varphi,\,\lambda,\,h)$ measures, at time $t$, a count rate per unit interval $N^{\cal D}(\vec{r},t)$, from the production (from CRs) of secondary particles in the atmosphere (atmospheric shower):
\begin{equation}
\label{eq:count-rate}
\!\!N^{\cal D}({\vec{r}},t) \!=\!\!\! \int_{0}^{\infty} \!\!\!\!{\cal T}(R,\vec{r}, t) 
  \times \!\!\!\!\sum_{i={\rm CRs}} \!\!\!{\cal Y}^{\cal D}_i(R,h)\, \frac{dJ_{i}^{\rm TOA\!\!\!}}{dR}(R,t)  \;dR,
\end{equation}
with $R=pc/Ze$, and $i$ running on CR species:
\begin{itemize}
   \item ${\cal T}(R,\vec{r}, t)$ is the transmission function in the geomagnetic field, which depends on the detector location and can vary with time (Sect.~\ref{sec:Rcut});
   \item ${\cal Y}^{\cal D}_{i}(R,h)$ is the yield function at altitude $h$, i.e. the detector response (in count m$^2$~sr) to a unit intensity of primary CR species $i$ at rigidity $R$ (Sect.~\ref{sec:Yield});
   \item $dJ_{i}^{\rm TOA}/dR$ is the top-of-atmosphere (TOA) modulated differential flux (or intensity) per rigidity interval $dR$, at rigidity $R$, time $t$, and for the CR species $i$ (in m$^{-2}$~s$^{-1}$~sr$^{-1}$~GV$^{-1}$).
   It is obtained from the interstellar flux $J^{\rm IS}$ (Sects.~\ref{sec:is_flux} and \ref{sec:ref_is}) modulated by a solar modulation model (Sect.~\ref{sec:force-field}).
\end{itemize}

In the most general case, ${\cal T}$ and ${\cal Y}$ above are entangled, due to the complex structure of the geomagnetic field, and the dependence of the transmission factor and the yield function on the primary particle incident angle (see Sect.~\ref{sec:Rcut}). A common practice is to consider the two terms independently, average the yield function over a few incident angles, and take a simple rigidity (or equivalently energy) effective vertical cutoff for the transmission function \citepads[see, e.g.,][for definitions]{1991NCimC..14..213C}. In this paper, unless stated otherwise, this is what we assume, and the effective vertical cutoff rigidity $R^{\rm eff}_c$ is referred to as the rigidity cut-off $R_c$ for short. 

Note that a full review of the subject|including count rate detector calculations and measurements, geomagnetic and magnetospheric variations, yield functions, the theory of CR meteorological effects, etc.|is given in the comprehensive monographs of \citetads{1974crvs.book.....D,2004ASSL..303.....D,2009crme.book.....D}.

\subsection{Interstellar flux}
\label{sec:is_flux}
At high energy ($\gtrsim 50$~GeV/n), the effect of solar modulation is negligible, and the IS spectra is directly obtained from CR data measurements. The recent PAMELA \citepads{2011Sci...332...69A} and CREAM \citepads{2010ApJ...714L..89A} data hint at a hardening of the spectrum above a few hundreds of GeV/n. However, preliminary AMS-02 results (shown at ICRC 2013 in Rio) seem to indicate otherwise. In any case, the CR contribution to count rates in NMs (resp. $\mu$ detectors) above 1 (resp. 10) TeV/n is negligible, whereas CRs above 100 (resp. 500) GeV/n contribute to $\lesssim10\%$ of the total (see Fig.~\ref{fraction_per_rbin}). Hence, the results in this paper are not very sensitive to the exact high energy dependence of the IS fluxes. Waiting for a clarification, we assume that a pure power law prevails up to the highest energies. 

At lower energy, fluxes are modulated by the solar activity (Sect.~\ref{sec:force-field}). Measurements at different times and/or different positions in the solar cavity (e.g., \citealtads{2008JGRA..11310108W,2009JGRA..11402103W}) allow to get the IS spectrum down to several hundreds of GeV/n, whereas other proxies can push this limit down to a few tens of MeV/n: actually (i) indirect measurements from CR ionisation in the ISM \citepads{1987A&A...179..277W,1994MNRAS.267..447N,1998ApJ...506..329W}; (ii) the impact of CR on molecules formation \citepads{2009A&A...501..619P,2012ApJ...745...91I,2012MNRAS.425L..86N}, and (iii) $\gamma$-ray emissions in molecular clouds \citetads{2012PhRvL.108e1105N}, seem to favour a low-energy flattening/break. A recent and exciting development is provided by the Voyager~1 spacecraft, which is witnessing what is believed to be the first direct measurement of the local interstellar spectrum in the $10-100$ MeV/n energy range \citepads{2013arXiv1308.6598W,2013arXiv1308.1895W,2013arXiv1308.4426W}. 

\subsection{Force-Field approximation for solar modulation}
\label{sec:force-field}
The force-field approximation was first derived by \citetads{1967ApJ...149L.115G,1968ApJ...154.1011G}. A simpler derivation is provided, e.g., in \citetads{1987A&A...184..119P} and \citetads{1998APh.....9..261B}, and the force-field approach limitation is discussed in \citetads{2004JGRA..109.1101C}. It provides an analytical one-to-one correspondence between TOA and IS energies, and also fluxes. For a given species (mass number $A$ and charge $Z$), at any given time, we have ($E$ is the total energy, $p$ the momentum, $T_{/n}$ the kinetic energy per nucleon, and $J\equiv dJ/dT_{/n}$ is the CR intensity with respect to $T_{/n}$):
\begin{eqnarray}
\label{eq:forcefield}
  \frac{E^{\rm TOA}}{A}&=&\frac{E^{\rm IS}}{A} - \frac{|Z|}{A} \phi\;, \\
  J^{\rm TOA} \left( E^{\rm TOA} \right)&=&
  \left( \frac{p^{\rm TOA}}{p^{\rm IS}} \right)^{2} \times J^{\rm IS}  \left( E^{\rm IS} \right), \nonumber
\end{eqnarray}
where the solar modulation parameter $\phi(t)$ has the  dimension of a rigidity (or an electric potential). Equation~(\ref{eq:forcefield}) amounts to both an energy and flux shift of the IS quantities (toward smaller values) to get TOA ones. We recall that $\Phi=|Z|/A\times \phi$ is sometimes used instead of $\phi$ (used throughout the paper).
\begin{table*}[!th]
\caption{Best-fit parameters $C_0$, $C_1$, and $C_2$ in Eq.~(\ref{eq:is_flux}) for all IS CR flux elements $j$ from H to Fe. The last two columns (calculated at 10~GV) are the fraction of the flux, and of the contribution of each species to NM count rates Eq.~(\ref{eq:frac_secies}). See text for details.}
\label{tab:jis_fit}
\begin{center}
\begin{tabular}{lccccc}\hline
  CR        &   $C_0$  & $C_1$ & $C_2$   & $\displaystyle f_j=\frac{J_j}{\sum_i J_i}$ & $\displaystyle \langle f_j\rangle_A=\frac{A_j J_j}{\sum_i A_i J_i}$ \vspace{0.15cm}\\
&(m$^2$ s sr GV)$^{-1}$& -&-&(\%)&  (\%) \vspace{0.1cm}\\\hline
  H$^\ddagger$ & $23350 \pm 184~~  $ &  $2.10 \pm 0.10 $ & $-2.839 \pm0.003 $ &   83.7    &   49.1    \\ 
\!\!\!\deut{}  & $838.5 \pm 29.5~  $ &  $3.62 \pm 0.08 $ & $-2.950 \pm0.060 $ &   2.21    &   2.59    \\ 
\!\!\!\het{}   & $512.7 \pm 2.0~~  $ &  $6.70 \pm 0.01 $ & $-3.045 \pm0.003 $ &   1.08    &   1.90    \\ 
  He$^\ddagger$& $3657. \pm 38.5~  $ &  $1.77 \pm 0.04 $ & $-2.782 \pm0.003 $ &   14.6    &   34.4    \\ 
  Li           & $18.86 \pm 0.86~  $ &  $4.58 \pm 0.07 $ & $-3.200 \pm0.400 $ &   0.027   &   0.11    \\ 
  Be           & $22.09 \pm 0.18~  $ &  $6.57 \pm 0.08 $ & $-2.948 \pm0.003 $ &   0.054   &   0.29    \\  
  B            & $72.77 \pm 0.32~  $ &  $5.77 \pm 0.01 $ & $-3.086 \pm0.002 $ &   0.132   &   0.85    \\ 
  C            & $116.5 \pm 0.50~  $ &  $4.26 \pm 0.01 $ & $-2.791 \pm0.002 $ &   0.438   &   3.08    \\
  N            & $45.7  \pm 0.45   $ &  $5.19 \pm 0.02 $ & $-2.971 \pm0.004 $ &   0.112   &   0.92    \\
  O            & $95.5  \pm 0.35   $ &  $3.87 \pm 0.01 $ & $-2.733 \pm0.002 $ &   0.413   &   3.88    \\ 
  F            & $35.73 \pm 0.03~  $ &  $5.63 \pm 0.02 $ & $-2.979 \pm0.005 $ &   0.084   &   0.94    \\ 
  Ne           & $16.75 \pm 0.10~  $ &  $4.29 \pm 0.02 $ & $-2.779 \pm0.003 $ &   0.065   &   0.76    \\ 
  Na           & $4.945 \pm 0.034  $ &  $5.02 \pm 0.02 $ & $-2.922 \pm0.003 $ &   0.013   &   0.18    \\ 
  Mg           & $20.25 \pm 0.10~  $ &  $4.03 \pm 0.02 $ & $-2.755 \pm0.003 $ &   0.083   &   1.17    \\ 
  Al           & $4.165 \pm 0.029  $ &  $4.65 \pm 0.02 $ & $-2.812 \pm0.004 $ &   0.015   &   0.23    \\ 
  Si           & $13.5  \pm 0.10   $ &  $3.86 \pm 0.02 $ & $-2.681 \pm0.003 $ &   0.066   &   1.08    \\ 
  P            & $1.084 \pm 0.014  $ &  $5.99 \pm 0.04 $ & $-2.938 \pm0.007 $ &   0.003   &   0.05    \\ 
  S            & $3.445 \pm 0.025  $ &  $4.87 \pm 0.02 $ & $-2.785 \pm0.004 $ &   0.013   &   0.24    \\ 
  Cl           & $1.428 \pm 0.015  $ &  $6.65 \pm 0.03 $ & $-3.052 \pm0.007 $ &   0.003   &   0.06    \\ 
  Ar           & $2.64  \pm 0.04   $ &  $6.24 \pm 0.04 $ & $-3.075 \pm0.007 $ &   0.005   &   0.11    \\ 
  K            & $2.192 \pm 0.005  $ &  $6.37 \pm 0.06 $ & $-3.110 \pm0.010 $ &   0.004   &   0.09    \\ 
  Ca           & $3.70  \pm 0.03   $ &  $5.24 \pm 0.05 $ & $-2.991 \pm0.005 $ &   0.009   &   0.20    \\
  Sc           & $1.106 \pm 0.019  $ &  $5.68 \pm 0.04 $ & $-3.120 \pm0.008 $ &   0.002   &   0.05    \\ 
  Ti           & $3.126 \pm 0.032  $ &  $4.96 \pm 0.03 $ & $-3.062 \pm0.005 $ &   0.006   &   0.17    \\ 
  V            & $1.357 \pm 0.014  $ &  $4.82 \pm 0.03 $ & $-2.995 \pm0.006 $ &   0.003   &   0.09    \\ 
  Cr           & $2.271 \pm 0.019  $ &  $4.51 \pm 0.03 $ & $-2.919 \pm0.005 $ &   0.006   &   0.19    \\
  Mn           & $1.132 \pm 0.014  $ &  $4.11 \pm 0.03 $ & $-2.776 \pm0.005 $ &   0.004   &   0.14    \\ 
  Fe           & $8.032 \pm 0.046  $ &  $3.37 \pm 0.03 $ & $-2.600 \pm0.001 $ &   0.047   &   1.55    \\
  Co           & $0.0055\pm 0.0037 $ &  $3.54 \pm 0.03 $ & $-2.610 \pm0.010 $ &$<10^{-3}$ &$<10^{-2}$ \\ 
  Ni           & $8.405 \pm 0.019  $ &  $4.50 \pm 0.10 $ & $-2.600 \pm0.020 $ &   0.002   &   0.08    \\
\hline
\end{tabular}
\\$^\ddagger$ ${\rm H}=$ \proton{}+\deut{}, and ${\rm He}=$ \het{}+\hef{}.
\end{center}
\end{table*}

\section{Determination of IS fluxes: from H to Ni}
\label{sec:ref_is}

Due to the interplay between the CR relative abundances and the yield function, the most important primary CR contributors to the count rates are protons, heliums, and heavier nuclei (in a small but non negligible fraction). In recent studies, in addition to proton and helium, the contribution of species heavier than He is accounted for as an effective enhancement of the He flux \citepads{2003JGRA..108.1355W,2011JGRA..116.2104U}.

In order to assess the uncertainties (on count rate calculations) associated with IS fluxes and heavy species, we propose a new fit based on recent TOA CR measurements. For the H and He fluxes, we also compare our results with previous parametrisations found in the literature.

\subsection{Fit function and best-fit values}
Following \citetads{2007APh....28..154S}, the parametrisation of the IS flux is taken to be:
\begin{equation}
 \frac{dJ^{\rm IS}}{dE_{k/n}} = c_0 \beta^{c_1} R^{-c_2}.
 \label{eq_is_flux_dekn}
\end{equation}
For practical purposes (all integrations are performed over rigidity), we use 
\begin{eqnarray}
\label{eq:is_flux}
 \frac{dJ^{\rm IS}}{dR} &=& C_0  \times \beta^{C_1} \times  R^{-C_2} \\
                        &=& c_0\frac{Z}{A} \times \beta^{(c_1+1)}  \times R^{-c_2}.\nonumber
\end{eqnarray}
The parameters and their errors are obtained from the {\sc minuit} minimisation package \citepads{1975CoPhC..10..343J}\footnote{\url{www.cern.ch/minuit}}, implemented in the {\sc root} CERN libraries\footnote{\url{http://root.cern.ch/drupal}}.

\paragraph{CR data selection}
The data we base the fit on are retrieved from the cosmic-ray data base\footnote{\url{http://lpsc.in2p3.fr/crdb}} \citepads{2013arXiv1302.5525M}. Many experiments have measured H and He, but before 1998, most of them were found inconsistent with one another. As a result, we chose to use the most recent data only, relying mostly on space-based experiments (which do not suffer from systematics related to interactions in the residual atmosphere), i.e. AMS-01 \citepads{2000PhLB..490...27A,2000PhLB..494..193A,2002PhR...366..331A}, and PAMELA \citepads{2010PhRvL.105l1101A,2011Sci...332...69A,2013ApJ...765...91A,2013ApJ...770....2A}, and only the most recent BESS balloon flights (BESS00: \citealtads{2007APh....28..154S}; BESS-TeV: \citealtads{2013AdSpR..51..234K}). Fewer experiments have measured heavier species, in particular up to Ni. We rely on HEAO3-C2 \citepads{1988A&A...193...69F,1990A&A...233...96E} and ACE-CRIS \citepads{2006AdSpR..38.1558D,2009ApJ...698.1666G,2013ApJ...770..117L} data. We also fit the \deut{} and \het{} fluxes: due to the scarcity of data, we have no choice here, but to rely on many experiments, i.e. AMS01 \citepads{2002PhR...366..331A,2011ApJ...736..105A},
BESS93 \citepads{2002ApJ...564..244W}, BESS94, 95, 97, and 98 \citepads{2005AdSpR..35..151M}, BESS00 \citepads{2013AdSpR..51..234K}, CAPRICE94 \citepads{1999ApJ...518..457B}, CAPRICE98 \citepads{2004ApJ...615..259P},
IMAX92 \citepads{2000AIPC..528..425D,2000ApJ...533..281M}, and PAMELA \citepads{2013ApJ...770....2A} measurements.

\paragraph{Best-fit values}
The parameters $c_1$, $c_2$, and $c_3$ of Eq.~(\ref{eq_is_flux_dekn}) are simultaneously fitted to H and He data, and then up to Fe data, having a single modulation level for each data taking period. This is necessary to reduce the degeneracy between the chosen IS flux parametrisation and the modulation parameter (see Sect.~\ref{sec:degeneracy}). The best-fit parameters and their error are gathered in Table~\ref{tab:jis_fit}, and the value $\phi$ for each epoch are given in Table~\ref{tab:phi_fitted}. Note that the uncertainties for heavy nuclei are probably underestimated since the fit is based on a single set of data (HEAO3-C2) for energies above a few GeV/n. The next to last column represents (at 10 GV) the fraction of a given CR flux to the sum of all contributions. The last column gives, at the same rigidity, a gross estimate of the yield weighted contribution $\langle f_j\rangle_{\cal Y}$ of any species $j$ to NM count rates:
\begin{eqnarray}
   \langle f_j\rangle_{\cal Y}&=& \frac{{\cal Y}_j(R) \times J^{\rm TOA}_j(R)}{\displaystyle\sum_{\rm i=CRs} {\cal Y}_{i}(R) \times J^{\rm TOA}_{i}(R)}\,,\\
   \langle f_j\rangle_{\cal Y} &\approx&\langle f_j\rangle_A= \frac{A_j J^{\rm TOA}_j}{\displaystyle\sum_{\rm i=CRs} A_i J^{\rm TOA}_{i}}\,.\nonumber
   \label{eq:frac_secies}
\end{eqnarray}
Species heavier than Ni are not considered because they provide a negligible contribution to count rates\footnote{Their abundance ranges from $10^{-4}$ Fe for Zn to $10^{-7}$ Fe for U \citepads{1989ApJ...346..997B,2003ApJ...591.1220L,2009ApJ...697.2083R}.}.

\begin{table}
\caption{Best-fit modulation parameters for each experiment considered, along with the published values (if exists) for illustration.}
\label{tab:phi_fitted}
\begin{center}
\begin{tabular}{lllc}
\hline
Experiment& Period         & $\phi_{\rm fit}$ & $\phi_{\rm publi.}$\\
     -    &     -          &    (MV)      &  (MV)\\
\hline
HEAO3-C2  & 1979/10-1980/06 &    $648\pm13$    &600 \\
IMAX92    & 1992/07         &    $698\pm9 $    &750 \\
CAPRICE94 & 1994/08         &    $671\pm12$    &710 \\
ACE-CRIS  & 1997/08-1998/04 &    $227\pm10$    &325 \\
ACE-CRIS  & 1998/01-1999/01 &    $396\pm9 $    &550 \\
CAPRICE98 & 1998/05         &    $502\pm17$    &600 \\
AMS-01    & 1998/06         &    $648\pm9 $    &650 \\
BESS00    & 2000/08         &$\!\!\!1339\pm14$ &\!\!\!1300\\
BESS-TeV  & 2002/08         &$\!\!\!1004\pm11$ &\!\!\!1109\\
ACE-CRIS  & 2001/05-2003/09 &    $796\pm18$    &900 \\
PAMELA    & 2006/11-2006/12 &    $391\pm4 $    &  - \\
PAMELA    & 2007/11-2007/12 &    $426\pm4 $    &  - \\
PAMELA    & 2006/07-2008/12 &    $461\pm6 $    &500 \\
PAMELA    & 2006/07-2009/12 &    $522\pm5 $    &  - \\
PAMELA    & 2008/11-2008/12 &    $399\pm5 $    &  - \\
ACE-CRIS  & 2009/03-2010/01 &    $188\pm15$    &250 \\
PAMELA    & 2009/12-2010/01 &    $248\pm6 $    &  - \\
\hline
\end{tabular}
\end{center}
\end{table}

\paragraph{Relative importance of species}
\begin{figure}[!t]
\begin{center}
\includegraphics[width=\columnwidth]{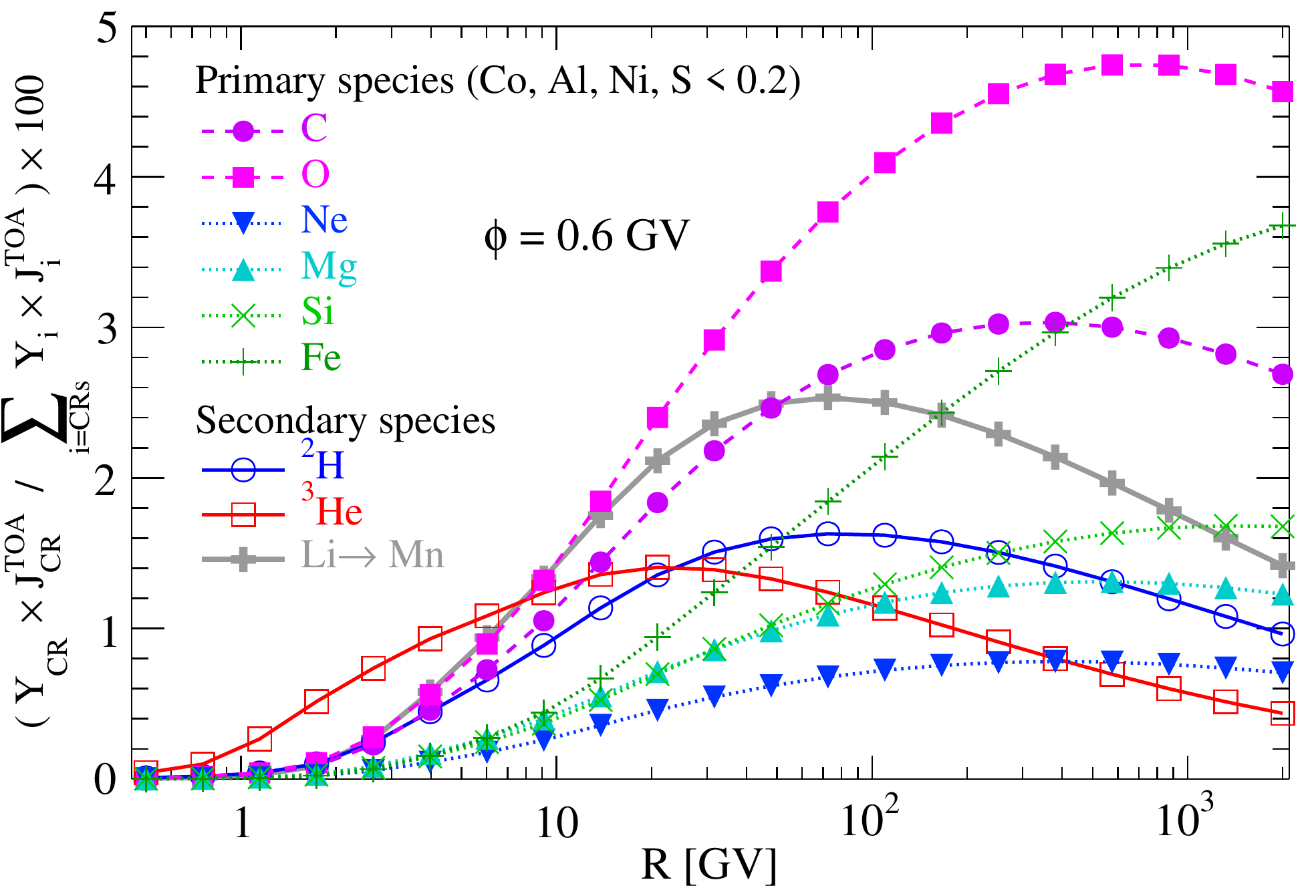}
\includegraphics[width=\columnwidth]{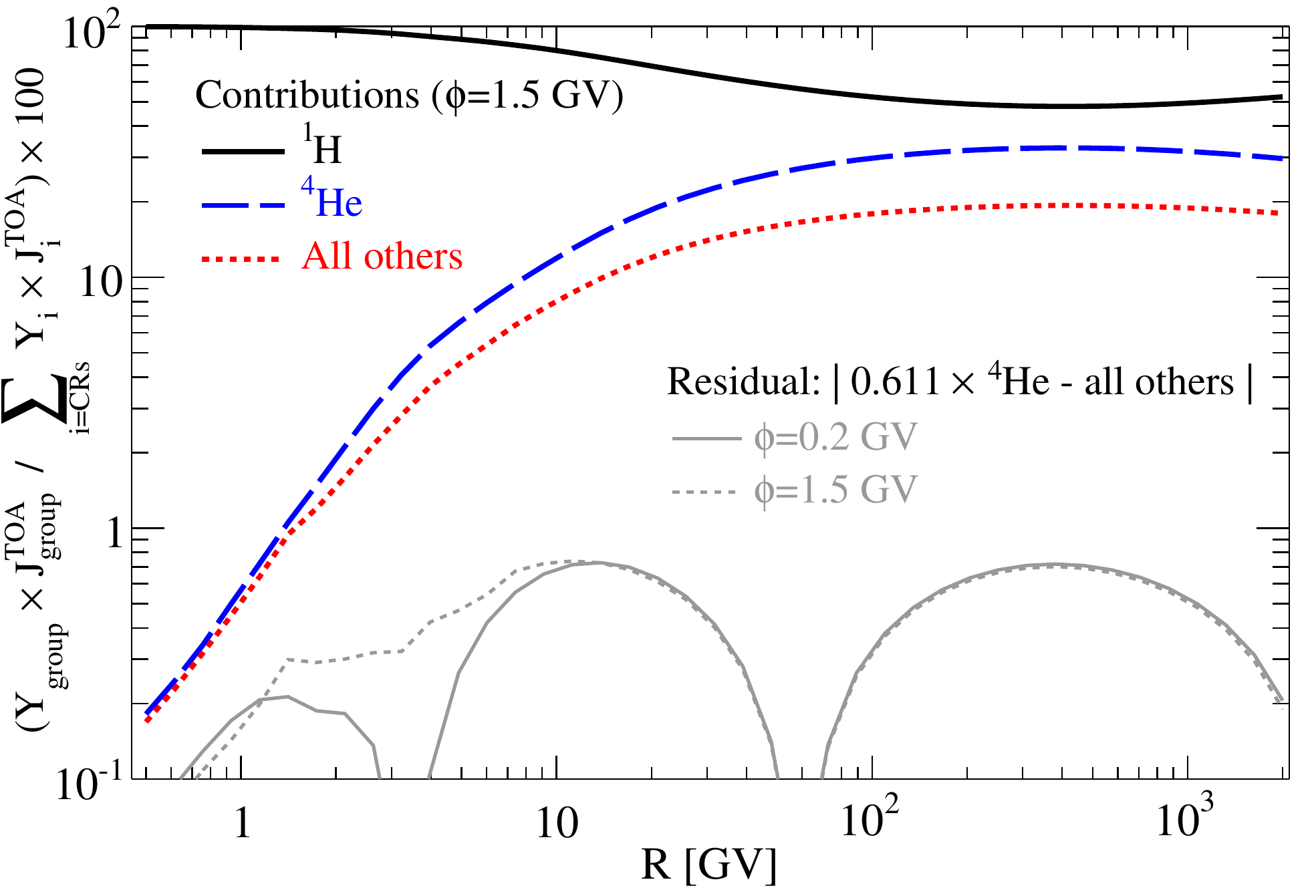}
\end{center}
\caption{Fractional contribution w.r.t. the total (all species) as a function of rigidity, see Eq.~(\ref{eq:frac_secies}). {\bf Top panel:} zoom in on contributions larger than 0.2\% for secondary species and for primary species heavier than He. {\bf Bottom panel:} contributions from \proton{} (solid black line), \hef{} (dashed blue line), and all others (dotted red line). To check whether accounting for the contribution of all other species as a scaling of \hef{} is a good approximation, the thin grey lines show their residual of the difference, i.e.  $|0.611\times$\hef{}~-~`all others'$|$ (solid and dashed lines correspond to minimal and maximal modulation periods).}
\label{fraction_cr}
\end{figure}
Focusing on the fourth column (slope $C_2$), CRs fall in two groups with either $C_2\approx 2.8$, or $C_2\approx 3.0$. They correspond respectively to the so-called primary species (CRs accelerated in sources and propagated in the Galaxy) and secondaries species (spallative products of primary species). Furthermore, heavier species suffer more inelastic interactions than lighter ones during propagation, providing a flatter IS flux at low energy. The last two columns on Table~\ref{tab:jis_fit} show that:
\begin{enumerate}[(i)]
   \item the contribution to count rates of heavy nuclei are significant up to Ni: low fluxes for heavy species are redeemed by the number of nucleons available in the yield function; 

   \item primary species contribute more than secondary ones: the contribution of heavier species $(Z\geq3)$ w.r.t. to He, in this simple estimate, is $(Z\geq3)/{\rm He}=0.480$, in agreement with the value 0.428 used in \citetads{2011JGRA..116.2104U}; 

   \item secondary species $Z\geq3$ contribute up to $\sim 4\%$ of the total, but due to a steeper slope w.r.t. primary species, their contribution decreases with rigidity; 

   \item the \deut{} and \het{} isotopes also contribute to 4\% of the total, and they should be dealt with separately from the rest because of their different A/Z value (they are not similarly modulated).
\end{enumerate}

Anticipating on the description of a realistic yield function (presented in Sects.~\ref{sec:Yield_detail} and \ref{sec:6NM64}), Fig.~\ref{fraction_cr} shows the result of the full calculation (without approximation) Eq.~(\ref{eq:frac_secies}), as a function of rigidity (for TOA fluxes modulated at $\phi=600$~MV). The top panel zooms in on the fractional contributions of secondary species, and primary species heavier than He. The numbers are in fair agreement with those given in the last column of Table~\ref{tab:jis_fit}, but with two noteworthy features: 
\begin{enumerate}[(i)] 
   \item the contributions are not constant with energy, peaking between $10-50$ GV for secondary species, while constantly increasing for primary species. The heavier the species, the larger the increase. This is explained by the increase of the ratio of heavy to light primary species with energy, due to spallation effects at low energy (see Fig.~14 of \citealtads{2011A&A...526A.101P}). As a result, at 100 GV, the Fe contribution is almost at the level of the C one; 
 
   \item \deut{} has a different A/Z ratio than all other species shown in the top panel, hence its contribution is shifted to lower rigidity. 
\end{enumerate}

The bottom panel of Fig.~\ref{fraction_cr} shows the contributions of \proton{}=H-\deut{} (50\%), \hef{}=He-\het{} (30\%), and the sum of all other contributions (20\%). The latter differs significantly from previous results:
\begin{enumerate}[(i)] 
   \item the full calculation of the contribution of species heavier than helium gives
\begin{equation}
  s_{Z>2} = 0.611^{+0.016}_{-0.009},
  \label{eq:heavy_scaledhe}
\end{equation}
instead of the value 0.428 obtained in the simple estimate and used in the literature;

   \item we check that this approximation is better than 1\% for all modulation levels, as illustrated by the difference between `true' (all species) and `scaled \hef{}' contributions, shown in grey. The uncertainty $^{+0.016}_{-0.009}$, i.e. $\sim 1.5-2.5\%$ on this factor, is obtained by propagating the errors on the CR IS flux parameters given in Table~\ref{tab:jis_fit}.
\end{enumerate}

\subsection{Degeneracy between $J_{\rm IS}$ and $\phi$}
\label{sec:degeneracy}

It has been shown that taking different parametrisations for the IS fluxes provides similar TOA fluxes and count rates, but with a shifted modulation parameters in time series \citepads{2005JGRA..11012108U,2010JGRA..11500I20H}. Indeed, unless either strong assumptions are made on the transport coefficients in the solar cavity, or IS data are available, or sufficient data covering all modulation periods with a good precision exist, the degeneracy is difficult to lift. 

\begin{figure}[!t]
\begin{center}
\includegraphics[width=\columnwidth]{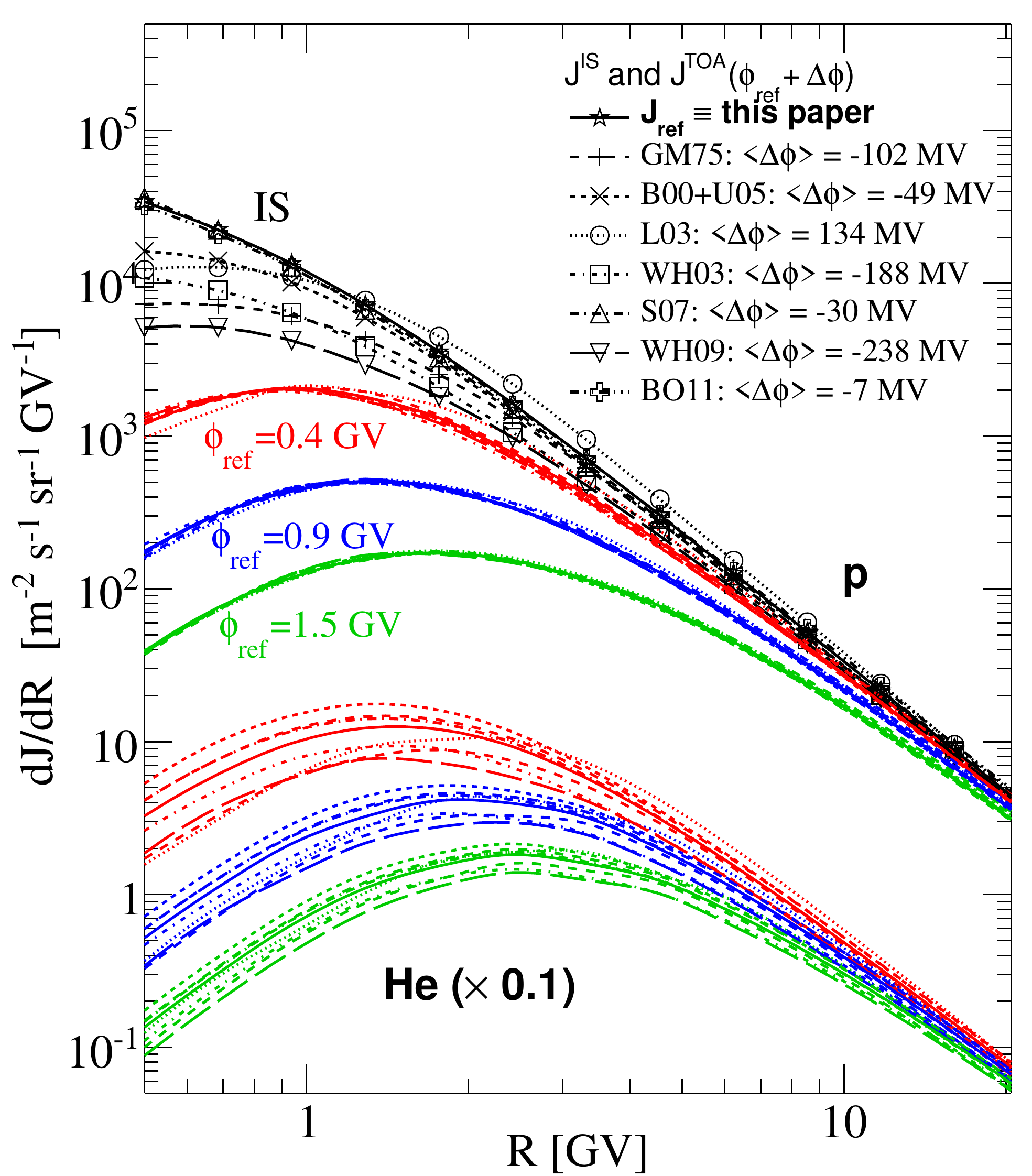}
\end{center}
\caption{Proton (IS, symbols; TOA, lines) and He (TOA only) fluxes for different modulation levels, and different parametrisations of the IS fluxes found in the literature (see App.~A): GM75 \citepads{1975ApJ...202..265G}, B00+U05 \citepads{2000JGR...10527447B,2011JGRA..116.2104U}, WH03 \citepads{2003JGRA..108.1355W}, L03 \citepads{2003JGRA..108.8039L}, S07 \citepads{2007APh....28..154S}, WH09 \citepads{2009JGRA..11402103W}, BO11 \citep{2010ITNS...57.3148O}. The parameter $\langle\Delta\phi\rangle$ gives the mean modulation shift in order for $J^{\rm TOA}_i$ to best match $J^{\rm TOA}_{\rm ref}(\phi_{\rm ref})$, as illustrated for three reference modulation level $\phi_{\rm ref}=0.4$~GV (red), 0.9 GV (blue), and 1.5 GV (green). See text for discussion.}
\label{fig:jis_jtoa}
\end{figure}
\begin{figure}[!t]
\begin{center}
\includegraphics[width=\columnwidth]{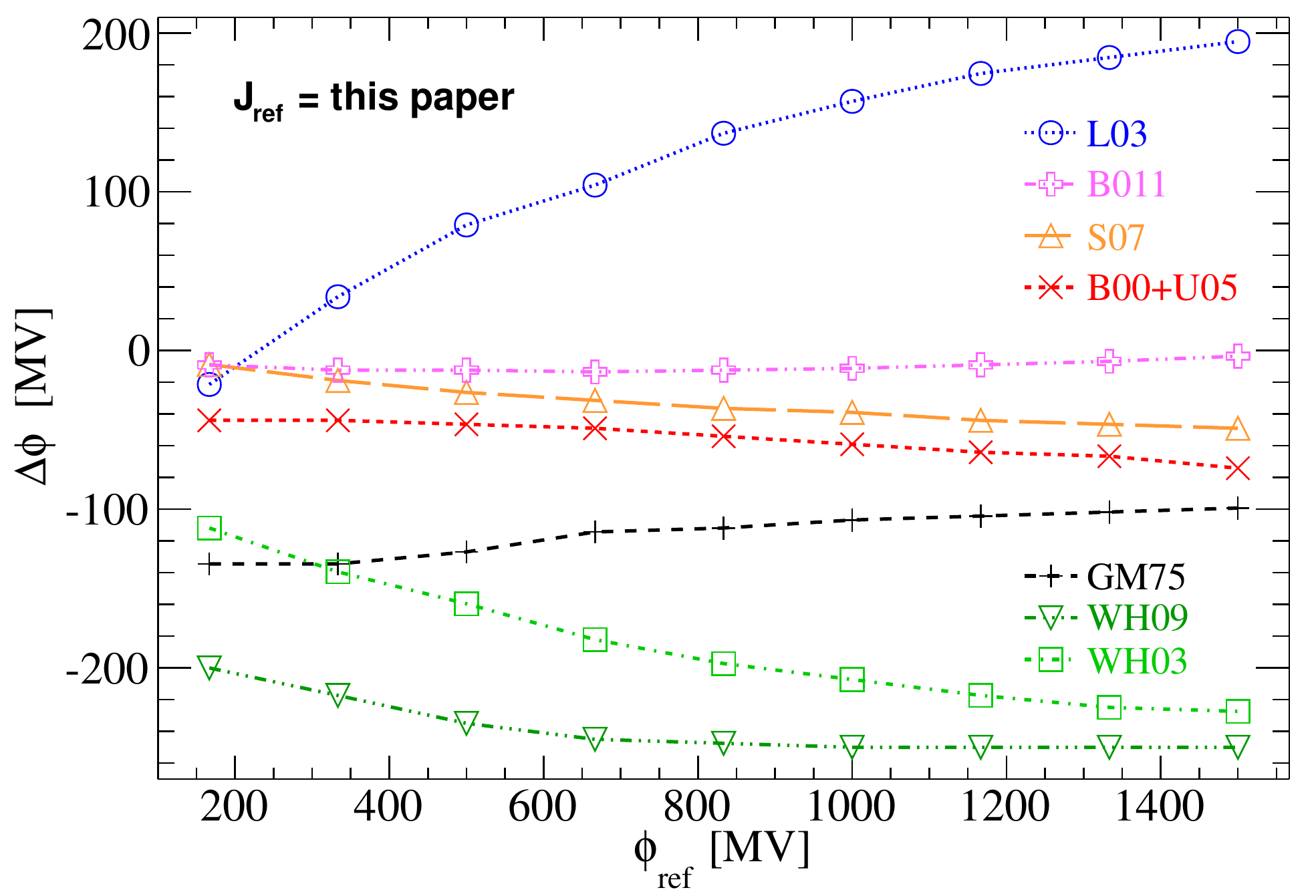}
\end{center}
\caption{Value of the modulation shift $\Delta\phi_i$ to add to the reference modulation level $\phi_{\rm ref}$, for $J^{\rm TOA}_i(\phi_{\rm ref}+\Delta\phi_i)$ to be as close as possible to  $J^{\rm TOA}_{\rm ref}(\phi_{\rm ref})$. The index $i$ runs on the IS flux parametrisations shown in Fig.~\ref{fig:jis_jtoa} (see also App.~A).}
\label{fig:jis_phishift}
\end{figure}

To illustrate this point, Fig.~\ref{fig:jis_jtoa} shows the IS proton flux (symbols) for several parametrisations behaving very differently at low energy. The reference IS flux (star) is the one fitted in the previous section. We then modulate protons for this reference flux (solid lines) at three different modulation levels ($\phi_{\rm ref}=0.4$~GV in red, 0.9 GV in blue, and 1.5 GV in green). Taking each IS flux parametrisation in turn, we search for the shift $\Delta\phi_i$ to apply to $\phi_{\rm ref}$ in order to minimise the difference between (i) the reference flux modulated at $\phi_{\rm ref}$ and (ii) a given IS parametrisation $i$ modulated at $\phi_{\rm ref}+\Delta\phi_i$. As can be seen on the figure, there always exists a value for which all the fluxes are very close to one another: this is what is meant by a degeneracy between the IS flux parametrisation and the modulation parameter value. The shift to apply slightly depends on $\phi_{\rm ref}$ itself (Fig.~3 in \citealtads{2010JGRA..11500I20H}). It is illustrated in Fig.~\ref{fig:jis_phishift} showing $\Delta\phi_i(\phi_{\rm ref})$ for all IS flux parametrisations. Except for WH09 \citepads{2009JGRA..11402103W}, the most recent parametrisations are in better agreement than the older ones. The BO11 model \citep{2010ITNS...57.3148O} is the most compatible with the present study.

\begin{figure}[!t]
\begin{center}
\includegraphics[width=\columnwidth]{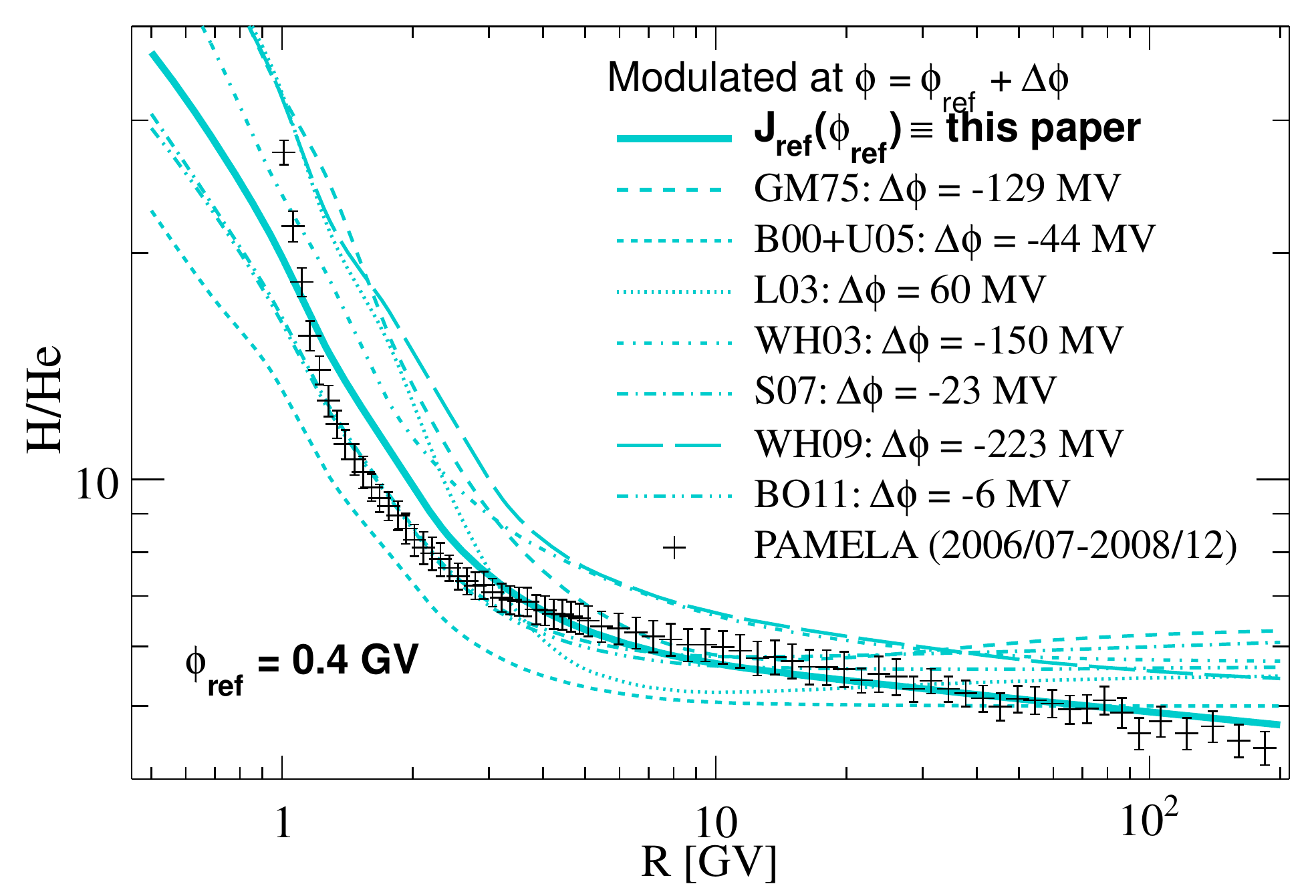}
\end{center}
\caption{p/He ratio (vs rigidity) modulated at $\phi_{\rm ref}+\Delta\phi_i$ for all IS parametrisations $i$ shown in Fig.~\ref{fig:jis_jtoa}. The reference modulation level is taken to be $\phi_{\rm ref}=400$~MV for PAMELA data ('+' symbols) taken from \citetads{2011Sci...332...69A}.}
\label{fig:phe_ratio}
\end{figure}
He fluxes ($\times 0.1$) are also shown in Fig.~\ref{fig:jis_jtoa}: the solid line is our best-fit IS flux, others He fluxes being obtained by scaling protons by 0.05 \citepads{2005JGRA..11012108U}. For a better view, p/He ratios for all parametrisations are plotted in Fig.~\ref{fig:phe_ratio} (for $\phi_{\rm ref}=400$~MV), against the recent PAMELA data \citepads{2011Sci...332...69A}. We find that the following scaling values give less scatter w.r.t. the data:
\begin{equation}
   f_{{\rm He}/{\rm p}} \equiv \frac{dJ_{\rm He}/dR}{dJ_{\rm p}/dR} =
     \begin{cases}
        0.0525 \quad \textrm{(GC75)},\\
        0.0475 \quad \textrm{(B00+U05)},\\
        0.0550 \quad \textrm{(L03)},\\
        0.0560 \quad \textrm{(WH03)},\\
        0.0525 \quad \textrm{(S07)},\\
        0.0605 \quad \textrm{(WH09)}.
     \end{cases}
\label{eq:f_hep}
\end{equation}

\subsection{Dealing with IS flux uncertainties in count rate and $\phi$ time series}
\label{sec:JIS_uncertainty}

The above degeneracy prevents us from obtaining substantial constraints on the IS flux but not on count rate calculations.

\begin{figure}[!t]
\begin{center}
\includegraphics[width=\columnwidth]{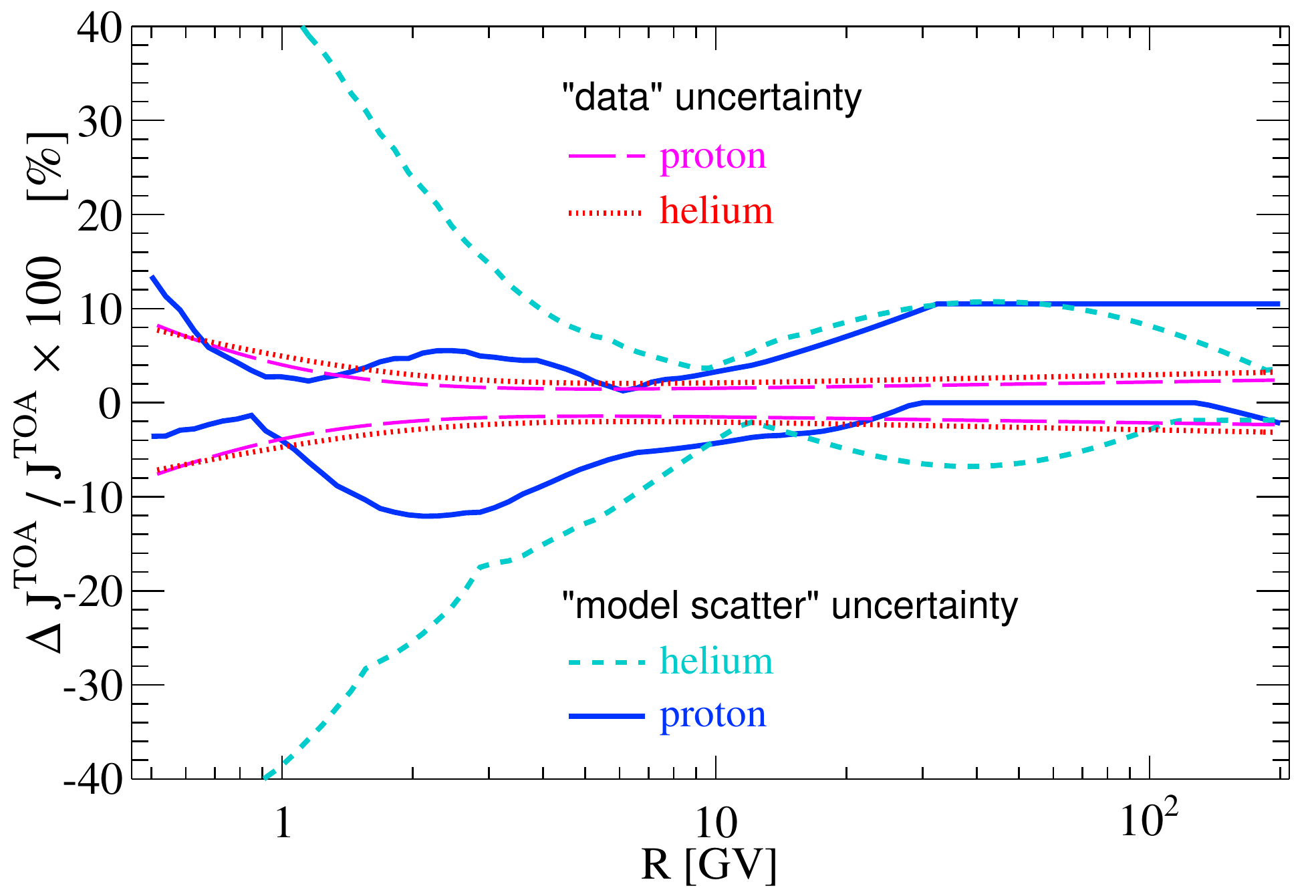}
\end{center}
\caption{Relative uncertainty (envelopes) on the p and He TOA fluxes. The long-dashed pink and dotted red lines \emph{``data uncertainty"} are calculated from the uncertainties on the fit on the CR p and He data (see Table~\ref{tab:jis_fit}). The solid blue and dashed cyan lines represent the `model' dispersion (related the choice of the IS flux and modulation model). This uncertainty is propagated to calculated count rates in Sect.~\ref{sec:rates_vs_rcutoff}.}
\label{fig:uncertainty_toa}
\end{figure}

\paragraph{TOA flux uncertainty} It can be estimated in two different approaches:
\begin{enumerate}
  \item {\em `data' uncertainty}: we assume that the reference IS flux model is the correct one, so that TOA uncertainties are directly obtained from the uncertainty on the fitted flux parameters (see Table~\ref{tab:jis_fit}). This gives a $\sim 2.5\%$ relative uncertainty, as shown in Fig.~\ref{fig:uncertainty_toa} for H (magenta dashed line) and He (red dotted line);

  \item {\em `model scatter' uncertainty }: lacking conclusive evidences to favour a particular IS flux shape, we can alternatively assume that all parametrisations $i$ are equally valid to provide similar (but not equal) effective TOA flux values. Plugging the appropriate effective modulation level $\phi_{\rm ref}+\Delta\phi_i$ to get, for each IS flux model $i$, its effective TOA flux (see Fig.~\ref{fig:jis_phishift}), we form the quantity 
  \[
      S_i(R) \equiv \frac{J_i^{\rm TOA}(\phi_{\rm ref}+\Delta\phi_i) - J_{\rm ref}^{\rm TOA}(\phi_{\rm ref})}{J_{\rm ref}^{\rm TOA}(\phi_{\rm ref})}.
  \]
The ``model scatter" uncertainty is obtained by keeping minimal and maximal values of $S_i(R)$ over several $\phi_{\rm ref}$ and IS flux parametrisations $i$ (we discard L03 which is too far away from the data). In Fig.~\ref{fig:uncertainty_toa}, the corresponding curves are shown for He (blue stars) and H (cyan circles). Note that some of the IS flux spectra used in this approach are probably already excluded by current data, so that the uncertainty range $\sim 10-30\%$ for H and He is (certainly too) conservative\footnote{A more consistent analysis, i.e. fitting the different IS flux parametrisations on the same data to evaluate a more realistic ``model scatter" uncertainty, is left for a future study. The benefit of keeping the IS fluxes as used in the literature so far, is to give a flavour of systematic differences related to their use.}.
\end{enumerate}
The uncertainty related to the contribution of heavier species should also be taken into account: the {\em `scaling approximation'} factor Eq.~(\ref{eq:heavy_scaledhe}) to account for CR heavier than He gives $s_{Z>2}=0.611 \pm 2.5\%$.

\paragraph{Impact on $\phi$ reconstruction}
Our ignorance of the real IS flux shape has a strong impact on the determination of the modulation level $\phi$. However, it  can be absorbed as a global (i.e., time-independent) shift $\Delta\phi$, once a given IS flux model is chosen. As seen from Fig.~\ref{fig:jis_phishift}, the shift can be quite large (from -250~MV to 200~MV). To obtain times series (and their uncertainties) and compare different results given in the literature, the procedure is as follows: (i) calculate $N_{\rm ref}(t)$ and $\phi_{\rm ref}(t)$ from a reference $J^{\rm IS}_{\proton{}, {\rm ref}}(R)$ and $(1+s_{Z>2}) \times J^{\rm IS}_{\hef{}, {\rm ref}}(R)$; (ii) the modulation level $\phi_i(t)$ for any given IS flux parametrisation $J^{\rm IS}_i$ is then simply related to the reference one:
\begin{equation}
   \begin{cases}
      N^{\rm calc}(t) = N_{\rm ref}(t)\pm \Delta N_{\rm ref}^{\rm TOA}(t),\\
      \phi_i(t) = \phi_{\rm ref}(t)\pm \Delta\phi_{\rm ref}^{\rm TOA}(t) + \Delta\phi_i^{\rm IS};
      \end{cases}
      \label{eq:recipe}
\end{equation}
In the above equations, $\Delta N_{\rm ref}^{\rm TOA}$ and $\Delta\phi_{\rm ref}^{\rm TOA}$ are evaluated by propagating uncertainties of TOA flux quantities (see Fig.~\ref{fig:uncertainty_toa}).

\section{Atmospheric propagation, yield function, and detectors}
\label{sec:Yield}

When entering the Earth atmosphere, CRs initiate cascades of nuclear reactions involving primary energetic particles (mainly hydrogen and helium but also heavier nuclei) and atmospheric nuclei such as oxygen or nitrogen. The so-called Extensive Air Showers (EAS) generate secondary particles along their path, to be detected by ground-based instruments.

In this section, we discuss the generation of secondary particles (Sect.~\ref{sec:atmopropag}) as an input to provide a new yield function parametrisation (Sect.~\ref{sec:Yield_detail}) for NMs (Sect.~\ref{sec:6NM64}) and muon detectors (Sect.~\ref{sec:muons}). We also discuss neutron spectrometers (Sect.~\ref{sec:BSS}) as a mean to study seasonal effects in NMs.

\subsection{Atmospheric propagation of secondaries (n, p, $\mu$)}
\label{sec:atmopropag}

The secondary atmospheric radiation field is composed of various hadronic components (mostly neutrons, protons, and pions). Charged pions undergo leptonic decays producing positive and negative muons. Key quantities are\footnote{Below we use the altitude $h$, but the atmospheric depth or grammage or pressure could have been equally used (conversion is made using the barometric formula).}
\begin{itemize}
  \item  $\dot{\varphi}_k(T_k,\vec{r},t)$: spectral fluence rate (cm$^{-2}$~s$^{-1}$~MeV$^{-1}$) of the $k$-type secondary particle at kinetic energy $T_k$, coordinates $\vec{r}=(\varphi,\,\lambda,\,h)$, and time $t$;
  \item $\varphi_{ik}^l(h,\,T_i\!\rightarrow\! T_k)$: spectral fluence (MeV$^{-1}$) of the $k$-type secondary induced at altitude $h$ by a $i$-type primary of kinetic energy $T_i$, and incidence within the zenith angle range $[\theta _l,\,\theta_{l+1}]$.
\end{itemize}

\begin{table}[!t]
\caption{\textsc{AtmoRad} main features.}
\label{tab:atmorad}
\begin{center}
\begin{tabular}{llrll}
\hline
Parameter    \!  & Qty &  \!\!\!Bins & Range & Unit\\\hline
Primary $i$\!  &   H  & \multirow{2}{*}{18} &  \multirow{2}{*}{$[0.1-251.2]$} & \multirow{2}{*}{GeV/n}\\\vspace{1mm}
                 &   He &  & &\\
Secondary $k$\!&   n  & 70 & $[10^{-3} - 10^{11}]$ & \multirow{2}{*}{eV}\\\vspace{2mm}
                 &  p, $\mu^\pm$ & 25 & $[10^{6}-10^{11}]$ & \\\vspace{2mm}
Incidence        & $[\theta_l,\theta_{l+1}]$\!\!\!& 3 & $\displaystyle\left\{0,\, \frac{\pi}{6}, \,\frac{\pi}{3},\,\frac{\pi}{2}\right\}$ & rad\\
Altitude         & h & 36 & $[0-30]$ & km asl\\\hline
\end{tabular}
\end{center}
\end{table}

Several works were dedicated to numerically estimate the spectral fluence rate $\dot{\varphi}_k$ during the solar activity cycle (e.g., \citealtads{1998AdSpR..21.1717R}; \citealt{roesler2002monte,sato2006analytical}; \citealtads{2013NIMPB.295...99N}). We rely here on the database of spectral fluence values $\varphi_{ik}^l$ presented in \citetads{2013ITNS...60.2411C}, in which Monte Carlo (MC) calculations were performed with GEANT4 \citepads{2003NIMPA.506..250A}.
For any coordinates $\vec{r}$ and solar modulation potential $\phi(t)$, the spectral fluence rate $\dot{\varphi}_k$ is obtained from $\varphi_{ik}^l$ read from the database as follows: 
 \begin{eqnarray}
\label{eq:fluence}
\!\!\!\!\!\!\!\!\dot{\varphi}_k(T_k,\vec{r},t) \!=\!S_T \!\!&\!\!\times\!\!&\!\! 
  \sum_{l=1}^3\Omega^{(\theta_l,\theta_{l+1})} \;\;\sum_i^{\rm CRs} \;\;\sum_{T_i^{\textrm{cut-off}}}^{T_i^{\rm max}} \Delta T_i\;\;\;\;\;\\
  \!\!&\!\!\times\!\!&\!\!  J_i^{\text{TOA}}(T_i,\phi (t)) \times 
    \varphi_{ik}^l(h, \,T_i\!\rightarrow\! T_k), \nonumber
\end{eqnarray}
with $S_T=\pi \cdot (R_E+h_{\text{max}})^2$, $R_E=6,378.14$~km (Earth radius), and $h_{\text{max}}=85$~km (highest atmospheric altitude). 

\begin{figure}[!t]
\begin{center}
\includegraphics[width=\columnwidth]{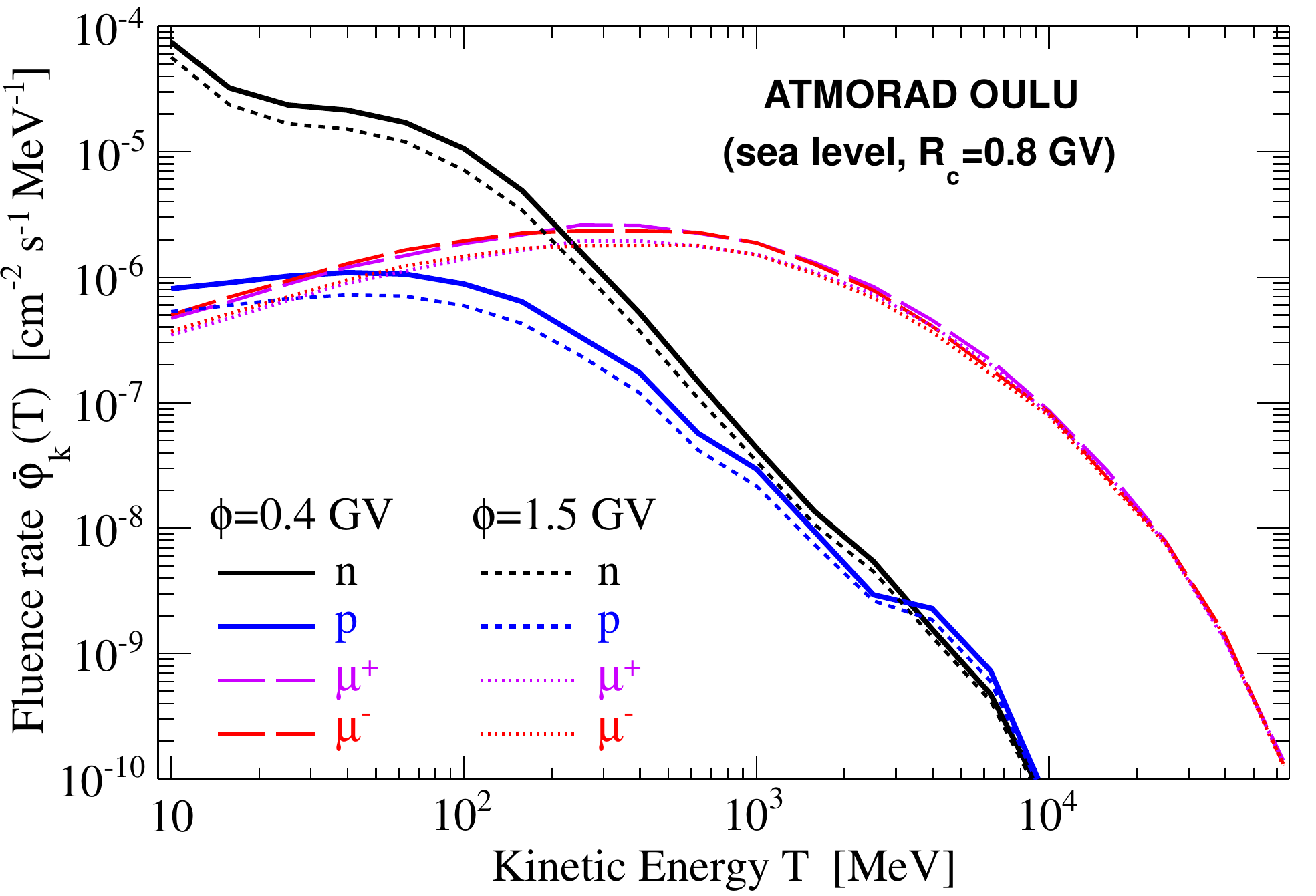}
\end{center}
\caption{Illustration of secondary spectral fluence rates (see Eq.~\ref{eq:fluence}) calculated with \textsc{AtmoRad} (sea level, $R_c=0.8$~GV) for two solar modulation periods.}
\label{fig:fluenceOulu}
\end{figure}

A \texttt{C++} routine named \textsc{AtmoRad} (ATMOspheric RADiation) developed at ONERA implements the various ingredients entering Eq.~(\ref{eq:fluence}). It handles both \texttt{QGSP\_BERT\_HP} and \texttt{QGSP\_BIC\_HP} reference users' physics lists. Table~\ref{tab:atmorad} lists the quantities (primary and secondary species) and bins (energy and altitude range, incidence angles) used.  The calculations were validated by extensive comparison with measurements, especially for the neutron component \citepads{2013JGRA..118.7488C}. In the following we use \texttt{QGSP\_BERT\_HP} physic's list. Figure~\ref{fig:fluenceOulu} is an illustration of the neutron, proton, and muon spectral fluence rates obtained at sea level with a cut-off rigidity $R_c=0.8$~GV (similar to conditions at the Oulu NM station). The solid and dotted lines correspond to a period of minimum and maximum solar modulation potential $\phi(t)$ equal to 0.4~GV and 1.5~GV, respectively. Muons are the most numerous particles above a few hundreds of MeV, but the relative contribution of various secondaries to count rates in a detector depends on its efficiency to each species.

\subsection{Yield function calculation and parametrisation}
\label{sec:Yield_detail}

As given in Eq.~(\ref{eq:count-rate}), the yield function ${\cal Y}_i^{\cal D}(T_i,h)$ of a ground-based detector ${\cal D}$ at altitude $h$ is its response (in count m$^2$~sr) to the unit intensity of primary CR $i$ at kinetic energy $T_i$. It can be described in terms of
\begin{itemize}
   \item ${\cal Y}_{ik}^{\cal D}(T_i,h)$: partial yield function from $i$-type primary species
   into $k$-type secondary species (in count~m$^2$~sr);
   \item ${\cal E}_k^{\cal D}$: detector efficiency to $k$-type secondary species.
\end{itemize}
The yield and partial yield functions are then given by
\begin{eqnarray}
\label{eq:partial_yield}
\!\!\!\!\!\!{\cal Y}_{i}^{\cal D}(T_i,h) \!&\!=\!&\! \sum_{k={\rm n,\,p,\,}\mu\dots}  {\cal Y}_{ik}^{\cal D}(T_i,h)\,,\\
\!\!\!\!\!\!\!\!\!{\cal Y}_{ik}^{\cal D}(T_i,h)\!&\!=\!&\!S_T\!\sum_{l=1}^3 \Omega^{(\theta_l,\theta_{l+1})}
\label{eq:yield}
\\
\!&\!\times\!&\! \int_0^\infty {\cal E}_k^{\cal D} (T_k) \times \varphi_{ik}^l(h, \,T_i\!\rightarrow\! T_k) dT_k. \nonumber
\end{eqnarray}

Note that only the primary CR ions \proton{} or \hef{} are evaluated below. For further usage, the resulting ${\cal Y}_{^1{\rm H}\rightarrow k}^{\cal D}$ and  ${\cal Y}_{^4{\rm He}\rightarrow k}^{\cal D}$ are parametrised with a universal form ($T_{i/n}$ is the kinetic energy per nucleon of $i$)
\begin{equation}
\label{eq:yield_fit}
\!\!\!\frac{{\cal Y}_{ik}^{\cal D}(T_i,h)}{\exp(hf_k)}\!=\!10^{\left(a_{ik} I -\frac{c_{ik}}{I^{d_{ik}}} - e_{ik}\right)}\,\, {\rm and}\,\; I\!\!=\! \log(T_{i/n})\!+\! b_{ik},\!\!\\
\end{equation}
where the best-fit coefficients ($a_{ik}$, $b_{ik}$, $c_{ik}$, $d_{ik}$, $e_{ik}$, and $f_{k}$) are calculated for the various detectors ${\cal D}$ considered (the fit is appropriate for altitudes up to $h=5,000$~m). The yield function for any other primary of atomic mass $A$ at rigidity $R$ is rescaled from the \hef{} yield (at the same rigidity), namely
\begin{equation}
\label{eq:yield_fit_scaling}
 {\cal Y}_{A}^{\cal D}(R,h) = \frac{A}{4} \times {\cal Y}_{^4{\rm He}}^{\cal D}(R,h)\,.
\end{equation} 
This assumption was tested with nitrogen, oxygen, and iron in \citetads{2011AdSpR..48...19M}, and was found to work well in the lower atmosphere (below 15 km).

\subsection{Response and yield function for 6-NM64}
\label{sec:6NM64}
Standardised neutron monitors (NM64 model) are widely used across the world to monitor CRs since the 1950s \citepads{2000SSRv...93...11S}. These detectors are especially powerful once integrated in a worlwide network \citepads{1995ICRC....4.1316B,2004ASSL..303.....D}. They provide count rates with very interesting time intervals (typically one minute) thanks to the high efficiency of the detectors. An elementary unit of a 6-NM64 consists of six BF$_3$ proportional counter tubes which are mounted in raw and surrounded by a cylindrical polyethylene moderator. The tubes and the inner moderator are inserted in a large volume of lead (the producer). The outer walls of the NM64, the so-called reflector are again made of polyethylene or wood. A more detailed description of the standard NM64-type neutron monitors can be found elsewhere \citepads{1964CaJPh..42.2443H}. 

\paragraph{NM response function}
NMs are optimised to measure the high-energy hadronic component of ground level secondaries above 100~MeV \citepads{2000SSRv...93...11S}. However, in spite of their name, they are also sensitive to other secondary radiations (protons, pions, and muons). The efficiency of NM64 to various species have been calculated in the literature from MC simulations with FLUKA \citepads{1999ICRC....7..317C} or GEANT4 \citepads{2011NIMPA.626...51P}. A detailed comparison of the efficiencies obtained in the literature is carried out in \citetads{2000SSRv...93..335C} and \citetads{2011NIMPA.626...51P}: a very good agreement was found, be it for incident protons and neutrons (the calculation for the latter were also compared to the only existing beam calibration data from accelerator of \citealtads{1997ICRC....1...45S,1999ICRC....7..313S}). Differences up to a factor of two (above GeV energies) nevertheless exist depending on which of the GEANT4 physics model or event generator is selected. As our fluence is calculated with GEANT4, we choose to directly use the efficiency given in \citetads{2011NIMPA.626...51P}, also calculated with GEANT4, and which is in very good agreement with the results of \citetads{2000SSRv...93..335C}. 

\begin{figure}[!t]
\begin{center}
\includegraphics[width=\columnwidth]{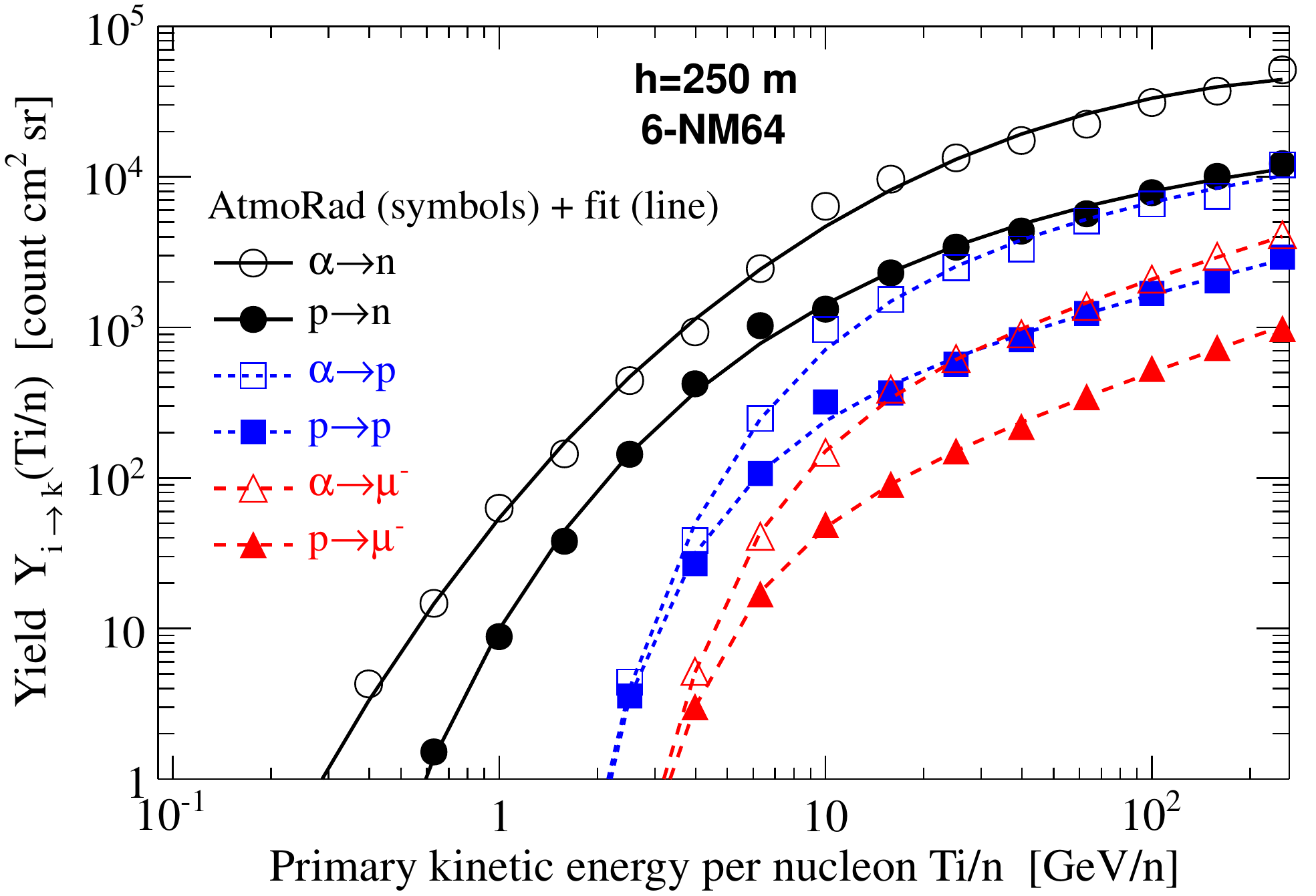}
\end{center}
\caption{Levenberg-Marquardt fit of the \textsc{AtmoRad} yield function for secondary neutrons, protons, and negative muons at +250~m for a 6-NM64. The corresponding fit parameters are gathered in Table~\ref{tab:NMOulu}.}
\label{fig:YieldCoeffs}
\end{figure}

\paragraph{Relative contribution of secondary species ($n$, $p$, and $\mu$)}
\begin{table}
\caption{Count rate fractional contribution from $k$ secondary particles at Oulu station (sea level and $R_c=0.8$ GV) during solar maximum and minimum periods, see Eqs.~(\ref{eq:countrate_tot}) and (\ref{eq:countrate_partial}).}
\label{tab:NMOulu}
\begin{center}
\begin{tabular}{cccccc}
\hline
            &                                 &         &         \vspace{-3.5mm}\\
$\phi$ [GV] & $N^{\textrm{6-NM64}}$ [s$^{-1}$]& \multicolumn{4}{c}{$N_k/N$ [\%]}\\
            &                &  n     &  p    & $\mu^+$ & $\mu^-$ \\\hline
    0.4     &       91       &  87.2  &  7.9  &   0.2   & 4.7     \\
    1.5     &       57       &  87.4  &  8.0  &   0.2   & 4.4     \\\hline
\end{tabular}
\end{center}
\end{table}
Although muons are the most numerous terrestrial particles (see Fig.~\ref{fig:fluenceOulu}), the efficiency of the 6-NM64 to muons is very low (3.5 order of magnitude below the hadrons at 1~GeV, see Fig.~5 of \citealtads{2000SSRv...93..335C}). Hence they do not contribute much to the total count rate in a NM
\citepads{2000SSRv...93..335C}. To back up this comment, we calculate the fraction of count rates from the secondary $k$ particles. Using Eqs.~(\ref{eq:fluence},\ref{eq:partial_yield},\ref{eq:yield}) in
\begin{equation}
\label{eq:countrate_tot}
N^{\textrm{NM}}({\vec{r}},t)= \sum_{k={\rm n,\,p,\,}\mu\dots}N^{\textrm{NM}}_k({\vec{r}},t),
\end{equation}
the contribution $N^{\cal D}_k$ can be expressed to be
\begin{equation}
\label{eq:countrate_partial}
N^{\textrm{NM}}_k({\vec{r}},t) = \sum_{i={\rm CRs}}\int_{0}^{\infty} {\cal E}^{\textrm{NM}}_k(T_k) \times \dot{\varphi}_k(T_k,\vec{r},t) \; dT_k.
\end{equation}  
Folding the fluence rate (calculated with \textsc{AtmoRad}) with the 6-NM64 efficiency (from \citealtads{2000SSRv...93..335C}), we gather in Table~\ref{tab:NMOulu} the total count rate and the fraction due to the $k$-th particle, at Oulu location. Note that the total count rate calculated in the table is slightly lower than the observed one\footnote{\url{http://www.nmdb.eu/nest/search.php}}: this difference amounts to an extra normalisation of the yield function that will be addressed in our next study (see also \citealtads{2011JGRA..116.2104U}). The variation between a minimum and maximum modulation is almost a factor of two. The main contributions at sea level come from secondary neutrons (87\%), protons (8\%) and $\mu^-$ (5\%), the $\mu^+$ contribution being negligible (0.2\%). This fraction slightly changes for high altitude stations: the steeper increase of the number of nucleons relative to $\mu$ with altitude\footnote{In the exponential term of the yield fit function Eq.~(\ref{eq:yield_fit}), $f_k^{\rm n,~p}\sim0.00068$~m$^{-1}$, to be compared to $f_k^\mu=0.00025$~m$^{-1}$
(see Tab.~\ref{fig:YieldCoeffs}).} leads to a smaller muon fraction ($\sim2.2\%$ at 2000~m and $\sim1.2\%$ at 3500~m).

\paragraph{Results for our NM yield function}
The {\em partial} yield functions ${\cal Y}^{\text{6-NM64}}_{ik}$ data points (symbols) calculated in this study from Eq.~(\ref{eq:yield}) are shown in Fig.~\ref{fig:YieldCoeffs} for different primary $i$ and secondary $k$ particles. Also shown are the best fits (lines) to these data relying on Eq.~(\ref{eq:yield_fit}), whereas the best-fit parameters are gathered in Table~\ref{tab:best-fit_yield}. As already underlined above, the altitude dependence is steeper for nucleons than for muons. The energy dependence is similar to the yield function obtained in previous studies (see below), with a sharp cutoff at low energy, and a shallow power-law dependence at high energy. 
\begin{table}[!t]
\caption{Best-fit parameters for a 6-NM64 yield function (in count m$^2$~sr) for a single tube with \textsc{AtmoRad} (see Fig.~\ref{fig:YieldCoeffs}), relying on the parametrisation Eq.~(\ref{eq:yield_fit}). }
\label{tab:best-fit_yield}
\begin{center}
\begin{tabular}{ccccccc}\hline\vspace{1mm}
&&&&&&\vspace{-4.5mm}\\
$\!\!i\!\rightarrow \!k\!\!$       & $a_{ik}$  & $b_{ik}$ & $c_{ik}$ & $d_{ik}$ & $e_{ik}$ &\!\!\!$f_{k}$ [m$^{-1}$]\!\!\! \\\hline
&&&&&&\vspace{-3.5mm}\\
$\!\!\!\!p\!\rightarrow \!n\!\!$       &  -0.105\! &  2.862\! &  66.98\! &  2.648\! & -5.432\! & \multirow{2}{*}{\!\!0.00067}\\\vspace{2mm}
$\!\!\!\!\alpha\!\rightarrow \!n\!\!$  &  -2.442\! &  5.484\! &  138.9\! &  0.834\! & -48.71\! & \\
$\!\!\!\!p\!\rightarrow \!p\!\!$       &  0.5281\! &  1.588\! &  142.0\! &  6.295\! & -1.367\! & \multirow{2}{*}{\!\!0.00069}\\\vspace{2mm}
$\!\!\!\!\alpha\!\rightarrow \!p\!\!$  &  0.2219\! &  1.803\! &  132.6\! &  4.753\! & -3.219\! & \\
$\!\!p\!\rightarrow \!\mu^-\!\!\!\!$       &  0.722\! &  0.686\! &  4.104\! &  4.742\! & -0.802\! & \multirow{2}{*}{\!\!0.00025}\\
$\!\!\alpha\!\rightarrow \!\mu^-\!\!\!\!$  &  0.626\! &  1.276\! &  70.48\! &  5.824\! & -1.340\! & \\\hline
\end{tabular}
\end{center}
\end{table}

\paragraph{Comparison to other yield functions}
\begin{figure}[!t]
\begin{center}
\includegraphics[width=\columnwidth]{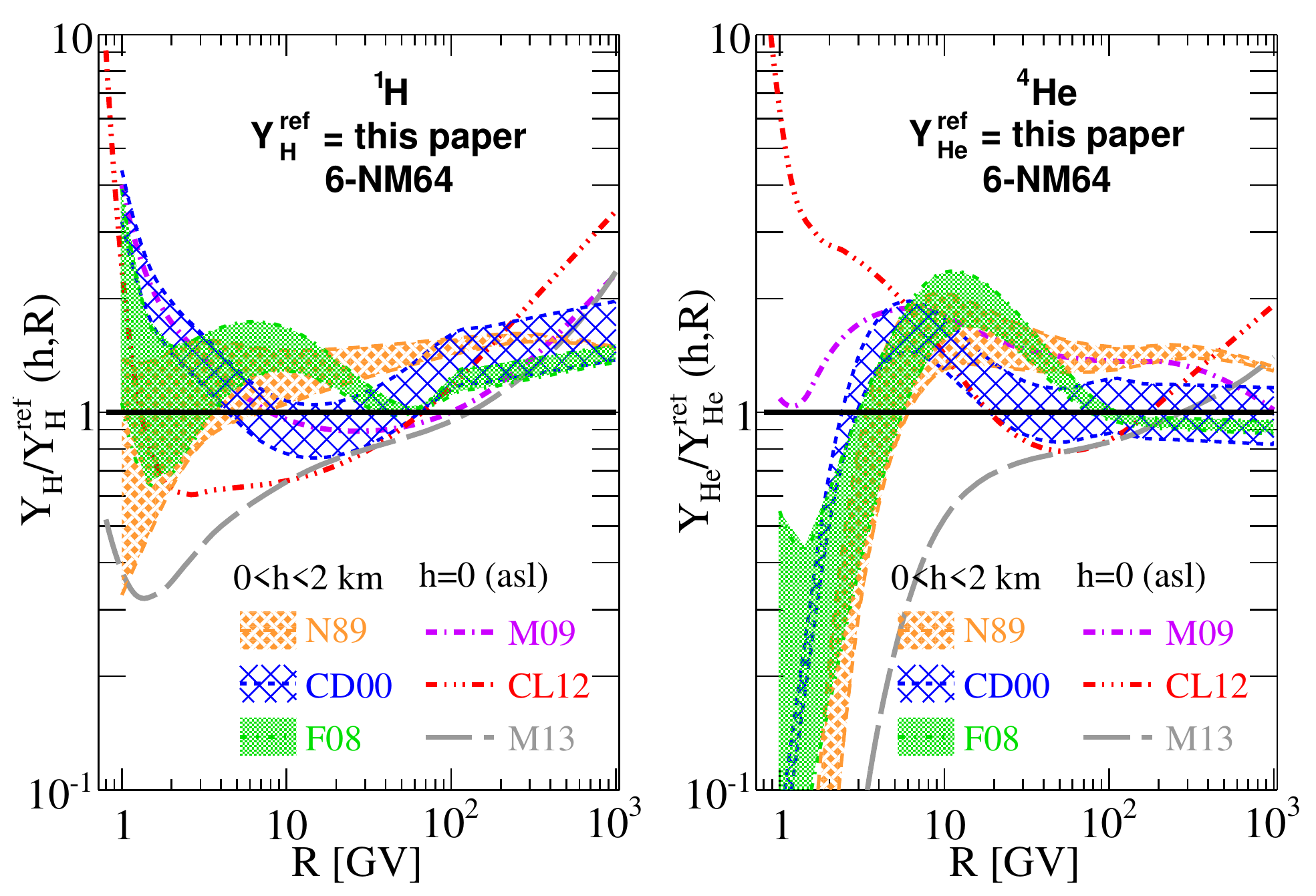}
\includegraphics[width=\columnwidth]{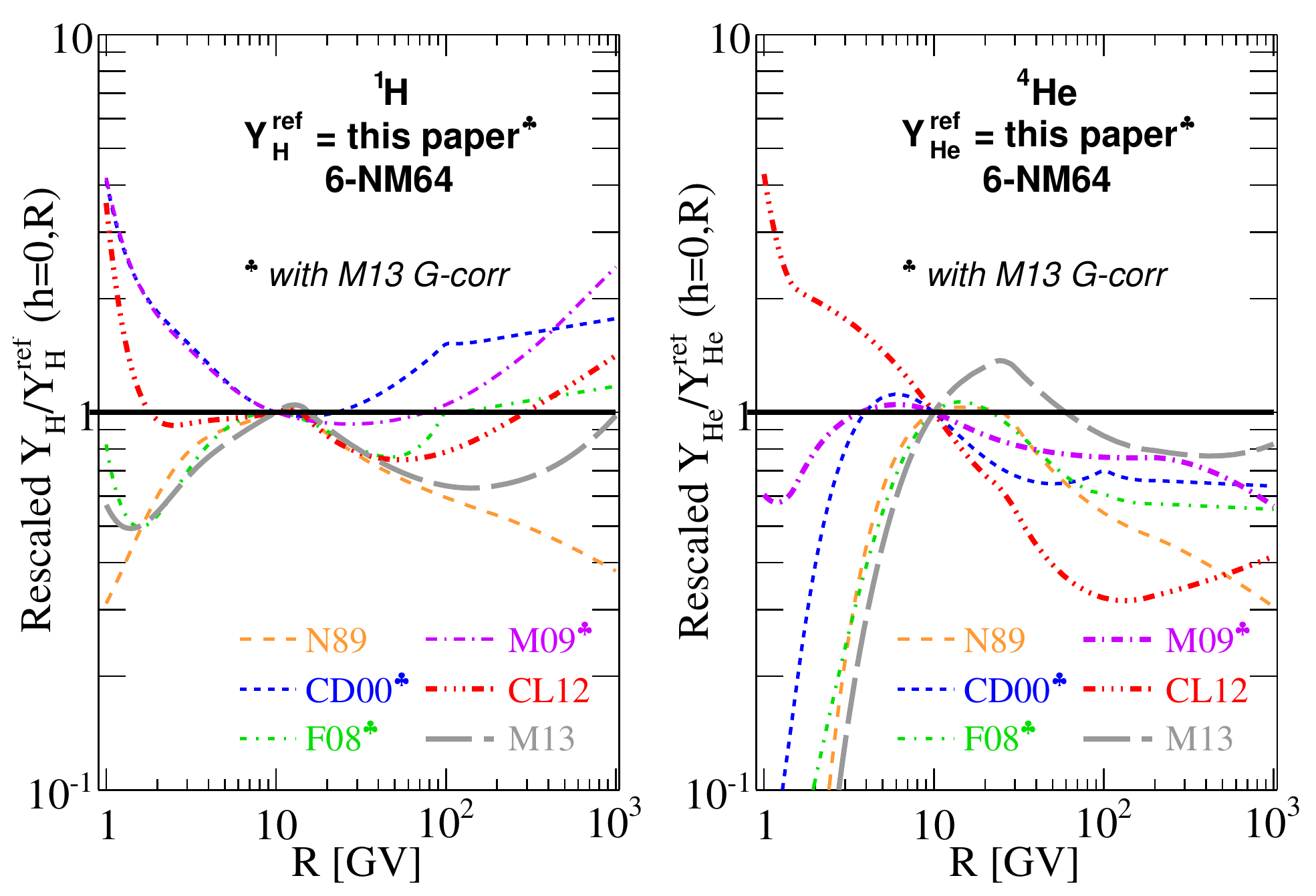}
\end{center}
\caption{Ratio of various yield functions from the literature  to ours (see App.~B) for \proton{} (left panels) and \hef{} (right panels): N89 \citepads{1989NCimC..12..173N,1990ICRC....7...96N}, CD00 \citepads{1999ICRC....7..317C,2000SSRv...93..335C}, F08 \citepads{2008ICRC....1..289F}, M09 (\citealt{matthia_thesis}; \citealtads{2009JGRA..114.8104M}), CL12 \citepads{2012JGRA..11712103C}, M13 \citepads{2013JGRA..118.2783M}. {\bf Top panel:} lines are evaluated at sea level altitude, and filled areas show the dispersion of results related to a different altitude dependence in the different parametrisation (when available) for $h\in[0-2]$~km. {\bf Bottom panel:} The yields are rescaled at 10~GV (so that the ratio is 1), and the geometrical factor correction proposed in \citetads{2013JGRA..118.2783M} is applied to theoretical calculations (see Eq.~\ref{eg:g-factor} in App.~B).}
\label{fig:ratio_yields}
\end{figure}
The {\em total} yield function ${\cal Y}^{\textrm{6-NM64}}_i$ (summed over all $k$ secondary particles) is compared to previous calculations in Fig.~\ref{fig:ratio_yields}. The figure shows the ratio of any given parametrisation to ours at sea level (lines), for protons (left panels) and helium (right panels). For parametrisations provided with the altitude dependence (N89, CD00, F08, and ours), the shaded areas in the top panels show the dispersion w.r.t.~the reference yield function altitude dependence, in the 0-2 km range. The different parametrisations rely on several MC generators/atmospheric models/NM responses \citepads{1999ICRC....7..317C,2008ICRC....1..289F,2009JGRA..114.8104M,2013JGRA..118.2783M} or latitude and altitude surveys \citepads{1989NCimC..12..173N,2012JGRA..11712103C}. In that respect, the various yield functions can be considered to be in fair agreement\footnote{Note that the uncertainty on the altitude dependence is disregarded as it is already encompassed in the dispersion arising from the various parametrisations.}. Because of the overall uncertainties in the modelling, the results are usually taken to be up to a global normalisation factor. This absorbs part of the difference if one is interested in count rate studies \citepads[e.g.,][]{2011JGRA..116.2104U}. The bottom panel shows the same quantity as in the top panel, but rescaled to 1 at 10 GV. Actually, \citetads{2013JGRA..118.2783M} recently proposed a correction factor $G$ to account for a hitherto forgotten geometrical factor (related to the NM effective size). These authors find that this correction is necessary to match existing latitude NM surveys (see also Sect.~\ref{sec:rates_vs_latitudesurvey}). In principle, this correction (see Eq.~\ref{eg:g-factor}) must be applied to any theoretical calculations, i.e. this work, CD00, F08, and M09 (it is by construction included in M13). These G-corrected yields at sea level altitude (renormalised at 10 GV) are shown in the bottom panel of Fig.~\ref{fig:ratio_yields}: a quite good agreement is now found in the $5-50$~GeV range, where most of the counts come from (see below). This scatter (of the yield functions) is propagated to calculated count rate uncertainties in Sect.~\ref{sec:rates_vs_rcutoff}.

\subsection{Cosmic-ray muon intensity}
\label{sec:muons}
\begin{table}[!t]
\caption{Best-fit parameters for the yield function (in m$^2$~sr) of a generic $\mu$ detector with \textsc{AtmoRad} (see Fig.~\ref{fig:YieldCoeffs_muons}), relying on the parametrisation Eq.~(\ref{eq:yield_fit}). Positive and negative muons are counted together.}
\label{tab:best-fit_yieldmuon}
\begin{center}
\begin{tabular}{ccccccc}\hline\vspace{1mm}
&&&&&&\vspace{-4.5mm}\\
$\!\!i\!\rightarrow \!\mu\!\!$       & $a_{ik}$  & $b_{ik}$ & $c_{ik}$ & $d_{ik}$ & $e_{ik}$ &\!\!\!$f_{k}$ [m$^{-1}$]\!\!\! \\\hline
&&&&&&\vspace{-3.5mm}\\
$\!\!p\!\rightarrow \!\mu\!\!$       &  0.9116\! &  2.068\! &  664.1\! &  5.818\! & 2.755\! & \multirow{2}{*}{\!\!0.00025}\\
$\!\!\alpha\!\rightarrow \!\mu\!\!$  &  -0.1315\! &  1.789\! &  49.75\! &  2.495\! & -3.702\! & \\\hline
\end{tabular}
\end{center}
\vspace{-4mm}
\end{table}
\begin{figure}[!t]
\begin{center}
\includegraphics[width=0.9\columnwidth]{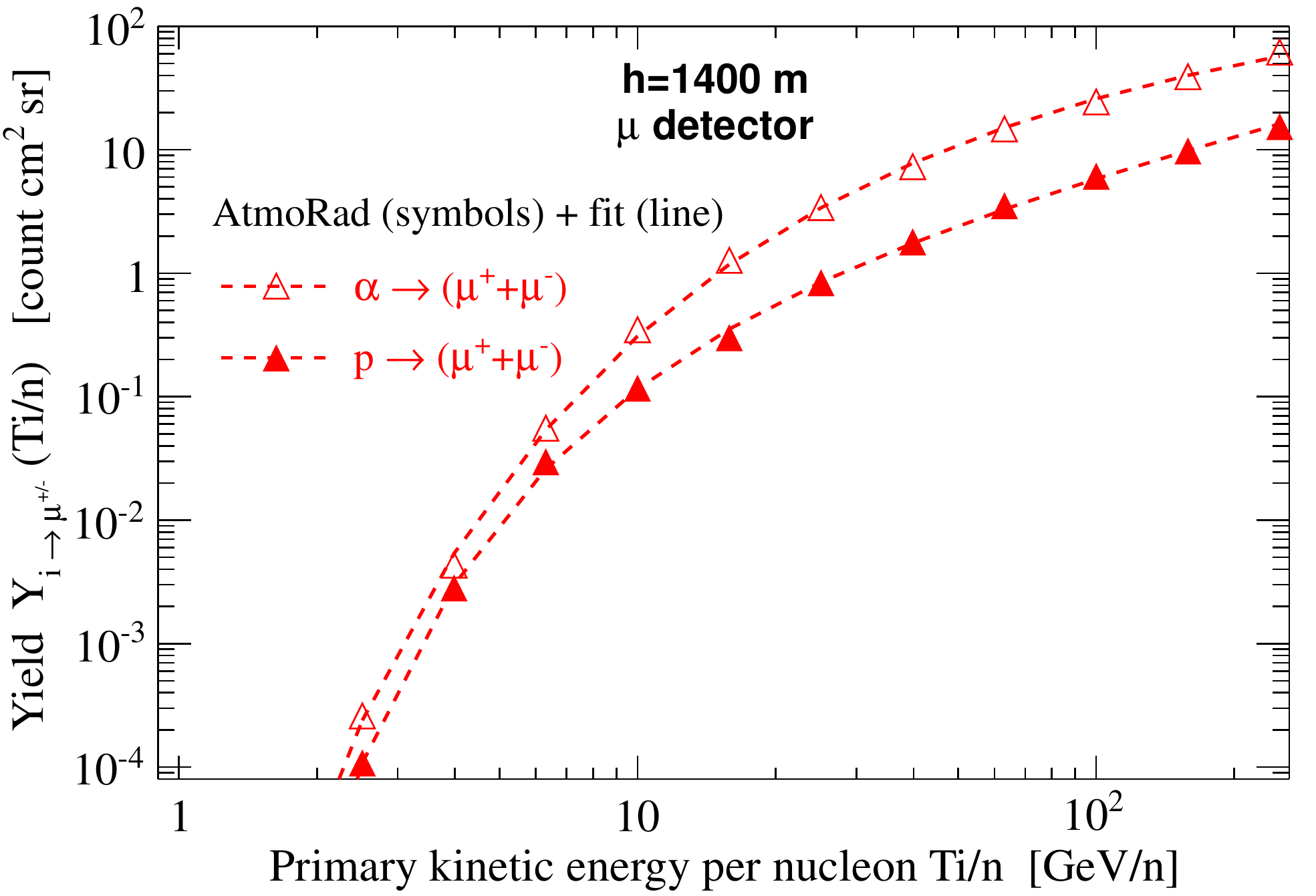}
\end{center}
\vspace{-4mm}
\caption{Levenberg-Marquardt fit of the \textsc{AtmoRad} yield function for muons at +1400~m (Auger altitude). The corresponding fit parameters are gathered in Table~\ref{tab:best-fit_yieldmuon}.}
\label{fig:YieldCoeffs_muons}
\vspace{-3mm}
\end{figure}

At sea level, muons are the most abundant charged particles, and they can be used in principle to monitor solar activity. Experimental aspects related to the detection of atmospheric muons are discussed, e.g., in \citetads{2012GI......1..185C}. Muon telescopes generally consist of layers of charged particle detectors and absorbing material, with the capability to determine the direction of $\mu$ arrival. The quantity of material crossed by the $\mu$ sets the detector threshold, which increases with the zenith angle for multi-directional telescopes. Some astroparticle physics detectors have also shown exquisite sensitivities to muons, as exemplified, on the one hand, by the measurements by L3 magnetic muon spectrometer at the LEP collider at CERN \citepads{2004PhLB..598...15L}, or on the other hand, by the huge array of surface detectors at the Pierre Auger Observatory \citepads{2011JInst...6.1003P}. In particular, The Pierre Auger Observatory, thanks to its 3,000 km$^2$ collection area, provides interesting data in the context of solar activity monitoring. The Auger scaler data (corrected for pressure), publicly available\footnote{\url{http://auger.colostate.edu/ED/scaler.php}}, are 15 minutes averages of the scaler rates, recorded since 2005. The threshold of the scaler mode is very low with a very high efficiency, so that in practice, it allows a muon counter equivalent mode (the scaler data variability were found to be well correlated with NM variations, \citealtads{2011JInst...6.1003P}).

In order to compare the behaviours of NMs and muon detectors, we calculate from \textsc{AtmoRad} the yield function of a perfect muon detector of 1~m$^2$. In \textsc{AtmoRad}, muons were validated by a cross-comparison with the \textsc{expacs} code, itself validated on CAPRICE~97 data \citepads{1999PhRvL..83.4241K}. The best-fit parameters relying on Eq.~(\ref{eq:yield_fit}) are gathered in Table~\ref{tab:best-fit_yieldmuon}. This parametrisation should provide a fair estimate of the expected variability, e.g., for the Auger scaler data. Above 10  GV, we checked that it is in very good agreement with the results of \citetads{2002JGRA..107.1376P} for protons. It is used in the rest of the paper to illustrate the results to expect from a {\em generic} muon detector.

\subsection{Neutron spectrometers to study NM count rates}
\label{sec:BSS}

Recently, Bonner Sphere Spectrometers (BSS) were deployed at ground level and mountain altitudes in order to characterise the CR-induced neutron spectrum over long-term periods for dosimetry or microelectronics reliability purposes (\citealt{ruhm2009continuous}; \citealtads{,2013ITNS...60.2418H}). Unlike NMs, BSS are only sensitive to the neutron component. However, BSS are far less efficient than NMs and dynamics of one spectrum per hour can be reached at best (in high altitude stations). A BSS designed to cover a wide range of energies (from $10^{-2}$~meV to GeV) generally consists of a set of homogeneous polyethylene (PE) spheres with increasing diameters $d$. A high pressure $^3$He spherical proportional counter placed in the centre allows high detection efficiency. Additionally, spectrometers include some PE spheres with inner tungsten or lead shells in order to increase the response to neutrons above 20~MeV. These extended spheres (HE) behave like small NMs. 

After an unfolding procedure \citepads{2012ITNS...59.1722C}, the neutron spectral fluence rate $\dot\varphi_{n}^{\text{BSS}}(T_n,\vec{r},t)$ can be derived from BSS data (i.e., count rates $M_d(\vec{r},t)$ for each of the $d$-Bonner sphere). The neutron component is very sensitive to local changes induced by meteorological and seasonal effects. BSS measurements allow to quantify such variations and to correlate them with variations expected/observed in NM count rates: we recall that neutrons amount to $\sim 87\%$ of the total count rate in NMs (see Table~\ref{tab:NMOulu}). Considering the local neutron count rate $N_n^{\textrm{X-NM64}}(\vec{r},t)$ of a $X$-NM64 at the BSS coordinates, we have
\begin{equation}
\label{eq:BSStoNM}
N_n^{\textrm{NM}}(\vec{r},t)=\frac{X}{6} \int _0^\infty {\cal E}_n^{\textrm{NM}}(T_n) \,\,\dot\varphi_{n}^{\text{BSS}} (T_n,\vec{r},t)\,\,dT_n.
\end{equation}
BSS measurements are used to study the seasonal snow effects of NMs in Sect.~\ref{sec:SeasonnalEffect}.

\section{Count rates: variations and uncertainties}
\label{sec:impact_factors}

In this section, count rates are calculated from Eq.~(\ref{eq:count-rate}), which involves the yield function ${\cal Y}^{\cal D}_i (h, R)$, the modulated fluxes $J^{\rm TOA}_i(t)$ for all CR species $i$, and the geomagnetic transmission ${\cal T}(R,\vec{r}, t)$. To validate our code, we compare count rate variations (vs $R_c$) to existing latitude surveys (Sect.~\ref{sec:ratevar_vs_rc}). We then propagate, on count rates, IS flux and yield function uncertainties (Sect.~\ref{sec:rates_vs_rcutoff}), and geomagnetic transmission function uncertainties (Sect.~\ref{sec:Rcut}). We conclude the section with time dependent effects (on count rates) unrelated to solar modulation  (Sect.~\ref{sec:SeasonnalEffect}). 

Here, the altitude is set to $h=0$~m, but this value is not important. Indeed, although partial yield functions depend strongly on $h$ (see Table~\ref{tab:best-fit_yield}), the altitude and energy dependences are not coupled. This remains true for the total yield function (hence count rates) of $\mu$ detectors, and mostly true for NMs\footnote{A coupling at the percent level exists because the NM total count rates receive a small contribution from $\mu$ (see Tab.~\ref{tab:NMOulu}), whose altitude and energy dependences are different from that of the main nucleonic contributions.}: the altitude dependence acts as a global factor that disappears when relative rate variations and relative errors are considered (as done below).

\subsection{Count rate variation $N(R_c,\,\phi)$ vs $R_c$}
\label{sec:ratevar_vs_rc}

\subsubsection{Relative contribution per rigidity bin}

\begin{figure}[!t]
\begin{center}
\includegraphics[width=\columnwidth]{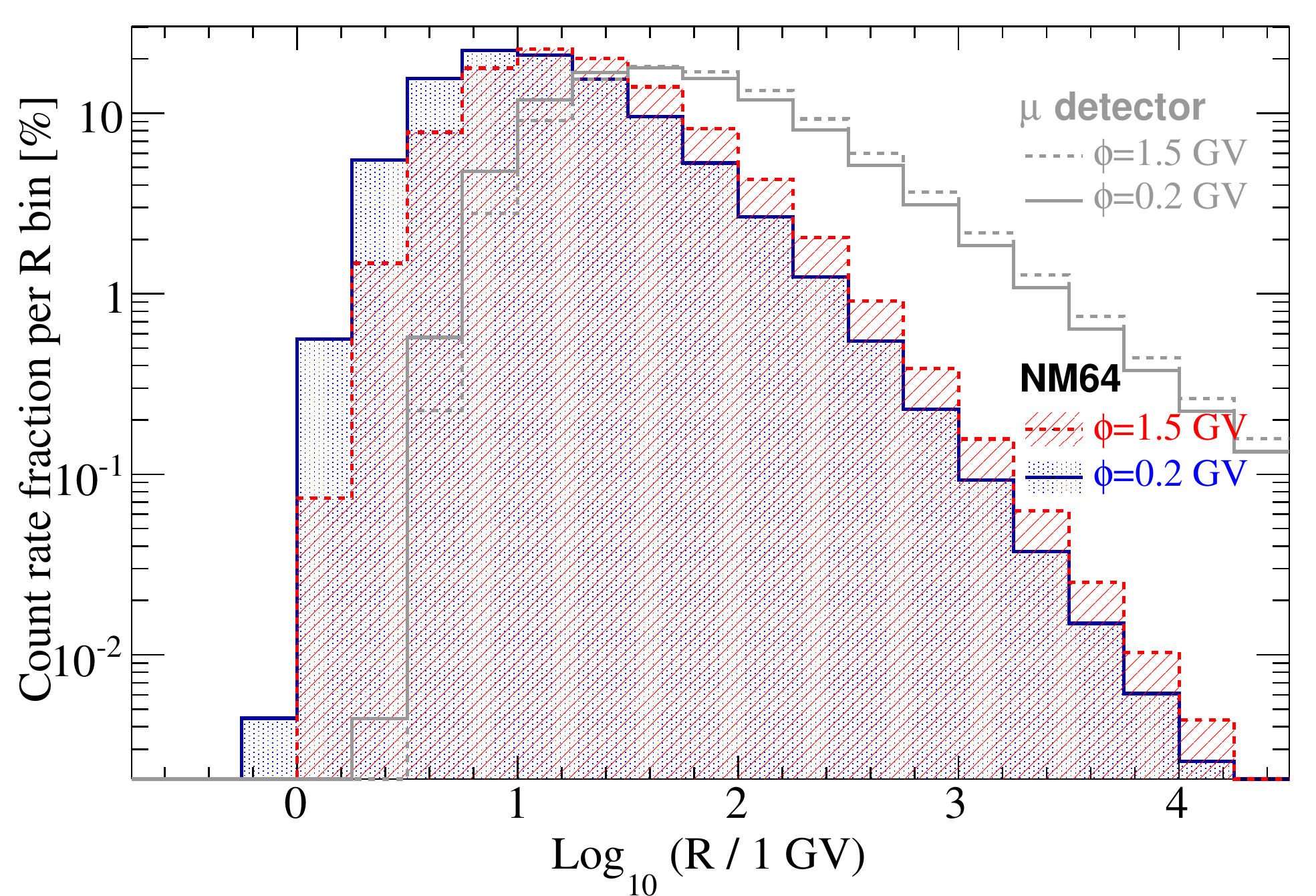}
\includegraphics[width=\columnwidth]{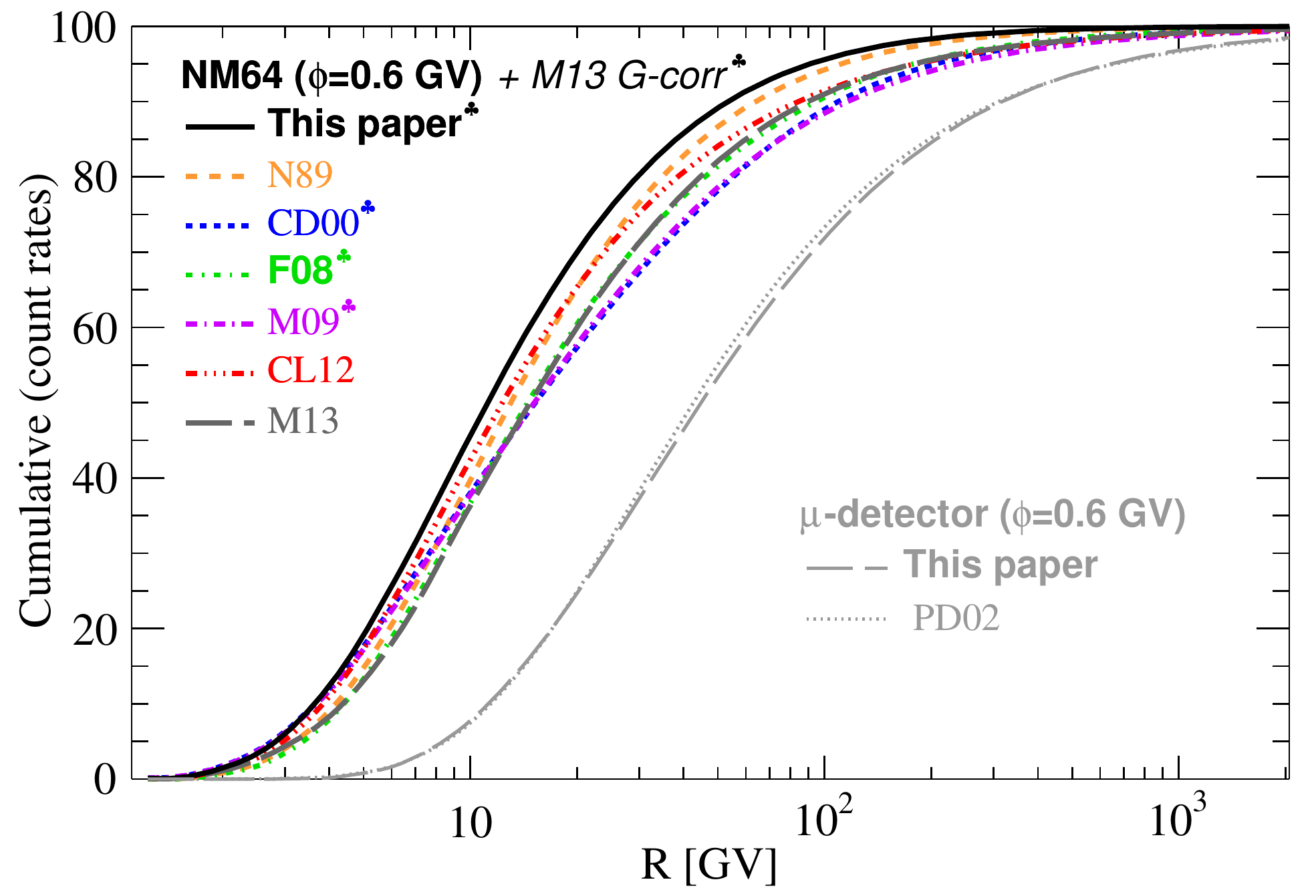}
\end{center}
\vspace{-5mm}
\caption{Origin of count rates for a polar response function (no rigidity cut-off). {\bf Top panel:} relative contribution per rigidity bin at two modulation periods for NM64 detectors (blue and red shaded bars) and $\mu$-like detectors (thick solid and dashed grey lines). For instance, primary CRs between 10 and 20 GV contribute to 20\% of the total counts in NMs, but only 10\% in $\mu$ detectors. {\bf Bottom panel:} cumulative of the count rates for several yield functions for NM64 (compared in Fig.~\ref{fig:ratio_yields}, see also App.~B) and $\mu$-detector (PD02 is the $\mu$ yield function from \citealtads{2002JGRA..107.1376P}). NM64 yield functions indicated with a club$^\clubsuit$ ($G$-corr) are corrected for the $G$ factor of \citetads{2013JGRA..118.2783M}|see text and Eq.~(\ref{eg:g-factor}) for details.}
\label{fraction_per_rbin}
\vspace{-3mm}
\end{figure}
The top panel of Fig.~\ref{fraction_per_rbin} shows (considering the contribution of all CR species) for a polar response function (i.e. $R_c=0$) the fractional contribution per rigidity bin of the integrand ${\cal Y}\times J^{\rm TOA}$. The two shaded areas correspond to a period of minimal (blue shaded area) and maximal (red hatched area) modulation level. For CRs below 1 GV and above $\sim 50$~TV, the contribution is less than 1\% of the total: this mitigates the impact of having large differences at low and high energy between various yield function parametrisations (see Fig.~\ref{fig:ratio_yields}). The CR rigidity range contributing most to the count rates is shifted to higher energy when the modulation level is increased, or when sub-polar NM detectors ($R_c>0$) are considered.

The bottom panel of Fig.~\ref{fraction_per_rbin} proposes a complementary view, that is the cumulative of the count rates with $R$ (also for a polar response function). For MC-based yield function parametrisations (this paper, CD00, F08, M09), we take into account the $G$-correction of \citetads{2013JGRA..118.2783M}. All parametrisations give quite similar results, where 50\%, (resp. 80\% and 90\%) of the count rates are reached when integrating up to 10~GV (resp. 30 and 70~GV). The value of the highest energy contributing is very sensitive to the high-energy slope of the yield function. For instance, we chose for CD00 and F08 (see App.~B) a high energy extrapolation ${\cal Y}\propto E^{0.5}$. Would a ${\cal Y}\propto E$ chosen instead, 80\% of the total count rate would be shifted from $R=30$~GV to 80~GV. It is thus important for future MC-based calculations to push the rigidity range up to 1 TV.

On both the top and bottom panels of Fig.~\ref{fraction_per_rbin}, the result for a muon detector is shown in grey lines. The solid and dashed lines correspond respectively to periods of minimal and maximal modulation level. With respect to NMs, the mean energy contributing to count rates is shifted to higher energy, in a region where the impact of the solar modulation is smaller. Hence, the relative count rate variation $\Delta N/N$ to a change of the modulation level $\Delta\phi$ is smaller for $\mu$ detectors than for NMs.

\subsubsection{Comparison to latitude survey data}
\label{sec:rates_vs_latitudesurvey}

Latitude survey experiments onboard planes, trucks, or ships cruising between equatorial and polar regions is another tool to derive yield functions and/or to compare with direct count rate calculations \citepads{2009crme.book.....D}. Monthly long ship surveys are generally performed during solar minimum periods|the most stable in terms of modulation changes|, in order to be only sensitive to rigidity cutoff effects \citepads{2000SSRv...93..285M}. 

\paragraph{Data available}
The solar cycle has an 11-year periodicity, and several surveys were carried out at minimum activity since the 50's: 1954 \citepads{1956CaJPh..34..968R,1968JGR....73..353K}, 1965 \citepads{1965ICRC....1..553C,1968JGR....73..353K}, 1976 \citepads{1979ICRC....4..352P,1980SAJPh...3...73S}, 1986 \citepads{1989JGR....94.1459M}, 1997 \citepads{2000JGR...10521035I}, but none that we are aware of in the last solar minimum period. The data from 1965 are discarded since they were found to differ from the similar 1954 and 1976 survey data \citepads{1979ICRC....4..352P}. Data from 1976, 1986, and 1997, are also found to be in agreement (see, e.g., Fig.~4 of \citealtads{2013JGRA..118.2783M}), with 1997 data thoroughly  corrected from meteorological and geomagnetic effects \citepads{2000JGR...10521035I,2000JGR...10521025V,2000JGR...10521047D}. 

\begin{table}
\caption{NM latitude surveys and corresponding CR data available at the same epoch.}
\vspace{-1mm}
\label{tab:nm_cs_crs}
\small
\begin{center}
\begin{tabular}{rcrlc}
\hline
Survey date &  Ref.   & CR data & Exp.  & Ref.  \\\hline
\multirow{2}{*}{12/75-11/76}&\multirow{2}{*}{[Pot79]}& 07/77 & Balloon & \multirow{2}{*}{[Web83]}\\\vspace{2mm}
             &            & 10-11/77               & Voyager1\\
05/86-10/87  &[Mor89]\!\! & 01-12/87               & Voyager2$^\dagger$&[Seo94]\\\vspace{2mm}
12/96-03/97  &[Iuc00]     & 07/97                  & BESS~97 &[Shi07] \\
2006-08      & -          &\!\!\!\!\!\!\!07/06-12/08$^\ddagger$\!\!&\!\!PAMELA\!\!& [PAM11]\!\!\! \\
\hline
\end{tabular}
\\
{\footnotesize 
$^\dagger$ $\phi$ estimated (from data at 23~AU) to be 500 MV  \citepads{1994ApJ...432..656S}.\\
$^\ddagger$ Representative of 1986/1987 \citepads{2012JGRA..11712103C}.
}
\end{center}
\vspace{-4mm}
\end{table}
To compare these data with calculations based on CR fluxes and yield functions, the knowledge of the modulation level to apply is critical. With no CR data available in 1954, we have to base our calculation on 1976, 1986, and 1997 CR measurements. We list in Table~\ref{tab:nm_cs_crs} the epoch of these surveys and the closest (in time) CR data available (retrieved from {\sc crdb}\footnote{\url{http://lpsc.in2p3.fr/crdb}}).
\begin{figure}[!t]
\begin{center}
\includegraphics[width=\columnwidth]{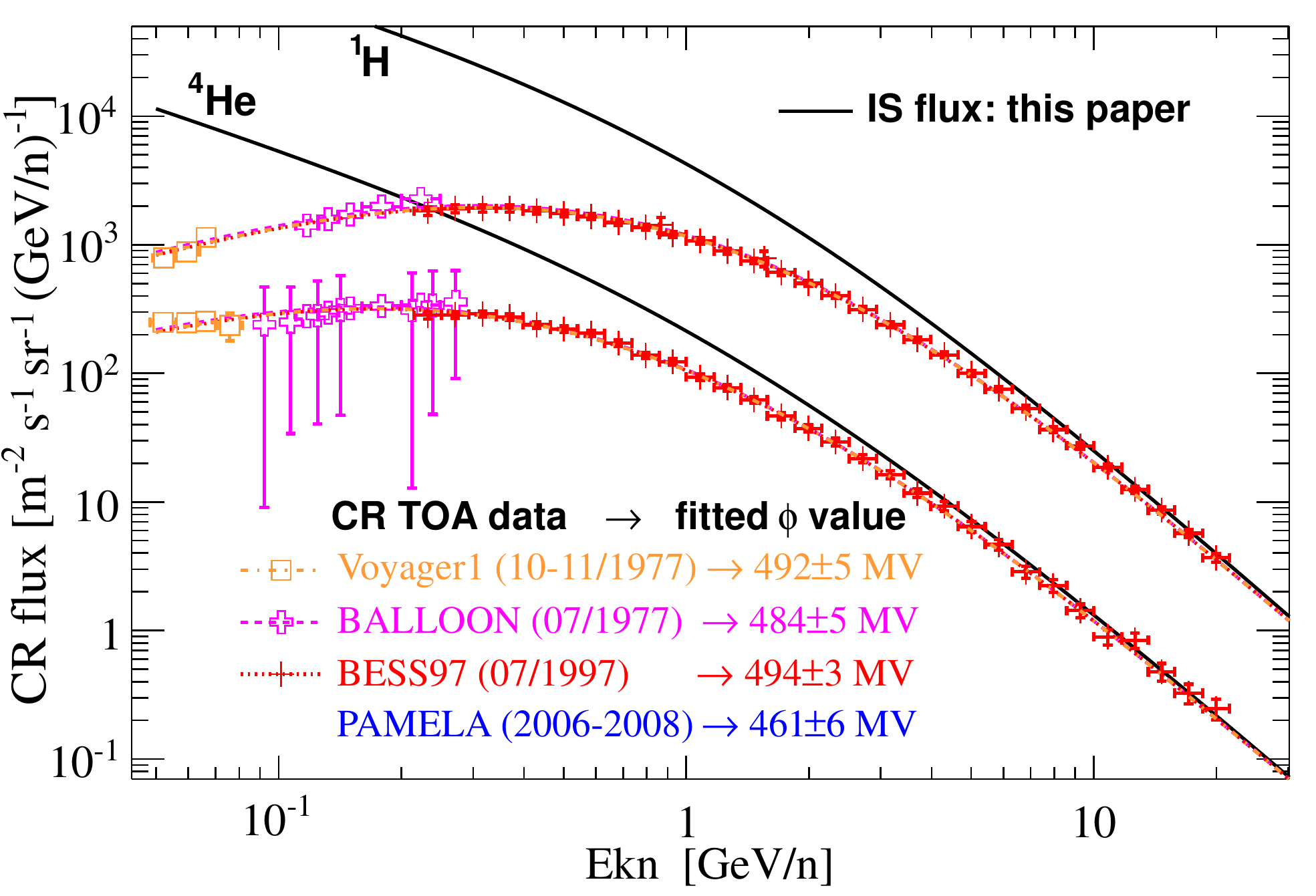}
\includegraphics[width=\columnwidth]{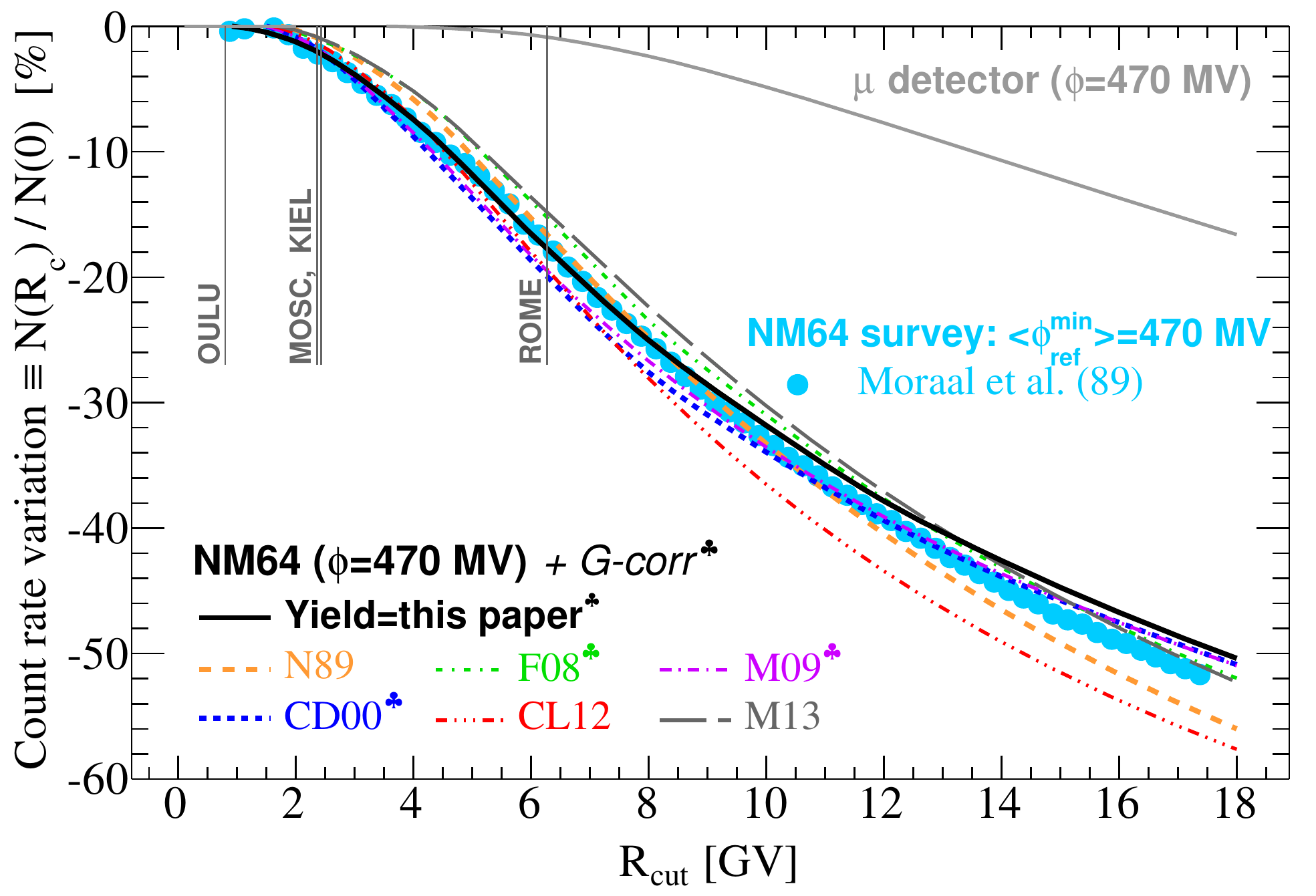}
\end{center}
\vspace{-5mm}
\caption{{\bf Top panel:} CR data at epochs of NM latitude surveys (see Table~\ref{tab:nm_cs_crs}), and best-fitted $\phi$ values (and corresponding TOA fluxes). {\bf Bottom panel:} Count rate relative variation $N(R_c)/N(0)$ for NM64 (various parametrisations, see App.~B) and $\mu$-like detectors vs $R_c$. Yield functions indicated with a club$^\clubsuit$ ($G$-corr) are corrected for the $G$ factor Eq.~(\ref{eg:g-factor}) of \citetads{2013JGRA..118.2783M}. Circles are data from NM latitude survey data \citepads{1989JGR....94.1459M}. Vertical segments show the rigidity cut-off for a sample of NMs.}
\vspace{-3mm}
\label{fit_hhe_epoch_nmsurvey}
\label{deltan_vs_r}
\end{figure}

\paragraph{Modulation level at solar minimum}
The CR data listed in Table~\ref{tab:nm_cs_crs} allow us to determine consistently (i.e., given our IS flux parametrisation) the modulation level $<\phi_{\rm ref}^{\rm min}>$. This level applies for epochs of minimal modulation. Figure~\ref{fit_hhe_epoch_nmsurvey} shows the bets-fit $\phi$ required to match CR TOA fluxes. A simultaneous fit of H and He data is performed using the force-field approximation (Sect.~\ref{sec:force-field}) and our reference IS flux parametrisation|Eq.~(\ref{eq_is_flux_dekn}) and Table~\ref{tab:jis_fit}|. Note that the value of $\phi$ for PAMELA is directly reproduced from Table~\ref{tab:phi_fitted}. We find that all $\phi$ values are consistent with one another, and we take in the following $<\phi_{\rm ref}^{\rm min}>=470\pm20$~MV. This value is slightly higher than the one used in \citetads{2013JGRA..118.2783M}\footnote{The latter is based on \citetads{2011JGRA..116.2104U} calculation, who use U05 (see App.~A) IS flux. Taking into account the correspondence $\phi_{\rm U05} = \phi_{\rm ref} - 50$~MV in Fig.~\ref{fig:jis_phishift}, their value $\phi_{\rm U05}^{\rm min}=400$~MV translates to $\phi_{\rm ref\leftarrow U05}^{\rm min}=450$~MV in terms of our IS flux.}.

\paragraph{NM and $\mu$-detector latitude dependence}
Figure~\ref{deltan_vs_r} shows, as a function of $R_c$, a comparison of the normalised (at $R_c=0$~GV) count rate variations (for various yield functions) to survey data (only the 1986-1987 survey is shown for clarity). The solar modulation level is set to $\phi=470$~MV, appropriate for a period of minimal activity (see above). We find, in agreement with the conclusions of \citetads{2013JGRA..118.2783M}, that taking into account the $G$-correction factor (see Eq.~\ref{eg:g-factor}) gives a much better match to latitude survey data (the curves without this correction are not shown). NM-survey based yield functions (N89 and CL12) give a similar albeit slightly less good match. Note that the scatter observed from the use of the various yield functions is larger than the variation obtained by shifting the modulation $\phi_{\rm ref}^{\rm min}$ by $\pm20$~MV.

Concerning the variation of count rates with $R_c$, as already underlined, count rates decrease when $R_{c}$ increase|$R_c$ is the lower boundary of the integral Eq.~(\ref{eq:count-rate}). Over the whole $R_c$ range, the variation is less marked for a $\mu$-like detector ($\lesssim 20\%$, grey line) than for NMs ($\lesssim 50\%$). This is understood as the mean rigidity of CRs contributing to the count rate is higher for the latter than for the former (see bottom panel of Fig.~\ref{fraction_per_rbin}).

\subsection{Count rate relative uncertainty $\frac{\Delta N}{N}(R_c,\,\phi)$}

For an overview of the various sources of uncertainties involved in count rate calculations, we refer the reader to the reference textbooks of \citetads{2004ASSL..303.....D} for meteorological effects, and \citetads{2009crme.book.....D} for cut-off rigidity effects.

\subsubsection{Error from IS flux and yield function modelling}
\label{sec:rates_vs_rcutoff}

\begin{figure}[!t]
\begin{center}
\includegraphics[width=\columnwidth]{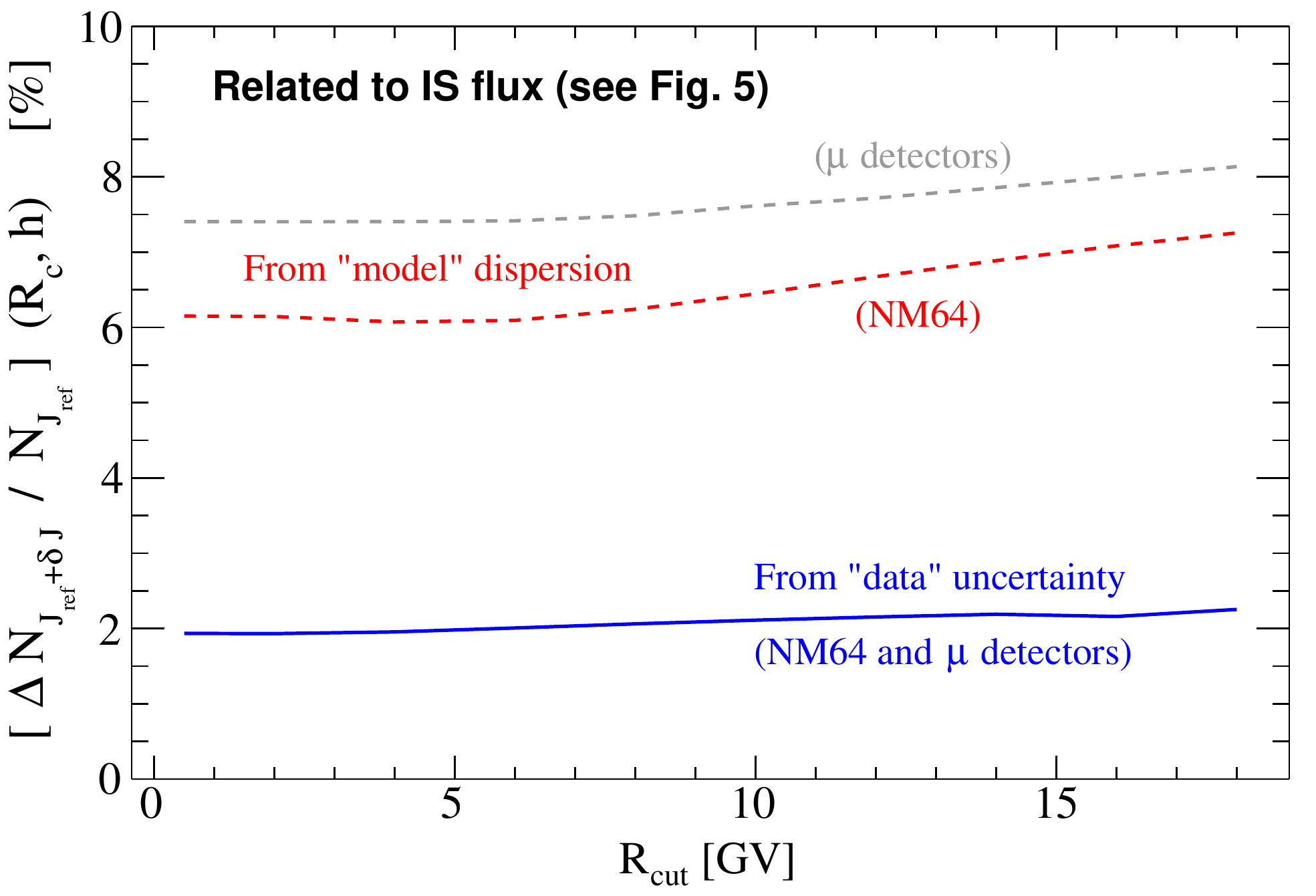}
\includegraphics[width=\columnwidth]{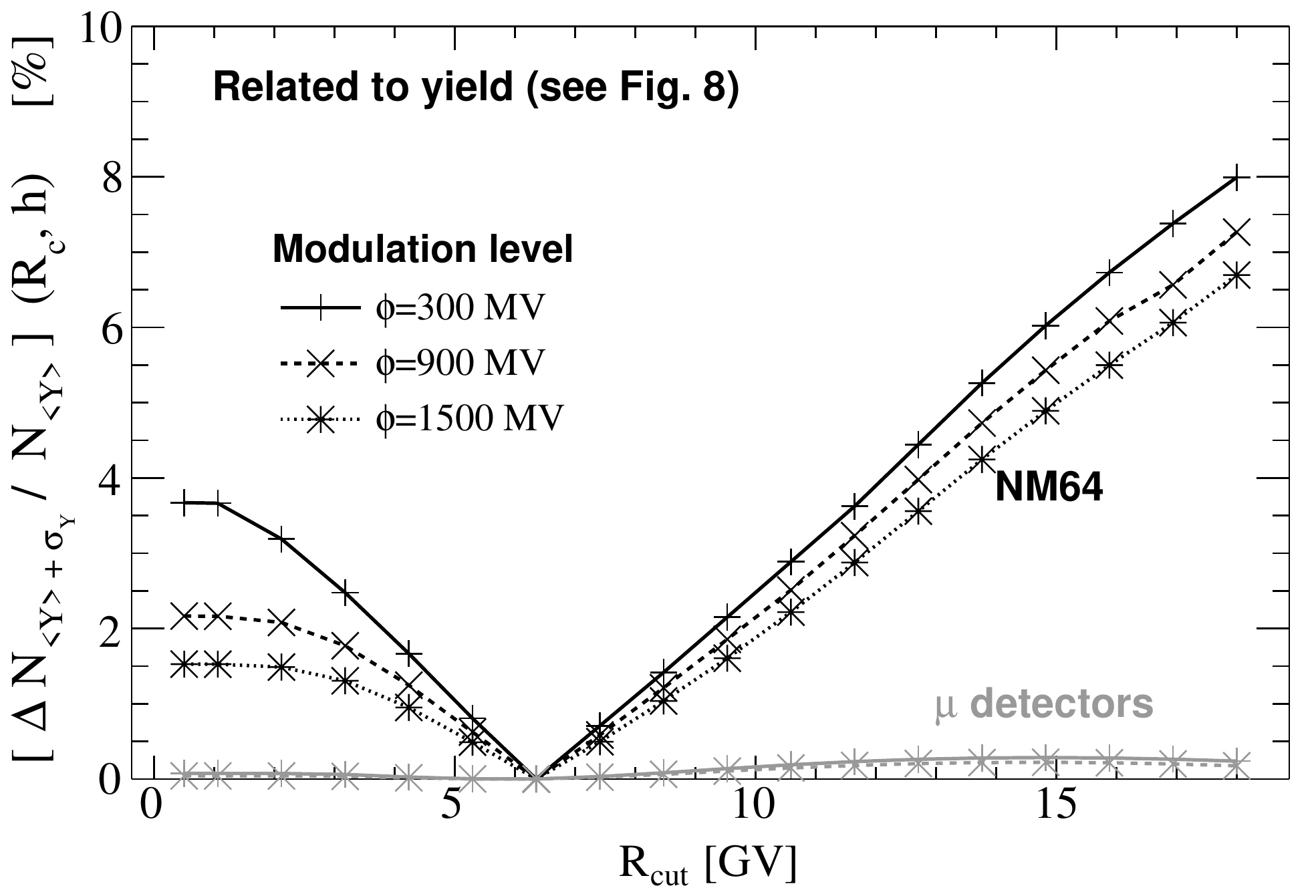}
\end{center}
\vspace{-5mm}
\caption{Relative uncertainty on count rates for NM64 and $\mu$ detectors. {\bf Top panel:} uncertainty propagated from IS flux uncertainties. {\bf Bottom panel}: uncertainty propagated from yield function dispersion (for NM64 in black symbols, $\mu$ in grey symbols). Note that the pinch observed is related to the arbitrariness in the choice of a reference $R_c$ to compare NM64 yield functions with one another.}
\label{rate_err_yieldjis}
\vspace{-3mm}
\end{figure}

The two panels in Fig.~\ref{rate_err_yieldjis} show the errors on the count rate calculation as a function of $R_c$ (for NM64 and $\mu$ detectors) propagated from the uncertainties on CR fluxes (top) and yield functions (bottom). For simplicity, the errors are symmetrised, i.e. we consider $({\rm err}_{\rm min}+{\rm err}_{\rm max})/2$.

For NM64 detectors, the solid blue line (resp. dashed red line) corresponds to the propagation of errors on the IS flux `data' (resp. `model') as discussed in Sect.~\ref{sec:JIS_uncertainty} and shown in Fig.~\ref{fig:uncertainty_toa}. These uncertainties are, to a very good approximation, independent of the solar modulation level and of the rigidity cut-off. It means that they contribute only to a global shift in count rate times series  (no time dependence, and no detector location dependence). This uncertainty is at the level of $2-6$\% for NMs, with a slightly larger range $2-8$\% for $\mu$ detectors (dashed grey line). Future CR data (e.g. AMS-02) will likely shrink these uncertainties at the percent level. 

On the same figure, the lines with symbols show the uncertainty related to the existing dispersion among the proposed NM64 yield functions in the literature (see Sect.~\ref{sec:6NM64}, Fig.~\ref{fig:ratio_yields}, and App.~B). We recall that yield functions are generally considered up to a normalisation. To get a meaningful result, we re-normalised all count rates to a reference value (set arbitrarily to $R_c=6.3$~GV), leading to a pinch in the curves. There is a mild dependence on the modulation level, but the overall uncertainty is estimated to be below 8\% over the whole $R_c$ range, and more particularly, at the $2-5$\% level for NMs located at $R_c<10$~GV. The dispersion is much smaller for $\mu$ detector ($<0.2\%$, grey lines and symbols): the latter is probably not conservative and may reflect the fact the only two parametrisations of their yield function are used for this study.

\subsubsection{Uncertainties from transmission function ${\cal T}$}
\label{sec:Rcut}

A key parameter for calculations is the transmission function of charged particles in the geomagnetic field \citepads{2009crme.book.....D}. Several factors can be taken into account to have an estimate of the associated uncertainties on the count rates. Indeed, the transmission depends on the geographical longitude and latitude $(\varphi ,\lambda)$, which can be calculated for a given state of the Earth magnetosphere \citepads{2000SSRv...93..305S}. The latter varies in time, and its full description requires both the long-term evolution of the geomagnetic field (International Geomagnetic Reference Field\footnote{\url{http://www.ngdc.noaa.gov/IAGA/vmod/igrf.html}}) and the short-term magnetospheric field model \citepads{2005JGRA..110.3208T,2009JGRA..114.0C21K,2013AnGeo..31.1745T}. A good summary of the past studies and findings is given in \citetads{2003AdSpR..32...95S,2009AdSpR..44.1107S}. 

The complicated structure of the geomagnetic field leads to a quasi-random structure of allowed and forbidden orbits, denoted `penumbra'. The effective vertical rigidity cut-off (see \citealtads{1991NCimC..14..213C}), used so far in this analysis ($R_c^{\rm eff}\equiv R_c$), consists in a weighted average value accounting for the allowed bands (between the upper and lower cut-off values). With the assumption that all regions contribute the same (flat spectrum hypothesis), it is given by \citepads{2008AdSpR..42..510D}:
\begin{equation}
  R_c^{\rm eff} \approx R_{\rm upper} - \sum_{i=R_{\rm lower}}^{R_{\rm upper}} \Delta R_i^{\rm allowed}.
\label{eq_rc_def}
\end{equation}
In this approach, it follows that the transmission function is described by the step function $H(R-R_c^{\rm eff})$. 

  \paragraph{Short and long time variation of $R_c$}
At each geomagnetic position and time $t$, an effective vertical cutoff rigidity $R_c$ can be calculated. For long term evolution, calculations with a fine position grid have been carried out at different epochs \citepads[e.g.,][]{2008ICRC....1..733S,2008ICRC....1..737S}. Over a 50 year (resp. 2000 year) evolution, increases or decreases of $R_c$ at the level of $\sim 5-10\%$ (resp. 30\%) are expected \citepads{2001ICRC...10.4063S,2003ICRC....7.4229F,2012IAUS..286..234M,2013AnGeo..31.1637H}. 

Short-term changes are more challenging computationally \citepads{2003ICRC....7.4241S,2006AdSpR..37.1206S,2008ICRC....1..769B}: small time step calculations are required and an evolving magnetospheric model must be considered. On short timescales, an enhanced geomagnetic activity leads to a temporary change of the effective vertical rigidity cutoff \citepads{1983JGR....88.6961F,1986JGR....91.7925F,2008AdSpR..42.1300K,2012IAUS..286..234M}. In particular, during geomagnetic storms, decreases of $R_c$ by a few GV for several hours are predicted, and confirmed from NM data \citepads{2009AcGeo..57...75D,2013AdSpR..51.1230T}.

\begin{figure}[!t]
\begin{center}
\includegraphics[width=\columnwidth]{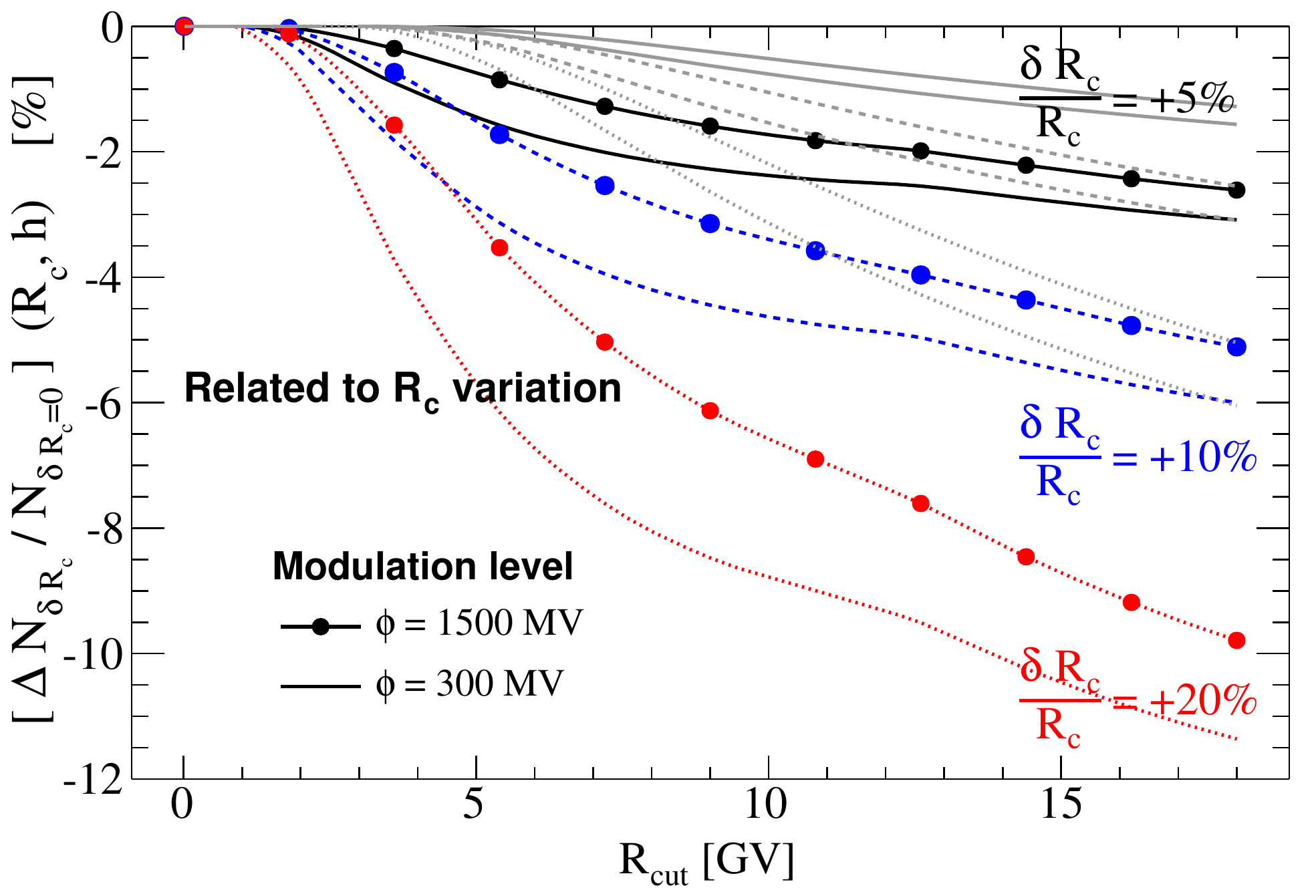}
\includegraphics[width=\columnwidth]{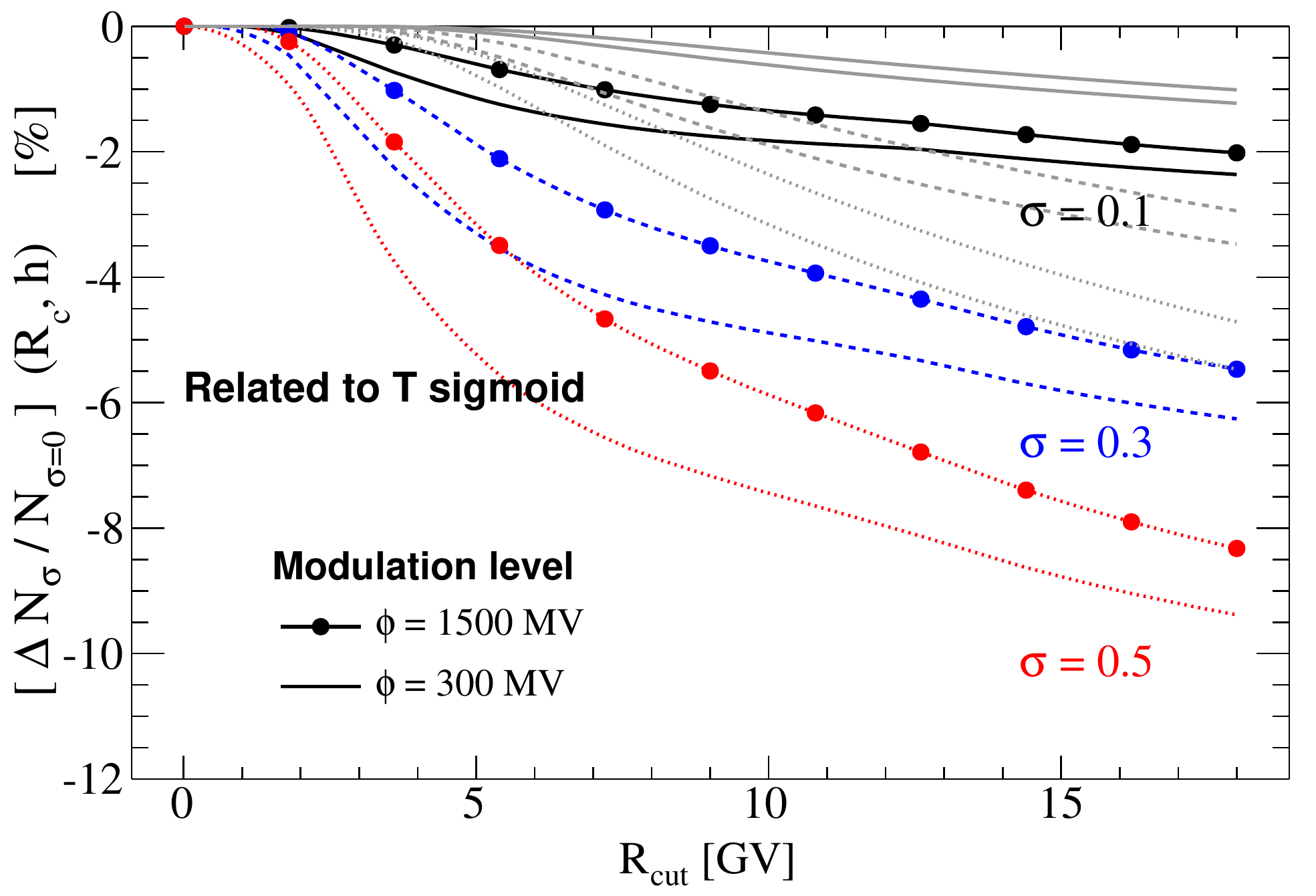}
\end{center}
\vspace{-5mm}
\caption{Relative uncertainty on count rates for NM64 and $\mu$ detectors (light grey curves) as a function of $R_c$, for two extreme values of the modulation level (lines, $\phi=300$~MV; symbols, $\phi=1500$~MV). {\bf Top panel}: count rate change when moving from $R_c$ to  $R_c+\delta R_c$. {\bf Bottom panel:} count rate change from the use of a sigmoid Eq.~(\ref{eq:sigmoid}) instead of a step function. }
\label{rate_err_rctransm}
\vspace{-3mm}
\end{figure}

The impact of changing $R_c$ to $R_c+\delta R_c$ is shown in the top panel of Fig.~\ref{rate_err_rctransm}. Whenever $R_c$ is increased, the count rates decrease, with a milder impact at epochs of high solar activity than for low activity. This is related to the steepness of the decrease of the count rate with $R_c$ shown in Fig.~\ref{deltan_vs_r}. For detectors located at $R_c<10$~GV, count rates over 50 years vary at most by -4\% for NM64 (blue dashed lines), and -1\% for $\mu$ detectors (dashed grey lines).

\paragraph{Allowed and forbidden rigidity: penumbra} A better description of the transmission function is the use of a sigmoid function, as done in the context of NMs  \citepads{2008AdSpR..42.1300K}, or the CR experiments HEAO-3 \citepads{1988A&A...193...69F} and AMS-01 \citepads{2006JGRA..111.5205B,2009AdSpR..43..385B}. The step function $H(R-R_c^{\rm eff})$ is the limit of a sigmoid of zero width. To see how good is this zero-width approximation, we compare count rates calculated with it and with the following sigmoid shape (centred on $R_c$):
\begin{equation}
\label{eq:sigmoid}
   {\cal T}(R) = \frac{1}{2}\left[1+{\rm erf}\left(\frac{R-R_c}{\sqrt{2}\;\Delta R}\right)\right].
\end{equation}
It is useful to define the width $\sigma$ of the sigmoid function to be 
\[
\sigma \equiv \frac{\Delta R}{R_c}.
\]
A typical range of values reproducing best AMS-01 data (depending on the position) is $\sigma\sim 0.1-0.2$ for $R_c\gtrsim 2$~GV, and $\sigma\sim 0.1-0.5$ for $R_c<2$~GV.

The bottom panel of Fig.~\ref{rate_err_rctransm} shows the ratio of count rates calculated for various values of $\sigma$ to count rates in the step function approximation. The sigmoid case gives less count rates than the step function case, an effect that increases with $R_c$ and with the width of the sigmoid. Moreover, the larger the modulation level $\phi$ (chained lines compared to lines), the smaller the effect. These behaviours are well understood if one keeps in mind how the contribution per rigidity band (described in Fig.~\ref{fraction_per_rbin}) varies: compared to the step function, the sigmoid allows less contributions above $R_c$ (where it matters most), and more contributions below $R_c$. For mild values of $\sigma=0.3$, the change below $R_c=10$~GV (where most stations lie) is \mbox{-4\%} at most for NM64 (blue dashed lines), and -2\% for $\mu$ detectors (dashed grey lines).

\paragraph{Obliquely incident particles: apparent cut-off rigidity} With the advance of more powerful computers, obliquely incident CRs|and the ensuing secondary particles reaching the detector|could be considered \citepads{1963JGR....68..345R,1997JGR...10226919C,2008AdSpR..42..510D}. Instead of weighting each vertical direction of the penumbra similarly as in Eq.~(\ref{eq_rc_def}), the vertical and non-vertical incident CRs are weighted according to the zenith angle dependent rigidity cutoff (assuming at first order that there is no azimuth dependence). This defines a so-called apparent vertical cut-off rigidity $R^{\rm app}_c$:
\begin{equation}
   \!\!\!\!R^{\rm app}_c (R_c)\equiv \frac{\int_0^{2\pi} d\phi  \,\int_0^{\pi/2} \,\,R_c(\theta,\,\phi) \times {\cal Y}(R_c,\,\theta,\,\phi) \, \, d\theta}{\int_0^{2\pi}  d\phi  \,\int_0^{\pi/2}\,\, {\cal Y}(R_c,\,\theta,\,\phi) \,  \,d\theta}.
\end{equation}
Its calculation is more demanding than that for $R_c^{\rm eff}$, though approximations to speed up the calculation exist \citepads{1997ICRC....2..389B}. In terms of the apparent cut-off rigidity, Eq.~(\ref{eq:count-rate}) for count rates becomes
\begin{eqnarray}
\!\!N^{\cal D}({\vec{r}}) \!&=&\!\!\! \int_{\rm sky}\!\!\!\!\! d\Omega \int_{R_c(\Omega)}^{\infty}{\cal Y}^{\cal D}_i(R,h,\Omega)\, J(R)  \;dR \nonumber\\
&=&\!\!\!\int_{\rm sky}\!\!\!\!\! d\Omega \int_{R_c^{\rm app}}^{\infty}{\cal Y}^{\cal D}_i(R,h,\Omega)\, J(R) \;dR.
\end{eqnarray}
All our previous calculations still apply, replacing the vertical effective rigidity cut-off $R_c$ by the apparent rigidity cut-off $R^{\rm app}_c$ values.

Folding an NM64 yield function (calculated with FLUKA and HEAVY packages) with rigidity cut-off maps, \citetads{1997JGR...10226919C} found $R_c^{\rm app}>R_c$. The difference is $\sim 10\%$ for $R_c \approx 0.1$~GV, then it decreases and stay at $\sim 3\%$ above $R_c\gtrsim 8$~GV (see their fig.~11). Further investigations by \citetads{2000JGR...10521047D,2008AdSpR..42..510D} rely on a parametrised angular distribution of the yield function compared with NM survey data. We fit their Fig.~5 and obtain
\begin{equation}
   \left(\frac{R_c^{\rm app}\!-\!R_c}{R_c}\right)\! \times \!100 \!=\! 
   \begin{cases}
     \!\frac{1.8}{R_c}\!-\!1.7\!+\!0.47R_c & \!\!\!{\rm if~} R_c>1,\\
      0 &\!\!\! {\rm otherwise}.
   \end{cases}
\end{equation}
\citetads{2008AdSpR..42..510D} results go in the same direction as \citetads{1997JGR...10226919C} ones, though the $R_c$ dependence of the relative error is slightly different. The maximum shift of $R_c$ to $R_c^{\rm app}$ is $\sim 6\%$ for $R_c=18$~GV, and the shift decreases with decreasing $R_c$. The impact of this change on the count rate calculations can be directly read off the top panel of Fig.~\ref{rate_err_rctransm}.

We underline that the effects related to the geomagnetic field are quite complex, and may be not all accounted for, as possibly illustrated by the not yet understood long-term decline of South pole neutron rates \citepads{2007JGRA..11212102B,2013JGRA..118.6847B}.

\subsection{Seasonal effects: pressure, temperature, snow coverage and water vapour}
\label{sec:SeasonnalEffect}

NM or muon detector count rates can be affected in many ways by meteorological and seasonal effects \citepads{2004ASSL..303.....D}. The quantities considered in this study are atmospheric pressure, temperature, the water vapour, and the snow coverage. For the latter, which is usually not included in public NM data, a comparison with the results of neutron spectrometers is used to assess the strength of the effect.

\subsubsection{Atmospheric pressure}
\label{sec:pe}
Atmospheric pressure effects are discussed in, e.g., \citetads{1968JGR....73.7503H}. Given $N_0$ particles observed at a reference atmospheric pressure $p_0$, their number $N$ at pressure $p$ is $N\approx N_0\exp\left[-\beta(p-p_0)\right]$ \citepads{1974crvs.book.....D}. The quantity $\beta$ is the barometric coefficient, which is obtained correlating secondary CR intensities and data for the atmospheric pressure. As illustrated in \citetads{2013NewA...19...10P}, $\beta\sim 0.72\%$~hPa$^{-1}$ for the Athen NM station. Considering a typical 20~hPa variation translates into a $\sim 13\%$ count rate variation. The barometric coefficient strongly depends on the station location and on the considered detector \citetads{2004ASSL..303.....D}. Publicly distributed data are generally corrected for this effect. However, uncertainties on the barometric coefficients $\Delta\beta\sim 0.02$\% hPa$^{-1}$ \citepads{2011AdSpR..47.1140C} lead to corrected data with count rate uncertainties $\Delta N/N|_{\text{Pressure}}\sim 0.2\%$.

\subsubsection{Atmospheric temperature}
\label{sec:te}
The well-known temperature effect is detailed, e.g., in \citetads{2000JGR...10521035I}. For NMs, it amounts to -0.03\%/$^\circ$C, that is an Antarctica-to-equator temperature effect  $\sim 1\%$, which is also the order of magnitude of the seasonal effect $\Delta N/N|_{\text{Temperature}}\sim 1\%$. 

For muons, the temperature effect, which is dominant over all other effects, is discussed, e.g., in \citetads{2011APh....34..401D}. The seasonal effect is $\sim 8\%$ with smaller variation of $1-2$\% on the background of seasonal trend, with a small dependence on $R_c$ \citepads{2000SSRv...93..335C}. It should be pointed out that this effect strongly depends on the zenith angle of incident muons. For $\mu$ data, real time correction for this effect is discussed in \citetads{2011BRASP..75..820B,2012ASTRA...8...41B}. 

\subsubsection{Water vapour}
\label{sec:wv}
As discussed in \citetads{1965ICRC....1..489B} and in \citetads{1966JGR....71.5183C}, an increase of atmospheric water vapour content attenuates the intensity of secondaries seen by a NM. To take into account this effect, they proposed a correction of the barometric coefficient $\beta$. The variation  $\Delta\beta$ is estimated between -0.09\%~hPa$^{-1}$ and -0.15\%~hPa$^{-1}$, leading to a seasonal effect $\Delta N/N|_{\text{Water Vapour}}\sim 0.2-0.3\%$. Recently and for the first time, this effect was investigated \citepads{2013JHyMe..14.1659R} in a detailed simulation based on the neutron transport code Monte Carlo (MCNPX). In agreement with \citetads{1965ICRC....1..489B}, the sensitivity of fast neutrons to water vapour effect was found to reach $\sim 10\%$ for sites with a strong seasonality in atmospheric water vapour, with a larger decrease of count rates in moist air than in dry air.

For muon detectors, pressure effects are also considered with barometric factor $\beta_\mu$, but lower than NM ones ($\beta_\mu = 0.03\%$~hPa$^{-1}$).

\subsubsection{Snow effect}
\label{sec:se}
The last effect discussed is the impact of snow on the neutron component (no effect is expected for the other secondary components). Recent works have shown that heavy snow fall impact strongly the 1~meV to 20~MeV neutron spectrum (i.e. thermal, epithermal and evaporation domains), while the cascade region ($>$ 20~MeV) is less affected \citepads{2012JGRA..117.8309R,2013JGRA..118.7488C}. As a matter of fact, hydrogen in water molecule is responsible for enhanced thermalisation and neutron absorptions. In the context of NMs, the snow both in the surroundings and above NM shelters affect count rates.
\begin{table}[!t]
\caption{Main characteristics of stations running BSS for continuous measurements at ground level (upper half). The next to last rows give the NM count rates (from neutrons) calculated from BSS measurement using Eq.~(\ref{eq:BSStoNM}) in winter ($N_n^{\rm min}$) and summer ($N_n^{\rm max}$). The last two rows give an estimate for the snow seasonal effect for NM64.}
\vspace{-1mm}
\label{tab:BSS}
\begin{center}
\begin{tabular}{cccc}
\hline
{\small Institute}             &     \multicolumn{2}{c}{{\sc hmgu}} &{\sc irsn/onera} \\
{\small Location}              &  {\em \small Zugspitze} &{\em \small Spitsbergen}&{\em \small Pic du Midi}\\\hline
h (m)                          &      +2,650      &        0       &   +2,885  \\
$R_c$  (GV)                    &       4.0        &        0.0     &    5.6    \\\vspace{2mm}
Date                           &       2004       &       6/07     &    5/11   \\
Spheres PE                     &        14        &         2      &    10     \\\vspace{2mm}
Spheres HE                     &        14        &         2      &    2      \\
$N_n^{\rm max}$ (s$^{-1}$)     &       501.9      &      77.5      &   400.3   \\\vspace{2mm}
$N_n^{\rm min}$ (s$^{-1}$)     &       470.3      &      75.8      &   364.6   \\
$\frac{\Delta N_n}{N_n}$ (\%)         &  -6.3    &      -2.2      &   -8.9    \\
$\frac{\Delta N}{N}|_{\rm Snow}$ (\%) &  -5.3    &      -1.8      &   -7.6    \\\hline
\end{tabular}
\end{center}
\vspace{-4mm}
\end{table}

\paragraph{BSS location and data} Data from three BSS are gathered to study the snow effect of the neutron component in NMs count rates. Two of them are run by the Helmholtz Zentrum M\"{u}nchen (HMGU) at mountain altitude at the summit of the Zugspitze in the German Alps\footnote{Environmental Research Station Schneefernerhaus and near North Pole
at Spitsbergen (Koldewey Station).}, as described in \citet{ruhm2009continuous,ruhm2009measurements} and \citetads{2012JGRA..117.8309R}. The last one is operated by the French Aerospace Lab. (ONERA) at the summit of the Pic du Midi de Bigorre\footnote{ACROPOL: high Altitude Cosmic Ray ONERA/Pic du Midi Observatory Laboratory.}\citealtads{2012JInst...7C4007C,2012ITNS...59.1722C}; \citealtads{Cheminet2013RPD,2013ITNS...60.2411C}). The main features of each experiment are highlighted in the upper half of Table~\ref{tab:BSS}.

\begin{figure}[!t]
\begin{center}
\includegraphics[width=\columnwidth]{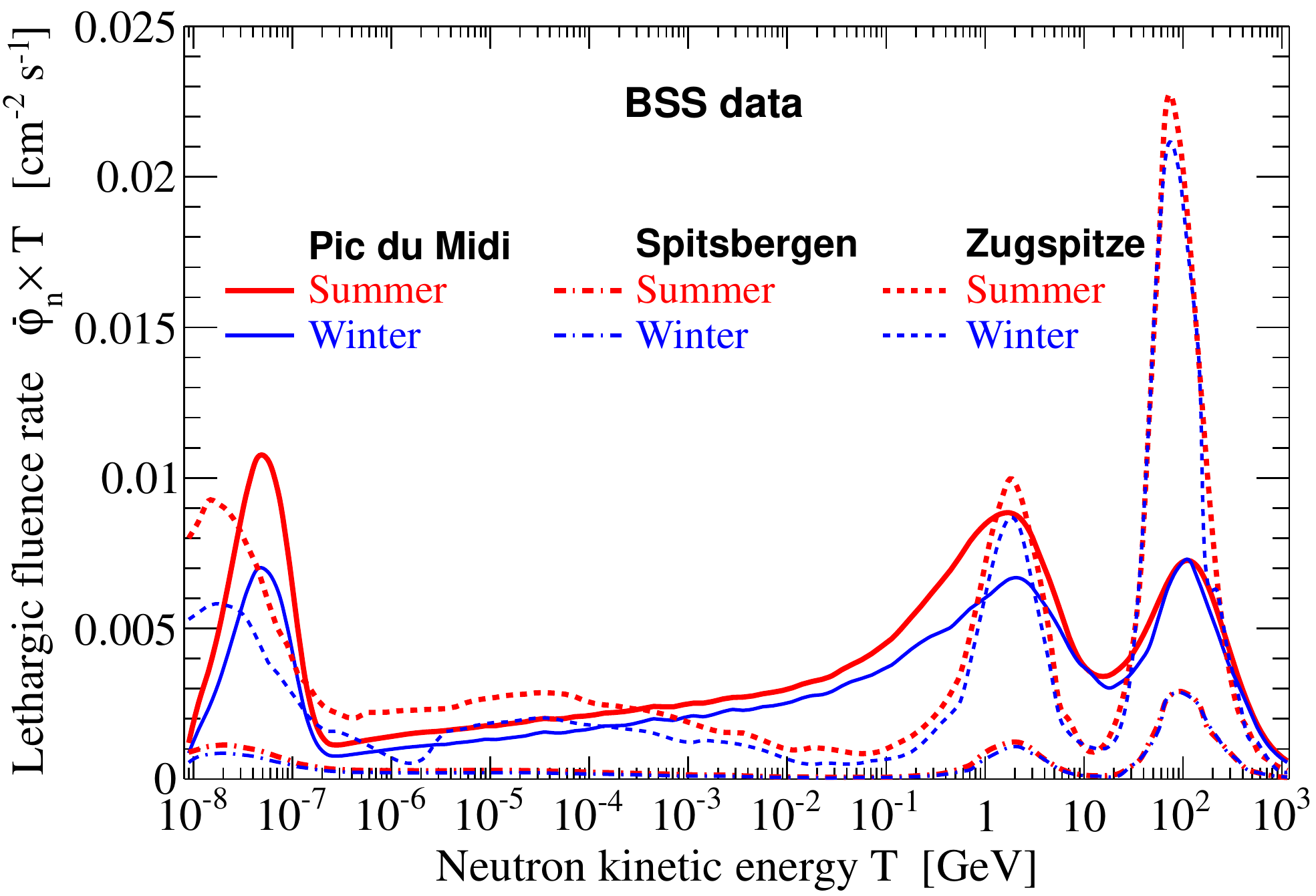}
\end{center}
\vspace{-5mm}
\caption{Neutron spectra obtained from BSS measurements at various locations. Data from Pic du midi are taken from \citetads{2013JGRA..118.7488C}, and from \citetads{2012JGRA..117.8309R} for the other stations.}
\label{fig:BSS}
\vspace{-3mm}
\end{figure}

As an illustration, Fig.~\ref{fig:BSS} shows typical spectra obtained for the three above-mentioned BSS in summer and winter. The decrease of intensity for low and intermediate energies is clearly visible in winter, when heavy snow falls occur in the northern hemisphere. 

\begin{figure}[!t]
\begin{center}
\includegraphics[width=\columnwidth]{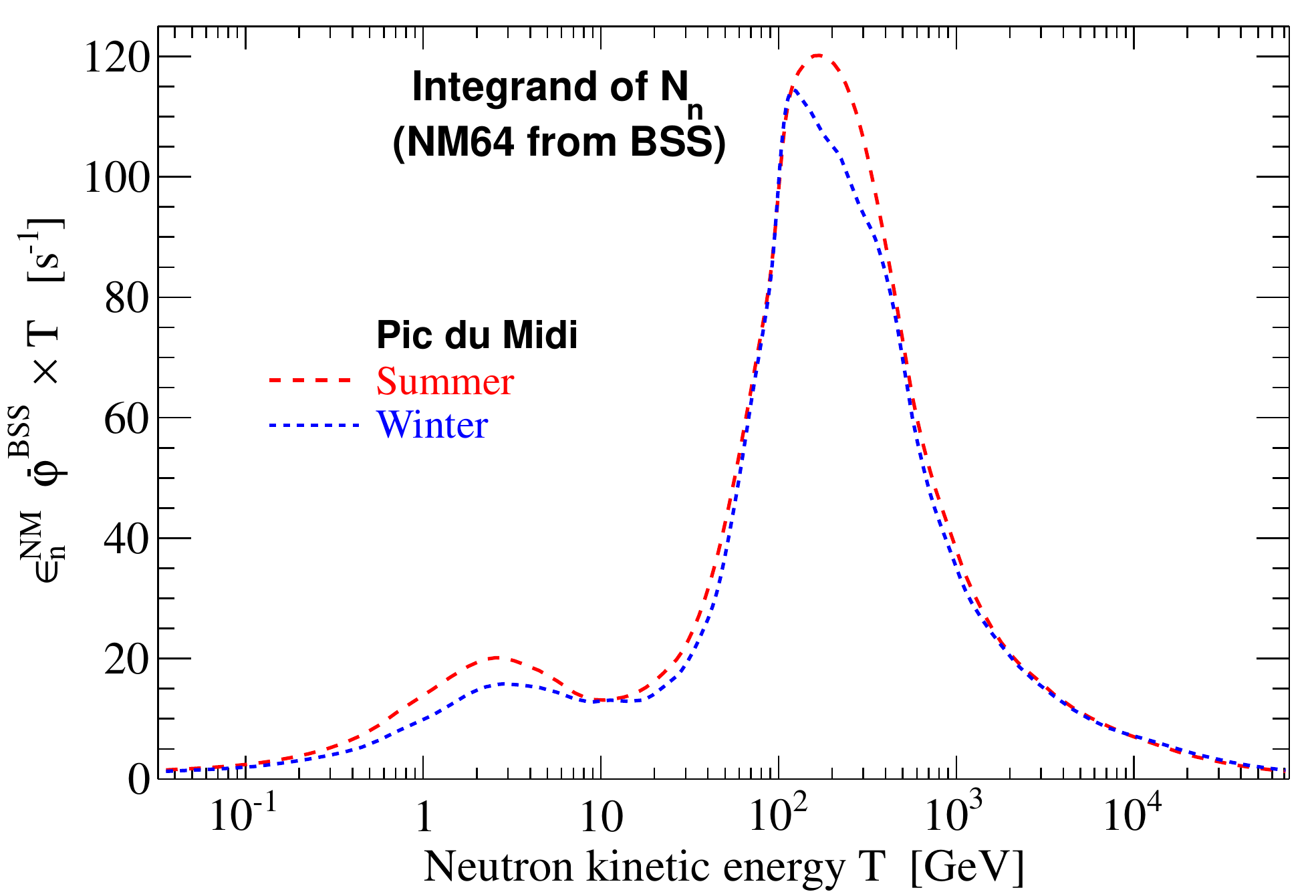}
\end{center}
\vspace{-5mm}
\caption{Integrand of Eq.~(\ref{eq:BSStoNM}) for the Pic du Midi case in summer and winter. The area under this function gives directly the expected count rates in NMs for the two seasons.}
\label{fig:NMSeasonal}
\vspace{-3mm}
\end{figure}

\paragraph{NM64 seasonal effect from BSS data}
As explained in Sect.~\ref{sec:BSS}, it is possible to derive the count rates $N_n$ due to neutron in NMs thanks to Eq.~(\ref{eq:BSStoNM}). We first show in Fig.~\ref{fig:NMSeasonal}, for the Pic du Midi case, the product of the NM64 response and the neutron spectra as a function of kinetic energy $E_n$. Although the majority of counts are due to cascade neutrons (above 20 MeV), evaporation neutrons are non negligible, and for both energy regions, the difference between summer and winter is significant.

For NM64, the value of the count rate variations $\Delta N/N|_{\text{Snow}}$ due to seasonal effect are calculated taking into account the fact that neutrons constitute 87\% of the total NM count rate. The results for the three stations are gathered in the lower half of Table~\ref{tab:BSS}: the effect varies between -1.8\% and -7.6\%. These estimations are consistent with data provided by, e.g.,  \citetads{1968Natur.219..926T} and \citetads{2011Ge&Ae..51..247K}, with a variation about -5\% recorded at NM station of Oulu, and -7\% in Rome. They are also in agreement with the results of \citetads{2012JGRA..117.8309R} based on Zugspitze and Spitsbergen data.

This confirms that snow has a significant and seasonal impact on NM count rates in stations that might know intense snow fall episodes (particularly at mountain altitudes). Indeed, recent effort are directed into having real-time and automated corrections for the snow effect in NM64 data \citepads{2011Ge&Ae..51..247K,2013JGRA..118.6852K}. After this correction, these authors estimate a residual error of $\sim 0.4\%$.

\subsection{Other effects}
\label{sec:calib}

NM count rates depend on the detector surroundings and the atmosphere state, but they also depend on the reliability and stability of the equipment. To improve further the usefulness of the NM network, inter-calibration of all stations is required. Portable calibration NMs were discussed in \citetads{2001ICRC...10.4083M}, built soon after \citepads{2003ICRC....6.3453M}, and several tests and validation carried out \citepads{2005ICRC....2..477K,2010AdSpR..46.1394K,2011ICRC...11..340K,2008JGRA..113.8101K,2013JPhCS.409a2171K}. Note that the target goal for the calibrator was to reach an accuracy of $\le 0.2\%$ (for spectral studies), which succeeded, as reported in \citetads{2010AdSpR..46.1394K}. However, during the tests, several effects were measured, that are of importance and amount somehow to uncertainties.

First, the instrumental temperature effect (not related to the atmospheric temperature effect) was recently reconsidered by \citetads{2008JGRA..113.8101K}, who measured a $\sim 0.05\%/^\circ$°C change for NM64. However, this should not impact count rates as long as the detectors are kept in a small temperature range. More worrisome is the fact that different local conditions lead to an unpredictable spread of $\sim 4\%$ \citepads{2010AdSpR..46.1394K}. Then the exact geometry of the detector \citepads{1964CaJPh..42.2443H}, whether it contains $n$ or $m$ tubes also slightly changes the efficiency of the detector (and {\em in fine} the yield function and count rates at various latitude): effects up to a few percent can exist \citepads{1964CaJPh..42.2443H,2003ICRC....6.3441K}, and in particular, differences up to $4\%$ were observed between the calibrator and a 3-NM64 \citepads{2013JPhCS.409a2171K}. The last two issues may explain the need of detrending NM data in order to reach a coherent picture of solar activity for the various stations \citepads{2013JGRA..118.5431O}.

Finally, anisotropy effects (e.g., diurnal and semi-diurnal variations) also exist, but are beyond the scope of this paper. Their amplitudes depend on many parameters (species measured, location of the detector, etc.), which complicates the study of ground-level events. We refer the interested reader to \citetads{2004ASSL..303.....D}.

\section{Conclusions: count rate variation and uncertainty iso-contours in the $(R_c,\,\phi)$ plane}
\label{sec:concl}

We have made a detailed study of count rates (and uncertainties) for neutron monitors and $\mu$ detectors, as a function of the rigidity cut-off and the modulation level $\phi$, in the context of the force-field approximation.

\subsection{Input parameters}
First, we have re-assessed (and compared with previous results from the literature) two key ingredients entering the calculation, namely IS fluxes and yield functions.
\begin{enumerate}[i)]
  \item{Results for IS fluxes:}
  \begin{itemize}
     \item we propose a new set of IS flux parametrisations for elements $Z=1-28$, see Eq.~(\ref{eq_is_flux_dekn}) and Table~\ref{tab:jis_fit};
     \item we improve the calculation of the factor accounting for heavy species ($Z\geq3$) as an extra contribution of \hef{} (for NM and $\mu$ detector count rate calculations). The required extra amount of \hef{} is found to be $0.611^{+0.016}_{-0.009}$ \hef{} (to be compared with 0.480 used in previous studies). We check that making the substitution is accurate at better than the percent level over the whole rigidity range;
     \item as previously studied in \citetads{2010JGRA..11500I20H}, it is always possible to recover the same TOA fluxes, starting with different combinations of IS fluxes and solar modulation parameter $\phi$ (degeneracy between $\phi$ and the IS flux). Equation~(\ref{eq:recipe}), to be used with Fig.~\ref{fig:jis_phishift}, provides a simple recipe to move from one $\phi$ time series to another, depending on the choice of the IS flux (all formulae are gathered in app.~A);
     \item we evaluate the uncertainty on TOA \proton{} and \hef{} fluxes (see Fig.~\ref{fig:uncertainty_toa}), directly from the fit (of our reference flux) to the data and their errors, or from the dispersion of TOA fluxes obtained with the use of several parametrisations of the IS flux (modulated at their appropriate value, as underlined above). We arrived at a 5\% uncertainty for the former, and a probably overestimated $\sim 10-20\%$ dispersion (energy dependent) for the latter.
  \end{itemize}
\item{Results for the yield functions:}
  \begin{itemize}
     \item we propose a new yield function parametrisation Eq.~(\ref{eq:yield_fit_scaling}) for primary CR protons and heliums, evaluated for NM64 in Table~\ref{tab:best-fit_yield} and for $\mu$ detectors in Table~\ref{tab:best-fit_yieldmuon};
     \item we provide a systematic comparison of available yield functions in the literature (see Fig.~\ref{fig:ratio_yields}, all formulae are gathered in App.~B). Differences of a factor of a few exist around a few tens of GV, these differences increasing at lower and higher rigidity. A better agreement at high energy is obtained when accounting for the geometrical correction factor of \citetads{2013JGRA..118.2783M};
     \item after renormalisation to a reference rigidity, the dispersion for the various yield functions can be used to estimate the uncertainty on count rates (see below). 
  \end{itemize}
\end{enumerate}

\subsection{Count rates and uncertainties}
Using these inputs, we have been able to characterise the count rate dependence on several parameters and related uncertainties.
  \begin{itemize}
    \item for polar stations, 90\% of the count rates are initiated by CRs above 5~GV for NM64 and above 10~GV for $\mu$ detectors (see Fig.~\ref{fraction_per_rbin});

    \item we validate NM64 yield functions against latitude surveys in two steps:
       \begin{enumerate}[a)]
         \item we derive the solar modulation level (from CR data, based on our reference IS flux) at the time of these surveys (minimum activity)|see top panel of Fig.~\ref{fit_hhe_epoch_nmsurvey}. We find $\phi_{\rm min}\approx470$~MV, a value slightly higher but in agreement with the value used in other works using these same surveys. 
         \item a comparison of various yield functions from the literature confirms that the geometrical correction factor proposed in \citetads{2013JGRA..118.2783M} is mandatory to better fit NM survey data. This effect and the energy dependence of MC yield function calculations in the 100~GV$-$1~TV should be further explored. When this correction is applied to all MC-based yield function (as opposed to yield functions derived from NM data surveys), a consistent picture emerges, with all modelling in fair agreement with one another. Rather unexpectedly, a slightly better fit to all the survey data (up to a rigidity cut-off of 10~GV) is given by the new yield function we propose.
       \end{enumerate}

    \item We propagate the uncertainties obtained for the IS flux and yield function to the calculated count rates:
       \begin{enumerate}[a)]
          \item a $\phi$ and $R_c$ independent uncertainty of 2\% (resp. $6-8\%$) is related to IS flux from data uncertainty (resp. from IS flux model dispersion), see top panel of Fig.~\ref{rate_err_yieldjis}. The scaling factor (for $Z>2$ species) uncertainty leads to another $0.6\%$. This applies to NM64 as well as to $\mu$ detectors;
          \item an $R_c$ dependent (and slightly dependent on $\phi$) uncertainty smaller than $2-4\%$ for $R_c<10$~GV (resp. $6-8\%$ for $R_c<18$~GV) is related to NM64 yield function dispersion, see bottom panel of Fig.~\ref{rate_err_yieldjis}. This uncertainty is smaller than 0.2\% for a $\mu$ detector. 
       \end{enumerate}

    \item We revisit the uncertainties related to the transmission function of CRs in the geomagnetic field. Focusing on effective vertical rigidity cut-off below $R_c=10$~GV (where most stations lie), we reach the following conclusions:
       \begin{enumerate}[a)]
          \item using a sigmoid function instead of a step function gives $\sim 4\%$ less count rates for NM64, and $\sim 2\%$ less for $\mu$ detectors. This effect is $R_c$ dependent, and maximal for large $R_c$ values (see bottom panel of Fig.~\ref{rate_err_rctransm});
          \item even in the step-function approximation, the count rate variation is expected to change due to long-term or short term geomagnetic variations. We evaluate that over 50 years a typical decrease of $\sim 4\%$ for NM64 (and $\sim 1\%$ for $\mu$ detectors) can occur (see top panel of Fig.~\ref{rate_err_rctransm}). The level of the variation depends on the geomagnetic position and $R_c$;
          \item the use of the apparent cut-off rigidity of \citetads{1997JGR...10226919C} and \citetads{2008AdSpR..42..510D} accounting for obliquely incident particles (in the geomagnetic field) is found to have an impact of $\lesssim 2-4\%$ on the  NM64 count rate and $\lesssim 1-2\%$ on $\mu$ detectors). As above, the effect depends on the geomagnetic position and $R_c$.
       \end{enumerate}

   \item We recap the various seasonal effects and their impact on count rates. First, muon detector data are dominated by temperature effects: the corresponding count rate variation is $\sim 8\%$ but corrections (which are seldom implemented in distributed data) are able to bring this variation down to $\sim 0.3\%$ \citepads{2011APh....34..401D}. Second, for NM64, all the following effects must be considered:
       \begin{enumerate}[a)]
          \item atmospheric pressure and temperature effects ($\sim 10\%$ level) are routinely corrected for in public data. The level of variation left after this correction is $\lesssim 0.5\%$ (pressure) and $\lesssim 1\%$ (temperature);
          \item water vapour is expected to lead to a $\sim 0.2-0.3\%$ effect;
           \item the effect of snow coverage in the surrounding of the detector is investigated by means of BSS measurements whose low energy spectrum is very sensitive to it. We obtain a $2-8\%$ seasonal variation for this effect (obviously strongly dependent on the climatic conditions at the station location), in agreement with direct measurements in NM stations. Recent efforts by \citetads{2011Ge&Ae..51..247K,2013JGRA..118.6852K} to provide real-time data corrected for this effect are an important step for the network of NMs around the world.
       \end{enumerate}

   \item Finally, some uncertainties are intrinsic to the detector itself, as thoroughly investigated by means of a calibrator \citepads{2010AdSpR..46.1394K}. These authors find a spread $\sim 4\%$ in their measurements  attributed to local conditions, but it may be even larger for some stations. However, such effects, along with differences attributed, e.g., to the exact geometry of the detector \citepads{1964CaJPh..42.2443H}, are not expected to change in time, and thus are probably not as problematic as seasonal effects.
   
   \end{itemize}
\begin{figure*}[!t]
\begin{center}
\includegraphics[width=0.995\columnwidth]{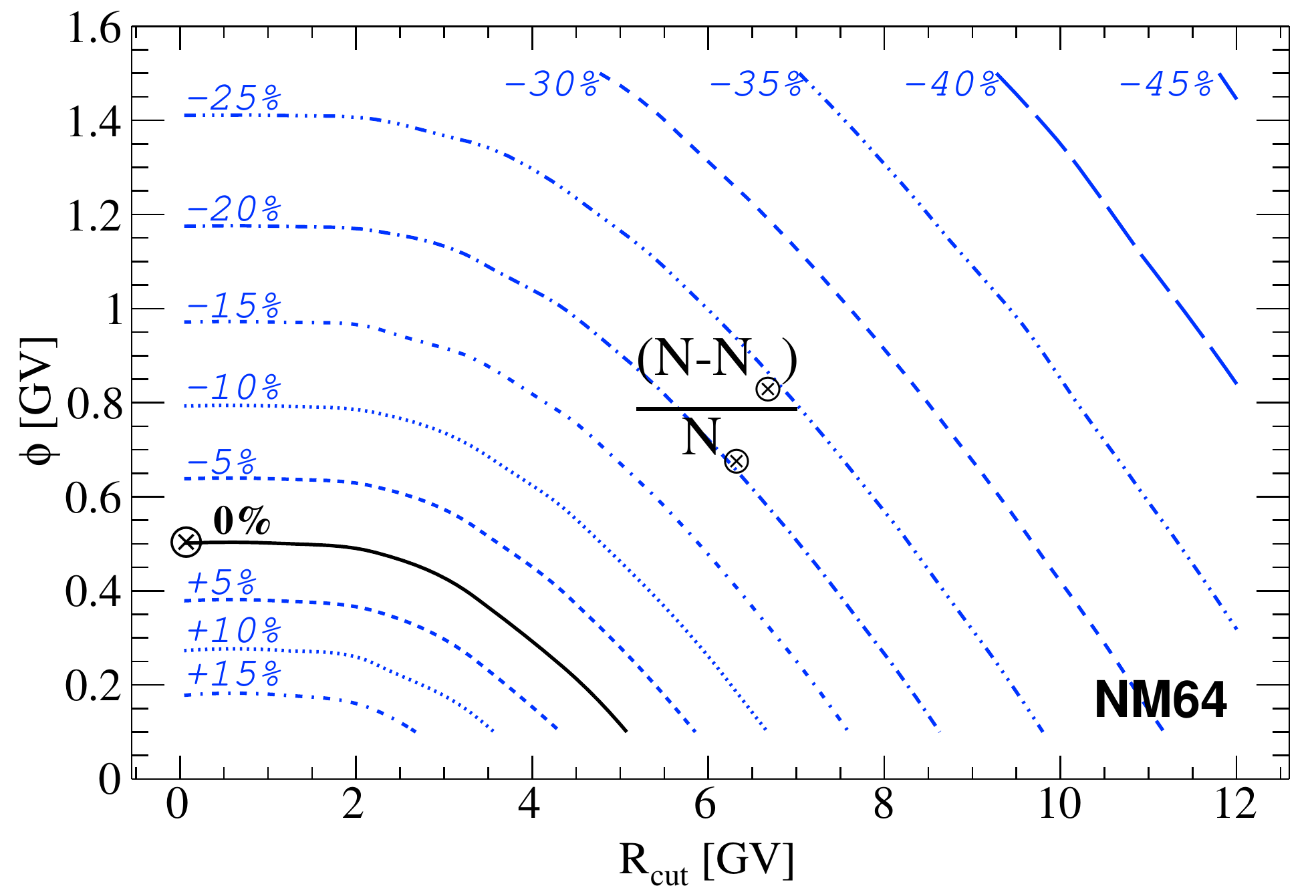}
\includegraphics[width=0.995\columnwidth]{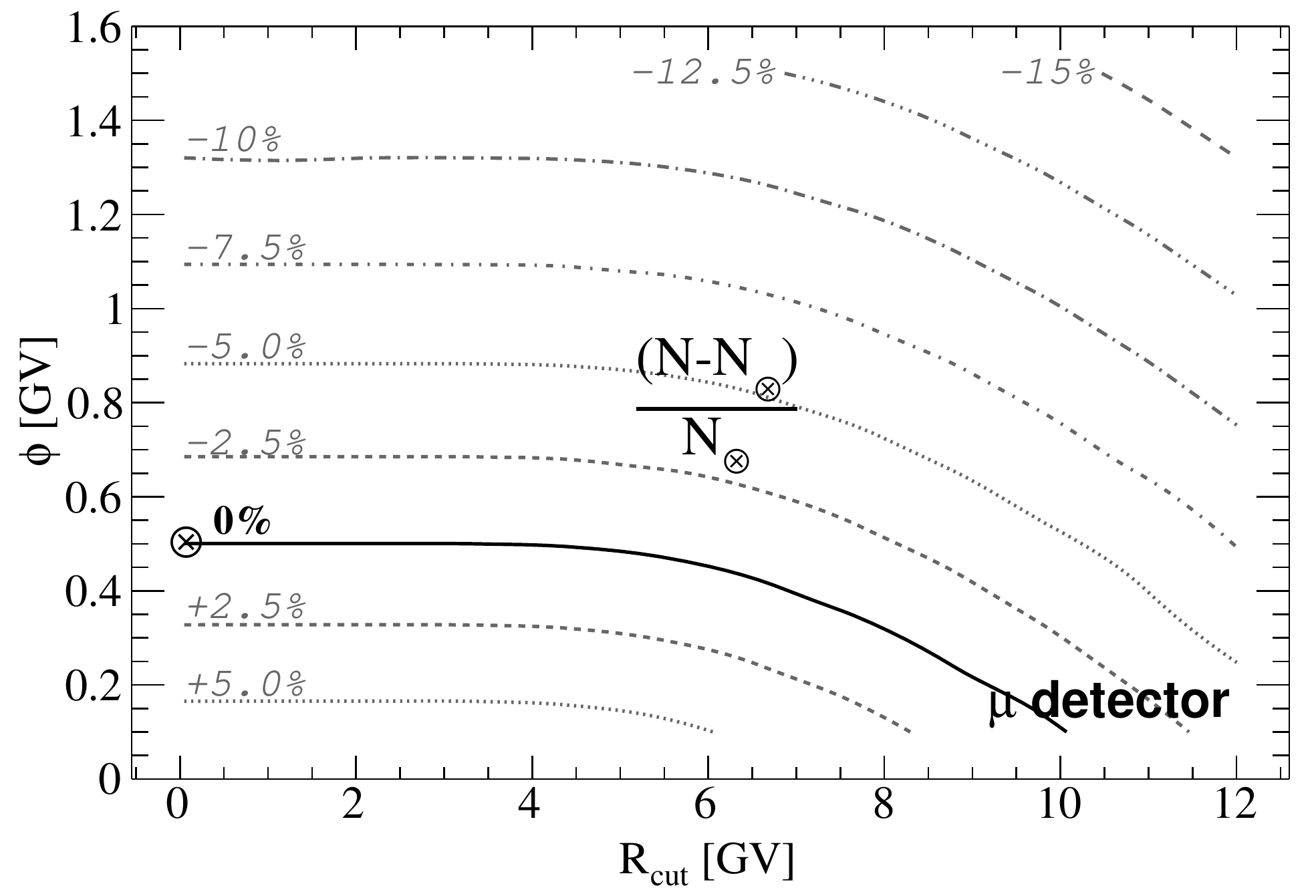}
\includegraphics[width=0.995\columnwidth]{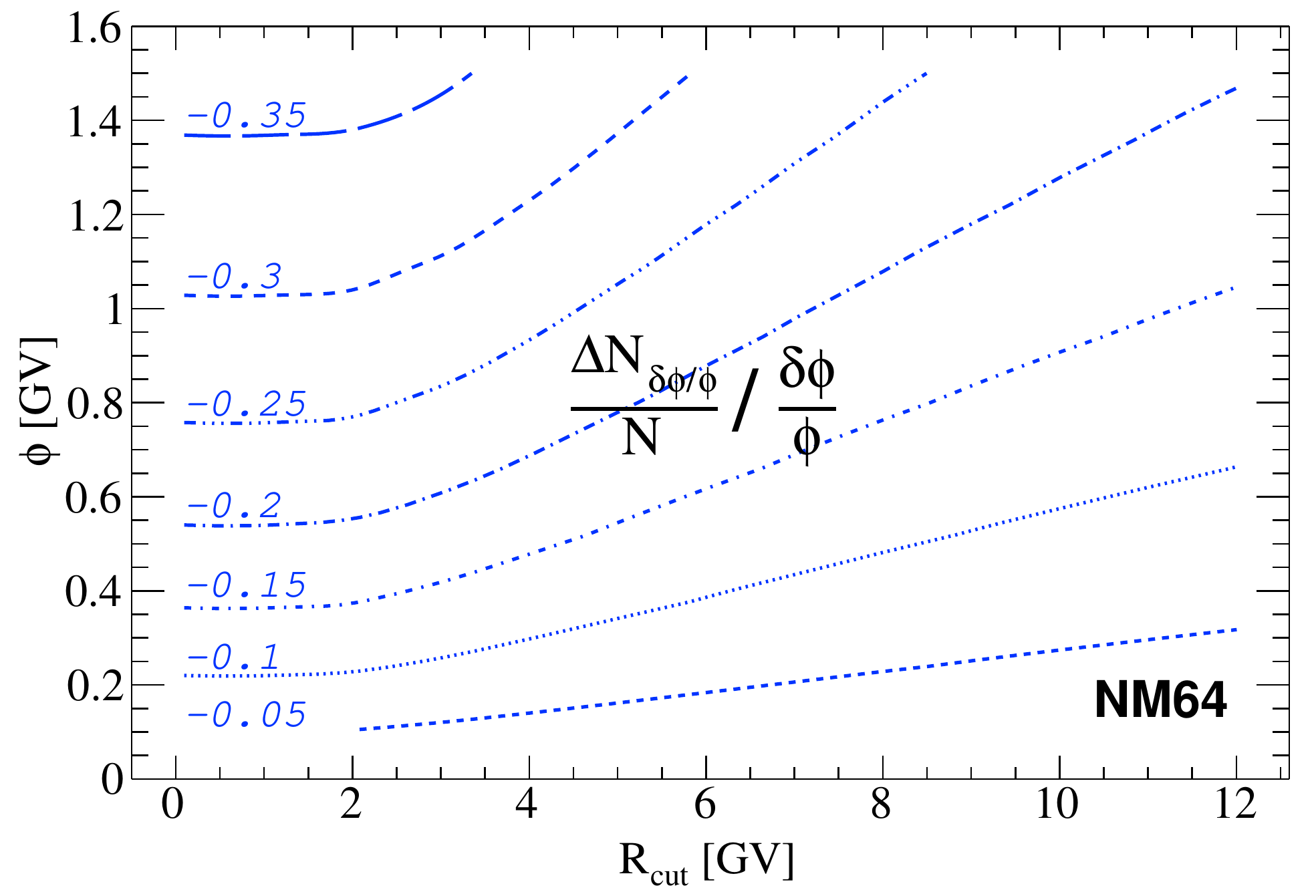}
\includegraphics[width=0.995\columnwidth]{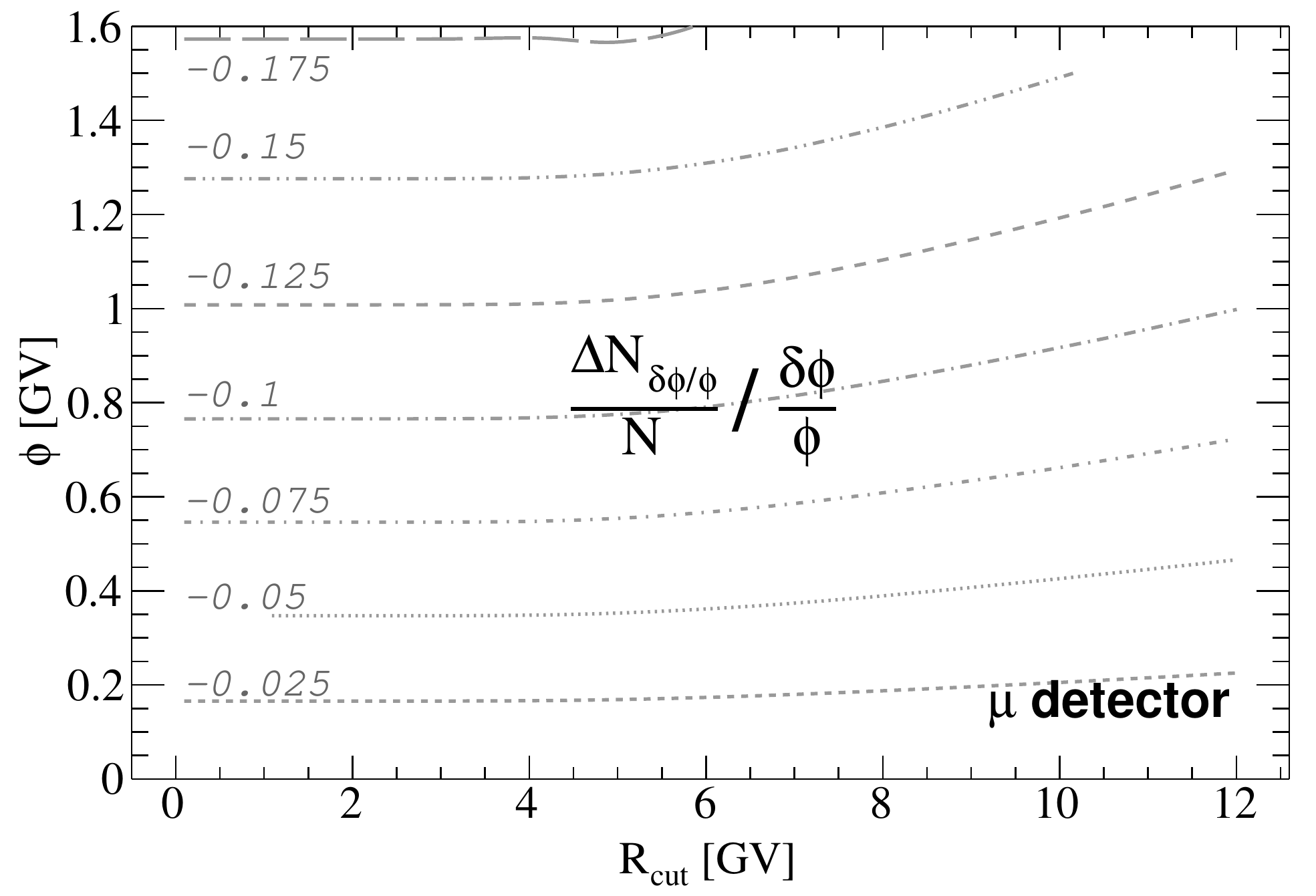}
\end{center}
\vspace{-4mm}
\caption{Left panels are calculated for NM64 and right panels for $\mu$ detectors. {\bf Top panels:} Count rate relative variation $\Delta N/N_{\otimes}$ with respect to a reference count rate $N_{\otimes}=N(\phi=0.5,\,R_c=0)$. The relative variation (in \%) are shown as iso-contours in the plane $(R_c,\,\phi)$, with the 0\%-isocontour (passing through the reference point $\otimes$) in black solid line. {\bf Bottom panels:} scaling factor $f$ to infer the count rate relative variation $[\Delta N_{\delta\phi/\phi}/N](R_c,\,\phi)$ for any modulation relative change $\delta\phi/\phi$ (for NM64, slight differences in the contours arise for values of $\delta\phi/\phi>20\%$ and factor $f<-0.3$). For instance, for a NM64 detector at $R_c=6.5$~GV and a solar modulation period $\phi=1.2$~GV, the scaling factor is $f=-0.25$, which reads: an increase of 5\% (resp. 10\%) in the modulation level $\phi$ translates in a decrease of $0.25\times5\%=1.25\%$ (resp. 2.5\%) in the detector count rate, and vice versa.}
\label{deltan_vs_phi}
\vspace{-3mm}
\end{figure*}

\begin{table*}[!t]
\caption{Impact of different effects/input ingredients ($1^{\rm st}$ and$2^{\rm nd}$ columns) on the relative count rate calculation for NMs and $\mu$-like detectors ($4^{\rm th}$ and $5^{\rm th}$ columns), and on the uncertainty associated to the derived modulation parameter value ($6^{\rm th}$ and $7^{\rm th}$ columns). The first two rows, in \textit{italic}, serve as a reference: they correspond to the maximum variation expected between periods of low and high solar activity, and low and high rigidity cut-off).
The $3^{\rm rd}$ column provides the figure or section where the effect is discussed in the paper. See text for description.}
\vspace{-3mm}
\label{tab:summary_table}
\begin{center}
\begin{tabular}{cccccccc}\hline
& & & & & & &\vspace{-3mm}\\
Ingredient      &         Effect           &\!\!\!\!\!\!\!\!Fig./Sect.\!\!\!\!\!\!\!\!\!\!\!\!\!\!& \multicolumn{2}{c}{$\displaystyle\frac{\Delta N}{N}$}& \multicolumn{2}{c}{$\Delta\phi^\ast$ [MV]} & Comment \\
                &                          &                       &      NM        &    $\mu$      &    NM   & $\mu$    &\vspace{2mm}\\
\!\!\!\!\!{\em Solar modulation}\!\!\!\!\!&  $\phi\in$~{\em [0.2,1.5]~GV} &{\em Fig.\ref{deltan_vs_phi}}&{\em [+15,-25]\%}&{\em [+5,-10]\%}&{\em -} &{\em -}& {\em w.r.t. $\mathit{\phi\!=\!0.5}$~GV}\\
\!\!\!{\em Cut-off rigidity}\!\!\!&  $R_c\in$~{\em [0,10]~GV}& {\em Fig.\ref{deltan_vs_r}}  &   {\em [+10,-20]\%}  &  {\em [0,-5]\%}   & {\em -} &{\em -}& {\em w.r.t. $\mathit{R_c\!=\!5}$~GV} \vspace{1.5mm}\\\hline
& & & & & & &\vspace{-2.5mm}\\
\multirow{3}{*}{TOA flux} & p and He CR data &Fig.\ref{rate_err_yieldjis}&   $\pm2\%$     &  $\pm2\%$    &  $\pm66$   & $\pm140$& $(t,\,R_c,\,\phi)$-independent\\
                & IS flux dispersion$^\P$    &Fig.\ref{rate_err_yieldjis}&   $\pm6\%$     &  $\pm8\%$    &  $\pm200$  & $\pm570$& $\Downarrow$\\
                & Heavy species              &Fig.\ref{fraction_cr}      &   $\pm0.6\%$   &  $\pm0.6\%$  &  $\pm20$   & $\pm40$ & Global norm. factor$^\diamond$\vspace{3mm}\\
\!\!\!Yield function\!\!\!  & Dispersion     &Fig.\ref{rate_err_yieldjis}&$\lesssim\pm4\%$&  $<0.2\%$    &$\lesssim120$&$\lesssim14$&$(R_c,\,\phi)$ dependent\vspace{3mm}\\
                & Sigmoid$(R_c,x\!=\!+\frac{\sigma}{0.1})$&Fig.\ref{rate_err_rctransm}&-$2x\%$&-$0.5x\%$& +$66x$& +$35x$   &For $R_c\gtrsim5$~GV\\
 Transfer      &\!\!\!\!\!$H(R_c\!\!+\!\!\Delta R_c)$: $x\!\!=\!\!\frac{\!\!(\Delta R_c\!/\!R_c)\!\!}{0.05}$&Fig.\ref{rate_err_rctransm}&-$2x\%$& -$x\%$& +$66x$  & +$71x$& For $R_c\gtrsim5$~GV\\
function        &- $R_c(t)$: $\frac{\Delta R}{R}\!\lesssim\!+0.2\%$/yr\!\!\!\!\!\!&\S\ref{sec:Rcut}&-0.4\%/yr\!\!\!\!\!\!\!\!\!&-0.1\%/yr\!\!\!\!\!\!\!\!\!&+13/yr\!\!\!\!\!&+7/yr\!\!\!& Depends on location\\
                &\!\!\!\!\!\!- $R^{\rm eff}_c\!\rightarrow\!R^{\rm app}_c$: $\!+3\%$&\S\ref{sec:Rcut}& -1.2\%& -0.3\%  &+40~ &+21~ & Depends on $R_c$\vspace{3mm}\\
                &  Pressure                  &    \S\ref{sec:pe}         &   $\pm0.2\%$   &  $\pm0.2\%$  &  $\pm6$    & $\pm14$ & After correction\\
Time-dep.       &  Temperature               &    \S\ref{sec:te}         &   $\pm0.5\%$   &  $\pm4\%$    &  $\pm15$   & $\pm290^\ddagger$& Not corrected\\
  effects$^\dagger$       &  Vapour water              &    \S\ref{sec:wv}         &   $\pm0.3\%$   &  $\pm0.1\%$  &  $\pm10$   & $\pm8$  &  Not corrected\\
                &\!\!\!\!\!\!\!\!Snow coverage ($T\!\!=\!\!1$~yr)\!\!\!\!\!\!\!&    \S\ref{sec:se}         &     -7\%       &      -       &  +230      &  -      & Not corrected\vspace{3mm}\\
%
  NM detector   &  Temperature               &    \S\ref{sec:calib}      &+0.05\%/$^\circ$C\!\!\!\!\!\!&     -        & -1.5/$^\circ$C\!\!\!\!\!\!     &    -    & $(t,\,R_c,\,\phi)$-independent\\                      
  effects       &  $n$NM6 vs $m$NM64         &    \S\ref{sec:calib}      &     few \%     &     -        & $\sim 100$ &    -    & $\Downarrow$\\
                &  Surroundings (hut)        &    \S\ref{sec:calib}      &     few \%     &     -        & $\sim 100$ &    -    & Global norm. factor$^\diamond$\\\hline     
\end{tabular}\\
\vspace{1mm}
{\small 
$^\ast$ The variation $\Delta\phi$ of the modulation level is calculated for a detector at $R_c=5$~GV and $\phi=0.5$~GV:\\
        refer to Fig.~\ref{deltan_vs_phi} to convert rate variations for any other $(R_c,\,\phi$) condition.\\ 
$^\P$ Very conservative estimate (some IS fluxes are based on old CR data).\\
$^\diamond$ Global normalisation factors can always be absorbed in the yield function normalisation.\\
$^\dagger$ Distributed data are either corrected or not corrected for these effects.\\
$^\ddagger$ After correction, $\Delta N/N\sim 0.3\%$ \citepads{2011APh....34..401D}, leading to $\Delta\phi\sim 22$~MV.}
\end{center}
\vspace{-6mm}
\end{table*}

\subsection{Abacus: count rates to solar modulation variations}
To conclude, we propose a last figure and a table for a panoptic view of all the effects we have approached in this study. Actually, these plots provide a direct link between solar modulation level and count rate variations (and vice versa) for both NM64 and $\mu$ detectors.

The top panels of Fig.~\ref{deltan_vs_phi} provide the relative count rate variation in the $R_c-\phi$ plane, with respect to a reference point $N_\otimes(R_c=0,\phi=0.5$~GV). In addition to providing a global view of the expected count rate variation between detectors at different $R_c$ and for different solar periods, it also gives a flavour of the precision required in order to be sensitive to changes in the $\phi$ parameter: the count rate variation over a full solar cycle is smaller for $\mu$ detectors than for NMs, but the latter are more sensitive to any uncertainty on $R_c$ (location in the geomagnetic field) than the former.

The bottom panels of Fig.~\ref{deltan_vs_phi} go further in that direction, as they directly provide, for any value $(R_c,\,\phi)$, how much variation $\Delta N/N$ to expect in the count rates, whenever the solar modulation changes by $\delta\phi/\phi$. This abacus usage is two-folds: first, on short term variations, it can directly be used to extract $\delta\phi/\phi$ from count rate variation in NM (or $\mu$) data; second, it can be used to estimate how much uncertainty is propagated in $\phi$ from the various uncertainties calculated on count rates.

This is what is gathered in Table~\ref{tab:summary_table}: for each input/effect discussed in the paper, we provide (in addition to the section/figure where it was dealt with) the typical uncertainty obtained on $\Delta N/N$, and the associated $\Delta\phi$ calculated for an NM64 or $\mu$ detector at $R_c=5$~GV and a solar modulation level of $\phi=500$~MV (using Fig.~\ref{deltan_vs_phi}). For $\phi$ calculations, the first thing to underline is that NM and $\mu$ detectors do not suffer the same amount of uncertainties, due to different sensitivities to the various effects explored. Moreover, there are no clear-cut ranking of these errors. Luckily, when interested in time series, time-independent normalisation effects can be absorbed in a normalisation factor \citepads[e.g.,][]{2011JGRA..116.2104U}: the latter accounts for differences in NM detector efficiency and their surroundings (last entries in table). The case of the IS fluxes  (first entries in table) is peculiar, since different choices $i$ lead to an overall shift $\Delta\phi_i$ of the time series. For NMs, the main source of uncertainties are the seasonal snow effects (strength depending on position, some stations not affected), and the yield function dispersion (applicable for all stations). All other effects cannot be simply disregarded as they typically have a $5-10\%$ on $\phi$. For $\mu$ detectors, the main effect is that of the temperature variation, but after corrections, it is at the level of other uncertainties ($5-10\%$). Overall, $\mu$ detectors 
seem to suffer slightly less uncertainties than NM64, but of course the latter benefit from a much larger time and position coverage than the former.

\subsection{Future works}

The approach we have followed in this study could easily be extended to other types of ground-based measurements, by simply using the appropriate yield function for each type (e.g., $^{10}$Be production in ice cores, \citealtads{2010JGRA..11500I20H}; ionisation measurements in the atmosphere, \citealtads{2008SSRv..137..149B}; etc.). In any case, one of the main challenge of such approaches is to obtain an accurate yield function. In that respect, the efforts to improve that of NMs should be pursued, given their role in the history of solar activity monitoring.

As underlined at the beginning of this study, our primary goal is to get time series of modulation parameters, taking advantage of the complementarity of NM count rates and TOA CR flux measurements. The above Table~\ref{tab:summary_table} provides a synthetic view of the difficulties. This table, and more importantly, the characterisation of the dependence of these uncertainties with $R_c$, `weather' conditions for the stations, etc., should help decide which stations to consider to minimise the uncertainties in the $\phi$ calculations. This is the aim of our next study.

\section*{Acknowledgements}
D.~M. thanks K. Louedec for useful discussions on the Auger scaler data, A. L. Mishev and I. G. Usoskin for clarifications on their yield function, and P. M. O'Neill for providing his BO11 flux model. We thank the anonymous referee for her/his careful reading that helped correcting several mistakes in the text.

\appendix
\label{sec:app}
\section{IS flux parametrisation (p and He)}
For completeness, we provide below all IS flux parametrisations $dJ^{\rm IS}/dT_n$ used in the paper. They are given for protons and heliums in unit of [m$^{-2}$~s$^{-1}$~sr$^{-1}$~(GeV/n)$^{-1}$]. The formulae are expressed in terms of:\\
 \indent - the rigidity $R$ (in GV)\\
 \indent - the kinetic energy per nucleon $T_n$ (in MeV/n)\\
 \indent - $\beta=v/c$.

\begin{itemize}
   \item GM75  \citepads{1975ApJ...202..265G}:
      \begin{flalign*}
           J_{\rm p}(T_n)&=9.9\,10^{11}\, \left(T_n+780\exp(-2.5\,10^{-4}\; T_n)\right)^{-2.65};&\\
           J_{\rm He}(T_n)&=0.0525 \times J_{\rm p}(T_n).&
      \end{flalign*}   

   \item B00+U05 \citepads{2000JGR...10527447B,2011JGRA..116.2104U}:
      \begin{flalign*}
           J_{\rm p}(R)&=\frac{1.9\,10^4\;R^{-2.78}}{1+0.4866R^{-2.51}};&\\
           J_{\rm He}(R)&=0.38 \times J_{\rm p}(R).&
      \end{flalign*}   

   \item L03 \citepads{2003JGRA..108.8039L}: using $x=\ln(T_n)$,
      \begin{flalign*}
           J_{\rm p}(T_n)&=\!10^{3}\!
              \begin{cases}
                e^{[22.976-2.86x - 1.5\,10^{3}/T_n]} \;\;(T_n\!\!>\!\!1000);\\
                e^{[0.823-0.08x^2 + 1.105x - 0.09202\sqrt{T_n}]}.\\
              \end{cases}&\\
           J_{\rm He}(T_n)&=0.055 \times J_{\rm p}(T_n).&
      \end{flalign*}   

   \item WH03 \citepads{2003JGRA..108.1355W}:
      \begin{flalign*}
           J_{\rm p}(T_n)&=\frac{10^{13}}{1.89\; T_n^{2.8} + 5.05\,10^{4}\; T_n^{1.58} + 9.33\,10^{7} \; T_n^{0.26}};&\\
           J_{\rm He}(T_n)&=0.056 \times J_{\rm p}(T_n).&
      \end{flalign*}   

   \item S07 \citepads{2007APh....28..154S}:
      \begin{flalign*}
           J_{\rm p}(R)&=1.94\cdot10^4 \times \beta^{0.7}R^{-2.76};&\\
           J_{\rm He}(R)&=0.71\cdot10^4 \times \beta^{0.5}R^{-2.78}.&
      \end{flalign*}   

   \item WH09 \citepads{2009JGRA..11402103W}: using $x=\ln(T_n)$,
      \begin{flalign*}
           J_{\rm p}(T_n)&=\!10^3\!
              \begin{cases}
                  e^{-51.68\ln(x)^2 + \frac{103.6}{x^{-1/2}}-\frac{709.7}{x}+\frac{1162.}{x^2}} \;\;(T_n\!\!>\!\!1000);\\
                  e^{-124.5-51.84\ln(x)^2 + \frac{131.6}{x^{-1/2}}-\frac{241.7}{x}+\frac{376.7}{x^2}}.\\
              \end{cases}&\\
           J_{\rm He}(T_n)&=0.0605 \times J_{\rm p}(T_n).&
      \end{flalign*}

   \item BO11\! (\citealtads{1996AdSpR..17....7B,2006AdSpR..37.1727O}; \citealt{2010ITNS...57.3148O}):\!\!\!
      \begin{flalign*}
           J_{\rm p}(T_n)&=4.502\cdot10^{12} \times \beta^{-1.7708}\;(T_n+938)^{-2.7748};&\\
           J_{\rm He}(T_n)&=2.472\cdot10^{11} \times \beta^{-2.2502}\;(T_n+938)^{-2.7796}.&
      \end{flalign*}   

   \item This paper $\equiv J_{\rm ref}$:
      \begin{flalign*}
           J_{\rm p}(R)&=2.335\cdot10^4 \times \beta^{1.1}R^{-2.839};&\\
           J_{\rm He}(R)&=0.7314\cdot10^4 \times \beta^{0.77}R^{-2.782}.&
      \end{flalign*}   
\end{itemize}

\section{Yield for NM and $\mu$ detectors}
Yield functions are given below for protons $Y_{\rm p}$, and in some cases for heliums $Y_{\rm He}$. If not available, a rescaled version of $Y_{\rm p}$ is used, see Eq.~(\ref{eq:yield_fit_scaling}). The yield functions are given in unit of [m$^{2}$~sr], and are expressed in terms of\\
  \indent - the grammage $g$ (in g~cm$^{-2}$)\\
  \indent - the altitude $h$ (in m) w.r.t. to sea level\\
  \indent - the rigidity $R$ (in GV)\\
  \indent - the kinetic energy $T$ (in GeV)\\
  \indent - the kinetic energy per nucleon $T_{/n}$ (in GeV/n)\\
  \indent - $\gamma=E_{\rm tot}/m$.

\subsection{NM64 yield parametrisations}
The yields are given for a 6NM64 neutron monitor. For an $x$NM64 device, the yield functions below are multiplied by $x/6$.

Note that \citetads{2013JGRA..118.2783M} proposed a correction factor $G$ to account for the geometrical factor of the NM effective size (in the context of yield functions calculated in MC simulations). This correction, fitted on their Fig.2, reads:
  \begin{equation}
    G = \max \left[1,\,1.4\times \log_{10}\left(\frac{T_{/n}}{2.39}\right)\right].
    \label{eg:g-factor}
    \end{equation}
To apply this correction, the below MC-based yield functions (i.e. CD00, F08, M09) can simply be multiplied by $G$. This correction is already accounted for in M13, and does not intervene in NM `count rate'-based formulae (N89, CL12).

\begin{itemize}
   \item N89  \citepads{1989NCimC..12..173N,1990ICRC....7...96N}: using $x=\frac{g}{1033}$,
      \begin{flalign*}
        {\cal Y}_{\rm p}(g,\gamma)&=10^4\, \exp\left(-2.2 \,x^{1.62} - \frac{12.7 \,x^{0.5}}{\ln(\gamma)^{0.42}}\right).&
      \end{flalign*}   

   \item CD00  \citepads{1999ICRC....7..317C,2000SSRv...93..335C}: the fit is adapted from \citetads{2012JGRA..11712103C},
      \begin{flalign*}
        {\cal Y}_{\rm p}^{\rm asl}(R)&=2\, 10^{-2}\, \left(0.45^{1.4}+R^{1.4}\right)^{-20.79}R^{30}.&
      \end{flalign*}
      A power-law extrapolation of slope $0.5$ is used above $R>100$~GV.

   \item F08  \citepads{2008ICRC....1..289F}:
      \begin{flalign*}
           {\cal Y}_{\rm p}(g,R)&=10^{\sum_{i=0}^4\sum_{j=0}^4 a_{ij} \, g^i \, (\log_{10}(R))^j}.&
      \end{flalign*}   
      with 
      \begin{equation}
      \begin{cases}
        a_{0j}=~~~~~~~~~~\{0.7983,\, 2.859,\, -2.060,\, 0.5654\};\\
        a_{1j}=10^{-2}\times\{-0.6985,\, 1.188,\, -0.9264,\, 0.2169\};\\
        a_{2j}=10^{-5}\times\{0.3593,\, -1.516,\, 1.522,\, -0.4214\};\\
        a_{3j}=10^{-9}\times\{-1.950,\, 7.969,\, -8.508,\, 2.491\}.
      \end{cases}
      \end{equation}
      A power-law extrapolation of slope $0.5$ is used above $R>100$~GV.

   \item M09  \citepads{2009JGRA..114.8104M,matthia_thesis}: using $I=\log_{10}(T)+1.94$ and $J=\log_{10}(T)+2.2$, 
      \begin{flalign*}
         {\cal Y}_{\rm p}^{\rm asl}(R)&=2\pi\, 10^{-4} \times 10^{-44.40-22.62I+1.007I^2+61.01\sqrt{I}};&\\
         {\cal Y}_{\rm He}^{\rm asl}(R)&=2\pi\, 10^{-4} \times 10^{-50.78-16.32J+0.409J^2+56.44\sqrt{J}}.&
      \end{flalign*}   
      Above $T_{/n}>500$~GeV/n, a power-law extrapolation is used (based on two points calculated at 450 and 500 GeV/n).

   \item CL12  \citepads{2012JGRA..11712103C}:
      \begin{flalign*}
        {\cal Y}_{\rm p}^{\rm asl}(R)&=2\, 10^{-2}\, \left(2^{1.45}+R^{1.45}\right)^{-4.696}R^{7.7};&\\
        \frac{{\cal Y}_{\rm He}}{{\cal Y}_{\rm p}}^{\rm asl}(R)&=2\, \left(0.45^{1.4}+R^{1.4}\right)^{-7.143}R^{10}.&
      \end{flalign*}

   \item M13  \citepads{2013JGRA..118.2783M}: using $I=\log_{10}(R)+0.469$ and $J=\log_{10}(R)+0.1$, we fit their data (Table~1) with
      \begin{flalign*}
         {\cal Y}_{\rm p}^{\rm asl}(R)&= 10^{-19.86-13.79I+0.963I^2+30.56\sqrt{I}};&\\
         {\cal Y}_{\rm He}^{\rm asl}(R)&=4 \times 10^{-19.83-13.81J+0.971J^2+30.58\sqrt{J}}.&
      \end{flalign*}   

   \item This paper: Eq.~(\ref{eq:yield_fit}) and Table~\ref{tab:best-fit_yield}. A power-law extrapolation is used above $R>200$~GV (based on two points calculated at 150 GV and 200 GV).
     
\end{itemize}

\subsection{$\mu$ detector yield parametrisations}
\begin{itemize}
   \item PD02  \citepads{2002JGRA..107.1376P}: the fit is from \citetads{2012JGRA..11712103C}.
      \begin{flalign*}
        Y_{\rm p}^{\rm asl}(R)&=1.07\cdot10^{-1}\, \left(1+R\right)^{-27.15}R^{28}.&
      \end{flalign*}   

   \item This paper (Eq.~\ref{eq:yield_fit} and Table~\ref{tab:best-fit_yieldmuon}): using $I=\log_{10}(T_{/n})+2.068$,
      \begin{flalign*}
        Y_{\rm p}(h,T_{/n})&=e^{0.00025h}\times 10^{(0.9116I-664.1I^{-5.818}-2.755)}.&
      \end{flalign*}   

\end{itemize}

\bibliography{nm_systematics}

\end{document}